\documentclass[lettersize,journal]{IEEEtran}
\usepackage[table]{xcolor} 
\usepackage{amsmath,amsfonts}
\usepackage{algorithmic}
\usepackage{algorithm}
\usepackage{array}
\usepackage[caption=false,font=normalsize,labelfont=sf,textfont=sf]{subfig}
\usepackage{textcomp}
\usepackage{stfloats}
\usepackage{url}
\usepackage{verbatim}
\usepackage{graphicx}
\usepackage{tikz}
\usepackage{siunitx}
\usepackage{multirow} 
\usepackage[colorlinks,linkcolor=blue]{hyperref}  
\usepackage{amsmath, amssymb, graphicx, color, booktabs, tabularx, url} 
\definecolor{lightgray}{gray}{0.9} 
\usetikzlibrary{shapes.geometric, shapes.misc, arrows.meta, positioning, calc, fit}
\usepackage{times} 
\usetikzlibrary{positioning}
\usepackage[numbers,sort&compress]{natbib}
\def\BibTeX{{\rm B\kern-.05em{\sc i\kern-.025em b}\kern-.08em
    T\kern-.1667em\lower.7ex\hbox{E}\kern-.125emX}}
\newcolumntype{C}{>{\centering\arraybackslash}X}

\begin{document}

\title{Deep Learning for Personalized Binaural Audio Reproduction}

\author{Xikun Lu,
        \and
        Yunda Chen,
        \and
        Zehua Chen,
        \and
        Jie Wang,
        \and
        Mingxing Liu, \\
        \and
        Hongmei Hu,
        \and
        Chengshi Zheng,
        \and
        Stefan Bleeck,
        \and
        Jinqiu Sang
        
\thanks{Corresponding author: Jinqiu Sang. E-mail: jqsang@mail.ecnu.edu.cn; Chengshi Zheng. E-mail: cszheng@mail.ioa.ac.cn; Hongmei Hu. E-mail: hongmei.hu@uni-oldenburg.de}
\thanks{Xikun Lu is with the Lab of Artificial Intelligence for Education, East China Normal University, Shanghai 200050, China.}

\thanks{Yunda Chen is with the Guangdong Key Laboratory of Intelligent Information Processing, College of Electronics and Information Engineering, Shenzhen University, Shenzhen 518060, China.}

\thanks{Zehua Chen is with the Department of Computer Science and Technology, Tsinghua University, Beijing 100190, China.}

\thanks{Jie Wang is with the School of Electronics and Communication Engineering, Guangzhou University, Guangzhou 511400, China.}

\thanks{Mingxing Liu and  Jinqiu Sang are with the School of Computer Science and Technology, East China Normal University, Shanghai 200050, China.}

\thanks{Hongmei Hu is with the Medizinische Physik, Carl von Ossietzky University of Oldenburg, Oldenburg 26129, Germany.}

\thanks{Chengshi Zheng is with the Institute of Acoustics, Chinese Academy of Sciences, Beijing 100045, China.}

\thanks{Stefan Bleeck is with the Institute of Sound and Vibration Research, University of Southampton, Southampton SO16 3HH, United Kingdom.}

}

\markboth{Journal of \LaTeX\ Class Files,~Vol.~14, No.~8, August~2025}%
{Shell \MakeLowercase{\textit{et al.}}: A Sample Article Using IEEEtran.cls for IEEE Journals}


\maketitle

\begin{abstract}

Personalized binaural audio reproduction is the basis of realistic spatial localization, sound externalization, and immersive listening, directly shaping user experience and listening effort. This survey reviews recent advances in deep learning for this task and organizes them by generation mechanism into two paradigms: explicit personalized filtering and end-to-end rendering. Explicit methods predict personalized head-related transfer functions (HRTFs) from sparse measurements, morphological features, or environmental cues, and then use them in the conventional rendering pipeline. End-to-end methods map source signals directly to binaural signals, aided by other inputs such as visual, textual, or parametric guidance, and they learn personalization within the model. We also summarize the field's main datasets and evaluation metrics to support fair and repeatable comparison. Finally, we conclude with a discussion of key applications enabled by these technologies, current technical limitations, and potential research directions for deep learning-based spatial audio systems.

\end{abstract}


\section{Introduction}

\IEEEPARstart{T}{he} human auditory system possesses a remarkable ability to perceive the location of sounds within an acoustic environment. This immersive experience is known as spatial sound perception \citep{moore2012introduction,blauert1997spatial}. This capability is vital for communication, navigation, environmental awareness, and creating engaging auditory environments \citep{rajguru2020spatial,xie2020spatial}. Consequently, spatial audio reproduction technology, which aims to reconstruct or simulate realistic three-dimensional (3D) sound fields, has attracted significant research interest. This technology has widespread applications in the domain of room simulation \citep{wendt2014computationally,kirsch2021low,ewert2022filter,kirsch2023universal,pulkki2019machine,kirsch2023computationally,pelzer2014integrating,brinkmann2019round,savioja2015overview,seeber2017interactive,schroder2011physically}, extended reality (XR), encompassing virtual reality (VR) \citep{schissler2016efficient}, augmented reality (AR) \citep{ranjan2015natural}, and mixed reality (MR) \citep{Larsson2010}.

Spatial audio reproduction techniques generally fall into three categories based on their implementation approaches and goals \citep{xie2020spatial}. The first category aims to physically reconstruct sound fields with high accuracy, including wave field synthesis (WFS) \citep{berkhout1993acoustic} and higher-order ambisonics (HOA) \citep{daniel2003further}. These methods can theoretically reproduce sound fields accurately over large areas but typically require many loudspeakers and complex processing, leading to high costs. The second category relies on psychoacoustic principles to approximate physical sound fields, such as traditional stereo and surround sound systems \citep{rumsey2012spatial}. These systems use fewer loudspeakers to create spatial sensations but provide precise 3D localization only within limited listening areas. The third category, which is the primary focus of this survey, is binaural reproduction. This approach, typically experienced through headphones, aims to simulate the sound pressure signals at a listener's eardrums. Binaural audio offers highly accurate spatial localization and immersion through headphones \citep{wightman1989headphone}, making it well-suited for VR, AR, gaming, and mobile applications.

\begin{figure*}[t]
\begin{center}
\includegraphics[width=1.0\textwidth]{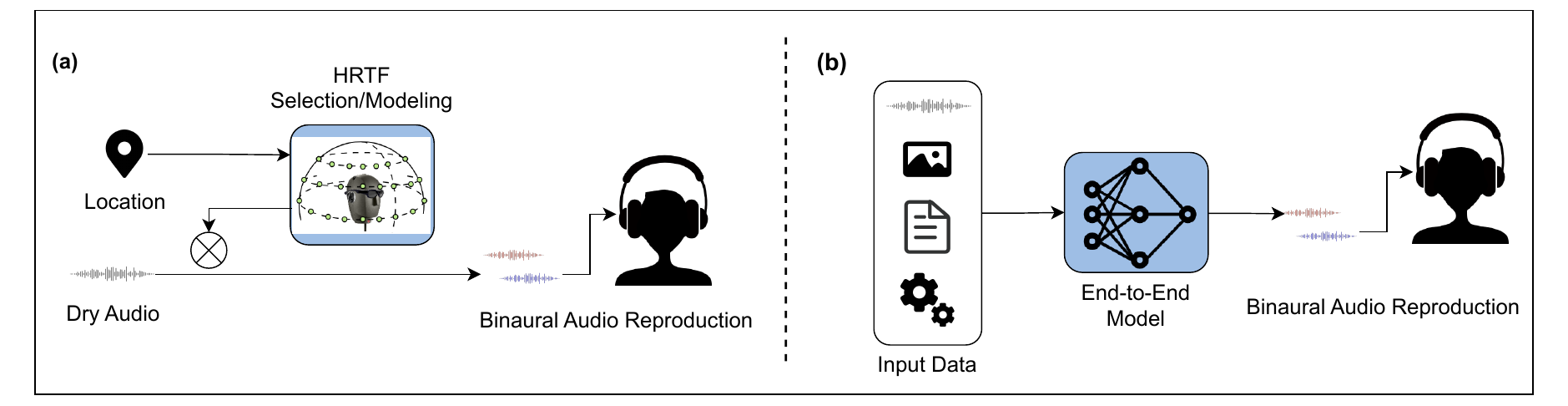}
\end{center}
\caption{Comparison of the two main paradigms in binaural audio reproduction: (a) HRTF-based filtering and (b) end-to-end binaural synthesis.}
\label{fig:paradigms} 
\end{figure*}

Binaural audio reproduction is achieved via two main technical approaches: (1) techniques based on head-related transfer function (HRTF) filtering, and (2) methods using end-to-end binaural synthesis. Both approaches have rapidly been transformed through modern data-driven methods. Figure~\ref{fig:paradigms} illustrates these two paradigms, which form the core structure of the technical section of this survey.

\subsection{HRTF-based Binaural Reproduction}

High-quality binaural reproduction through this approach (Figure~\ref{fig:paradigms}(a)) depends on accurately modeling the listener-specific acoustic filtering of the head, torso, and the pinnae. This filtering process adds crucial spatial cues to sound signals reaching the eardrums. These cues include interaural time differences (ITDs), interaural level differences (ILDs), and monaural spectral cues \citep{blauert1997spatial,annurev:/content/journals/10.1146/annurev.ps.42.020191.001031}. This acoustic transfer function is known as the HRTF \citep{xie2013head,moller1992fundamentals}. Synthesizing binaural audio typically involves convolving a source audio signal with the appropriate head-related impulse response (HRIR), which is the time-domain representation of the HRTF, for each ear.

A major challenge is that HRTFs vary significantly between individuals \citep{middlebrooks1999individual}. Using non-personalized HRTFs degrades the quality of the spatial quality experience, causing localization errors and reducing immersion \citep{wenzel1993localization, moller1996binaural,lee2018personalized}. Traditional methods for obtaining personalized HRTFs include acoustic measurement \citep{li2020measurement,algazi2001cipic}, numerical simulation \citep{katz2001boundary,brinkmann2019cross}, and spatial interpolation \citep{pulkki1997virtual,bruschi2023new,kistler1992model}. Acoustic measurements provide accurate results but are expensive, time-consuming, and require specialized anechoic environments \citep{li2020measurement}. Numerical simulations avoid physical measurements but require extensive computing resources and depend heavily on the accuracy of 3D anatomical models \citep{katz2001boundary,brinkmann2019cross}. Furthermore, traditional spatial interpolation methods, such as nearest-neighbor approaches \citep{pulkki1997virtual,bruschi2023new} or techniques based on functional models \citep{kistler1992model}, can operate with sparse HRTF measurements but often lack the accuracy or efficiency required for real-time personalization. These limitations hinder the widespread adoption of personalized binaural audio.

\subsection{End-to-End Binaural Synthesis}

The second approach (Figure~\ref{fig:paradigms}(b)) focuses on end-to-end binaural audio synthesis. These models are trained to directly convert various inputs into binaural audio output. These inputs can include mono or multi-channel audio, visual scene information, text descriptions, or other control parameters. This approach can bypass the explicit HRIR convolution step \citep{gebru2021implicit}, handle complex acoustic scenes, incorporate room acoustics (reverberation), and combine multi-modal information for richer spatial experiences. Although conceptually promising, traditional signal processing methods for synthesis from complex inputs struggle to generalize and capture realistic spatial sound characteristics \citep{richard2021neural}. In recent years, deep learning (DL) has transformed spatial audio reproduction by offering powerful tools to overcome limitations in both HRTF-based modeling and end-to-end synthesis. DL’s ability to learn complex, non-linear patterns from large datasets has driven significant innovation across the field, forming the central theme of this survey.

\subsection{Contribution and Scope}

The impact of machine learning (ML) on spatial audio has received considerable attention, leading to several relevant surveys. These reviews include surveys focusing on ML applications for HRTF personalization \citep{mcmullen2022machine, bruschi2024review, fantini2025survey}, as well as broader overviews encompassing data-driven approaches for spatial audio capture, processing, and reproduction alongside traditional methods \citep{cobos2022overview}.

While existing surveys effectively cover significant advancements in applying ML to HRTF modeling, a comprehensive survey dedicated to the full range of DL methods for binaural audio synthesis is still lacking. Specifically, there remains a gap in the literature that systematically examines DL-driven advancements across both main approaches: enhancing HRTF-based techniques and developing end-to-end binaural audio generation from diverse inputs.

This survey aims to fill this gap. Our work provides a structured overview of how DL is reshaping these two fundamental pathways for creating immersive binaural experiences. Notably, this survey offers one of the first comprehensive overview of DL-based end-to-end spatial audio synthesis, alongside a thorough examination of DL innovations in HRTF modeling. We organized and analyze recent advancements in:

\begin{itemize}
\item \textbf{HRTF personalized modeling (Section \ref{sec:ii})}: Including techniques for HRTF personalization using morphological features, environmental cues, and efficient spatial interpolation.
\item \textbf{End-to-end binaural synthesis (Section \ref{sec:iii})}: Covering methods driven by various inputs, including single-modal audio or multi-modal approaches incorporating visual, textual, or parametric guidance.
\end{itemize}

We also discuss relevant datasets and evaluation metrics within their respective sections. The survey then highlights key applications enabled by these DL-driven advancements (Section \ref{sec:iv}). By organizing our analysis around these core DL-driven pathways for binaural audio, this survey provides researchers with a comprehensive understanding of state-of-the-art strategies, identifies ongoing challenges, and suggests promising directions for future research (Section \ref{sec:v}).

\section{HRTF Personalized Modeling}
\label{sec:ii}

\begin{figure*}[t]
\begin{center}
\includegraphics[width=0.8\textwidth]{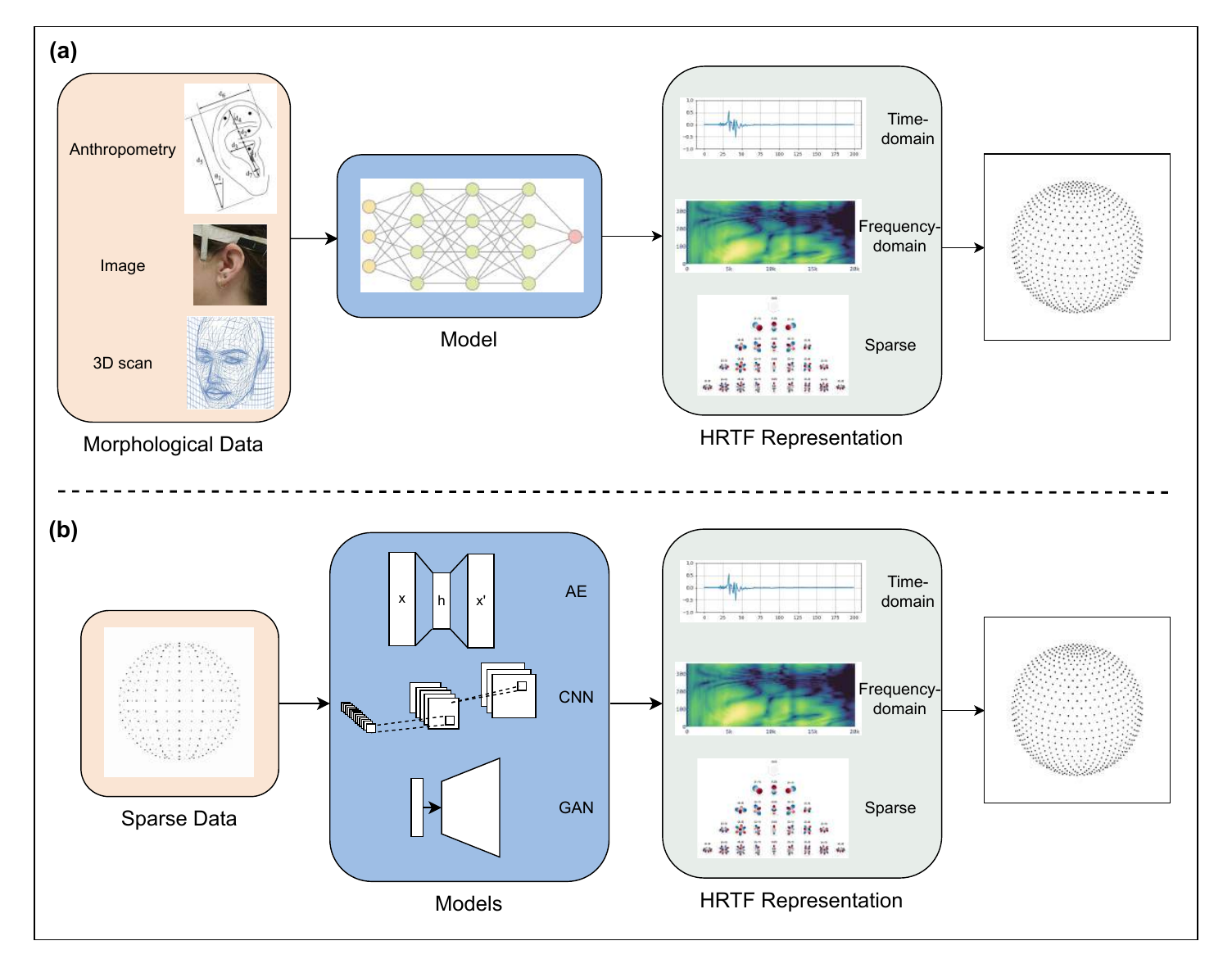}
\end{center}
\caption{Overview of DL applications in HRTF personalized modeling, illustrating approaches for (a) personalization using morphological data and (b) spatial interpolation from sparse measurements.}
\label{fig:dl_hrtf_tasks} 
\end{figure*}

Acquiring accurate and personalized HRTFs is fundamental to high-quality binaural reproduction. By leveraging large-scale data, DL methods effectively address the cost, time, individualization, and spatial resolution limitations of traditional HRTF modeling. This section provides a comprehensive overview of DL applications in HRTF personalization, covering data representation, personalization strategies, dataset fusion, and evaluation methodologies. Figure~\ref{fig:dl_hrtf_tasks} illustrates the main areas of DL's contributions.

\subsection{HRTF Data Representation}
\label{sec:hrtf_representation}

Effectively handling the high dimensionality and complex structure of HRTF data is a primary challenge for DL models, making the selection of an appropriate data representation crucial for model performance. Common representation methods include time-domain, frequency-domain, and sparse representations.

\subsubsection{Time-domain Representation}

This approach uses the HRIR sequence directly as input or output for DL models. HRIRs capture the complete temporal evolution of sound waves arriving at the eardrums, including initial onset delays and subsequent reflections from the listener’s anatomy \citep{blauert1997spatial,moller1992fundamentals}. 

\subsubsection{Frequency-domain Representation} 

The frequency-domain representation decomposes the HRTF into logarithmic magnitude and phase components. Log-magnitude spectra exhibit smoother patterns across frequencies and spatial directions, which better approximate the human ear's compressive response to sound intensity, potentially facilitating model learning \citep{katz2012perceptually}. Phase spectrum modeling is more challenging due to wrapping ambiguities that often require additional processing steps, which can introduce errors. Using complex-valued frequency representations preserves both magnitude and phase information without unwrapping. 

Beyond the full HRTF, researchers use specialized frequency-domain representations to isolate specific acoustic phenomena. The directional transfer function (DTF) captures only the direction-dependent HRTF components by removing the direction-independent common transfer function (CTF), which represents average spectral features \citep{middlebrooks1999individual, xie2013head}. This separation simplifies directional cue modeling. Similarly, the pinna-related transfer function (PRTF) isolates acoustic filtering effects specific to the listener’s pinna \citep{michele2010estimation,raykar2005extracting,guezenoc2020wide}. These specialized representations help focus modeling efforts on perceptually relevant spatial cues.

\subsubsection{Sparse Representation} 

To address the challenge of HRTF dimensionality, sparse representations capture the essential information using fewer parameters. Physics-based approaches employ spherical harmonics (SH) to project the spatial HRTF field onto basis functions, yielding coefficients that inherently represent spatial continuity \citep{romigh2015efficient}. The accuracy depends on the SH order, with higher orders required for capturing fine spatial details. Data-driven dimensionality reduction offers an alternative approach. For instance, Autoencoders (AE) learn non-linear mappings to a low-dimensional latent space \citep{chen2019autoencoding, baldi2012autoencoders}. Although they achieve significant compression, these methods may not preserve all perceptually relevant acoustic details.

\subsection{HRTF Personalization from Morphological Features}

By using readily accessible data such as anthropometric measurements, photographs, or 3D scans of a listener’s head and pinnae, DL models can predict personalized HRTFs that match an individual’s unique acoustic characteristics. This personalization relies on the strong physical relationship between an individual’s morphology and their HRTF. DL models excel at learning the mapping from these geometric features to the corresponding acoustic transfer functions. This section focuses on methods that directly predict HRTFs from such morphological features. Table~\ref{tab:feature_driven_hrtf} summarizes key studies in this area.

\begin{table*}[htbp] 
    \caption{Summary of HRTF Prediction Methods from Morphological Features.}
    \label{tab:feature_driven_hrtf} 
    \centering
    \setlength{\tabcolsep}{4pt} 

    \begin{tabularx}{\textwidth}{@{} l c c c c  c c c  c @{}} 
        \toprule
        \multirow{2}{*}{\textbf{Method}} & 
        \multirow{2}{*}{\textbf{Year}} & 
        \multicolumn{3}{c}{\textbf{Morphological Data}} & 
        \multirow{2}{*}{\parbox{3cm}{\centering\textbf{Datasets}}} & 
        \multirow{2}{*}{\parbox{3cm}{\centering\textbf{model}}} & 
        \multirow{2}{*}{\textbf{Data Representation}} & 
        \multirow{2}{*}{\parbox{2cm}{\centering\textbf{Physiological\\Parameters}}} \\

        \cmidrule(lr){3-5} 
        & & \textbf{Ant.} & \textbf{Ima.} & \textbf{3D sca.}  & & \\ 

        \midrule
        Hu et al. \cite{hu2008hrtf} & 2008 & \checkmark &  &  & CIPIC & BP ANN & HRTFs & 8 \\
        Chen et al. \cite{chen2019autoencoding} & 2019 & \checkmark &  &  & CIPIC & AEs+DNNs & AEs & 27 \\
        Wang et al. \cite{wang2020global} & 2020 & \checkmark &  &  & HUTUBS & 1D CNN & SH Cofficients & 25 \\
        Zhang et al. \cite{zhang2020modeling} & 2020 & \checkmark &  &  & CIPIC & DNN & SPCA Weights & 8  \\
        Miccini et al. \cite{miccini2020hrtf} & 2020 & \checkmark &  &  & HUTUBS & VAE+DNN & VAE \& CVAE & 27 \\
        Yao et al. \cite{yao2022individualization} & 2022 & \checkmark &  &  & HUTUBS & AE+DNN  & VAE & 12 \\
        Zhang et al. \cite{zhang2023modelling} & 2023 & \checkmark &  &  & CIPIC & FCNN+Attention  & HRTFs \& ITDs & 8 \\
        Sánchez et al. \cite{sanchez2025towards} & 2025  & \checkmark &  &  & HUTUBS & DDPM  & HRIRs & 27 \\
        Lee \& kim \cite{lee2018personalized} & 2018 & \checkmark & \checkmark &  & CIPIC & CNN-DNN  & HRTFs & 17 \\
        Miccini \& Spagnol \cite{miccini2021hybrid} & 2021 &  & \checkmark &  & HUTUBS & VAE+DNN  & CVAE & 3 \\
        Zhao et al. \cite{zhao2022magnitude} & 2022 & \checkmark & \checkmark &  & CIPIC & VGG-Ear+FC  & SH Cofficients & 17 \\
        PRTFNet \cite{ko2023prtfnet} & 2023 &  & \checkmark &  & HUTUBS & CNN  & PRTF  & — \\
        Javeri et al. \cite{javeri2023machine} & 2023 &  & \checkmark &  & HUTUBS &  3DR \& ASNN & HRTFs  & — \\
        Fantini et al. \cite{fantini2021hrtf} & 2021 &  &  & \checkmark & HUTUBS & GRNN  & DTF \& CTF  & 11 \\
        Zhou et al. \cite{zhou2021predictability} & 2021 &  &  & \checkmark & — & CNN-Reg/UNet-Reg  & HRTFs  & — \\
        Wang et al. \cite{wang2022predicting} & 2022 &  &  & \checkmark & HUTUBS & CNN-FC  & SH Cofficients  & — \\
        Zhao et al. \cite{zhao2024efficient} & 2024 &  &  & \checkmark & 3D3A & CNN-FC & HRTFs  & — \\
        \bottomrule
        \addlinespace[0.5em]
        \multicolumn{9}{@{}p{\textwidth}@{}}{
    \small \textbf{Abbreviations:} Ant(hropometry), Ima(ge), 3D sca(n). 

} \\
    \end{tabularx}
\end{table*}

\subsubsection{Predict from Sparse Anthropometry} 

A significant research direction focuses on predicting HRTFs from a limited set of easily measurable anthropometric parameters. Initial studies demonstrated that this approach is feasible using artificial neural networks (ANNs) \citep{hu2008hrtf}. To enhance performance with limited input data, researchers have employed representation learning techniques, such as AEs and their variants. These models reduce dimensionality by learning compact latent representations of HRTFs that can then be predicted from anthropometric data \citep{chen2019autoencoding,miccini2020hrtf,yao2022individualization}. Integrating signal processing knowledge through the spherical harmonic transform (SHT) \citep{wang2020global} or spatial principal component analysis (SPCA) \citep{zhang2020modeling} as intermediate steps has helped to incorporate structural information into the learning process.

Recently, more advanced architectures and powerful generative models have further improved  HRTF personalization. VAEs have been further refined for HRTF prediction from morphological data \citep{yao2022individualization}. A major advance is the emergence of diffusion models, which show promise for generating complete HRIR distributions from anthropometric inputs, potentially capturing finer details of HRTFs \citep{sanchez2025towards}. Alongside these generative approaches, neural network architectural elements like attention mechanisms have enhanced model effectiveness \citep{zhang2023modelling}. Attention mechanisms allow models to dynamically focus on the most important anthropometric features during prediction, leading to more accurate HRTF estimations. 

\subsubsection{Predict from Image} 

2D images, especially photographs of the pinna, provide richer geometric detail than sparse numerical measurements. Early studies in this area often combined image-derived features with traditional anthropometric parameters, demonstrating that the inclusion of visual data reduced spectral distortion compared to using anthropometry alone \citep{lee2018personalized,zhao2022magnitude}. Subsequent research has moved toward methods driven purely by image inputs. Conditional Variational Autoencoders (CVAE) have been used to predict HRTF from features extracted from pinna images \citep{miccini2021hybrid}. Specialized architectures like PRTFNet \citep{ko2023prtfnet} have been designed to predict PRTFs, which are crucial for vertical sound localization, from pinna images. Further advancements aim to predict both personalized HRTFs and corresponding headphone equalization filters from 2D images or short videos of the pinna \citep{javeri2023machine}, thereby improving the quality and accessibility of personalized spatial audio.

\subsubsection{Prediction from 3D Geometric Models} 

3D models from body scans provide the most complete information about an individual’s head and pinna morphology. Early research in this area focused on automatically extracting a predefined set of anthropometric features from these 3D models for HRTF prediction \citep{fantini2021hrtf}. However, such methods are limited by the manual selection of these features. A more direct approach involves predicting HRTFs directly from the raw 3D scan data. CNNs and U-Net architectures have been employed to process 3D pinna shapes, represented as point clouds or voxel grids. These models capture fine geometric details, leading to improved prediction accuracy, particularly for the medium and high frequency components of the HRTF \citep{zhou2021predictability}. Advances in geometric DL have enabled more efficient methods. For instance, some methods project 3D scan data onto mathematical basis functions before inputting them to the network \citep{wang2022predicting}. Specialized network designs have also been developed, such as cascading CNNs or models incorporating symmetries like vertical plane feature sharing, to reduce computational needs while maintaining high accuracy \citep{wang2022predicting, zhao2024efficient}. Recognizing the difficulty in obtaining high-quality 3D scans, recent research has also focused on improving data acquisition and preprocessing. Advanced DL frameworks aim to reconstruct 3D pinna models from more easily obtainable 2D images \citep{huang2023audioear,javeri2023machine}, while denoising models have been developed to improve the quality of scanned data, which can be noisy or incomplete \citep{di2024denoising}.

\subsection{HRTF Personalization Using Environment Cues}

The structural features of the human pinna provide a fundamental foundation for HRTF personalization. However, relying exclusively on physical measurements often proves insufficient for achieving optimal spatial audio quality and perceptual accuracy in real-world listening scenarios. These limitations, coupled with the practical difficulties in acquiring precise physical measurements, have led researchers to investigate alternative approaches that extract HRTF characteristics from acoustic cues embedded in the listener's interaction with their environment.

This emerging research direction estimates personalized HRTFs by analyzing how everyday sounds are modified by the listener’s presence and movements. The core concept is that acoustic interactions between listeners and their surroundings implicitly encode information about their HRTFs. Jayaram et al. \cite{jayaram2023hrtf} designed a U-Net architecture that learns to infer the listener's HRTF by observing changes in binaural audio recordings captured as listeners make natural head movements in response to ambient sounds. Similarly,  Thuillier et al. \cite{thuillier2025hrtf} used diffusion models to jointly estimate room impulse responses (RIRs) and HRTFs from binaural recordings, showing promise in preserving high-frequency individual differences in the estimated HRTFs. The growing availability of mobile devices with microphones makes such “in-the-wild” personalization increasingly attractive. This approach could make personalized spatial audio more accessible without requiring laboratory measurements or specialized equipment.

\subsection{HRTF Spatial Interpolation}

Beyond personalization based on individual features, DL addresses key challenges in HRTF modeling related to spatial resolution limitations and representation efficiency. Traditional interpolation methods struggle with accurately capturing the non-linear spatial-spectral characteristics of HRTFs, especially when upsampling from sparse measurements \citep{xie2012recovery,chen2008head}. This limitation stems from their underlying linear assumptions and reliance on manually designed features. DL-driven spatial interpolation research has advanced significantly along two main paths: discrete domain interpolation and continuous domain representation. Table~\ref{tab:Spatial_Interpolation} summarizes key studies in this area.

\begin{table*}[htbp] 
    \caption{Summary of DL-based HRTF spatial interpolation Methods.} 
    \label{tab:Spatial_Interpolation}
    \centering
    \begin{tabularx}{\textwidth}{@{} X c c c c c @{}} 
        \toprule 
        \textbf{Method} & \textbf{Year} & \textbf{Datasets} & \textbf{Model} & \textbf{Data Representation} & 
        \textbf{Sparse Locations} \\
        \midrule 
        \multicolumn{6}{c}{\cellcolor{lightgray}\textbf{Discrete-Domain}} \\
        Ito et al. \cite{9914751} & 2022 &  HUTUBS & AE + Aggregation & AEs & $9\sim196$ \\
        Zandi et al. \cite{zandi2022individualizing} & 2022 & ITA & CVAE & AEs & 60 \\
        Zurale \& Dubnov \cite{zurale2023spatial} & 2023 & BiLi & VQ-VAE + Transformer & AEs & 25 \\
        Chang et al. \cite{10.1121/10.0036032} & 2025 & CIPIC & VAE+DNN & AEs & — \\
        Zurale et al. \cite{zurale2022deep} & 2022 & CIPIC & DCNN & HRTFs & 72,18 \\
        Jiang et al. \cite{jiang2023modeling} & 2023 & CIPIC & U-Net & HRTFs & 3,4,6,8,12,23,105 \\
        Chen et al. \cite{chen2023head} & 2023 & HUTUBS & Spherical CNN &  SH Cofficients & 120 \\
        Thuillier et al. \cite{10418851} & 2024 & HUTUBS & SConvCNP & SH Cofficients & $0\sim100$ \\
        Zhao et al. \cite{zhao2025head} & 2025 & HUTUBS & CNN & HRTFs  & 6,14,26,38,50,74,86,110,146,170 \\
        Hogg et al. \cite{hogg2024hrtf} & 2024 & ARI & SR-GAN & HRTFs & 5,20,80,320 \\
        Hu et al. \cite{hu2024hrtf} & 2024 & SONICOM & AE-GAN & SH Cofficients & 8,12,18,27 \\
        HRTF-DUNet \cite{hu2025machine} & 2025 & SONICOM & Denoisy U-Net + AE-GAN & SH Cofficients & 3,4,8,18,27 \\
        \midrule
        \multicolumn{6}{c}{\cellcolor{lightgray}\textbf{Continuous-Domain}} \\
        Lee et al. \cite{10096144} & 2023 & HUTUBS & FiLM + HyperConv & HRTFs & 4,8,12,16 \\
        HRTF Field \cite{10095801} & 2023 & HRTF Datasets\textsuperscript{a} & SIREN / IGON & HRTFs & 5\%,10\%,15\%,20\%,25\%  \\
        Ma et al. \cite{ma2023spatial} & 2023 & HUTUBS & PINN & HRTFs & 315,675 \\
        NIIRF \cite{10448477} & 2024 & CIPIC / HUTUBS & INR & HRTFs & 10,20,30,50,100 \\
        Neural Steerer \cite{di2024neural} & 2024 & EasyCom & SIREN & HRIRs & 15\%,30\%,45\%,60\%,75\%,90\%  \\
        RANF \cite{masuyama2025retrieval} & 2025 & SONICOM & INR & HRTFs \& ITDs & 3,5,19,100  \\
        BiCG \cite{lu2025bicg} & 2025 & HRTF Datasets\textsuperscript{a} & IGON & ILDs \& ITDs & —\\
        \bottomrule 
        \addlinespace[0.5em]
        \multicolumn{6}{@{}l@{}}{\small \textsuperscript{a}\textbf{HRTF Datasets:} RIEC, 3D3A, Aachen, ARI, BiLi, CIPIC, Crossmod, HUTUBS, Listen, and SADIE.}
    \end{tabularx}
\end{table*}

\subsubsection{Discrete Spatial Interpolation}

Discrete domain methods work with HRTFs sampled at specific spatial locations. They learn the underlying spatial relationships to reconstruct denser HRTF sets from sparse measurements. Common DL architectures for this task include AEs, CNNs, and generative adversarial networks (GANs).

\textbf{Autoencoders (AEs).} AEs excel at interpolation tasks through their ability to learn compressed data representations (latent codes). For HRTFs, such architectures learn sparse representations that enable interpolation at unmeasured locations \citep{zurale2022deep}. Research has substantially improved AE-based HRTF interpolation. These improvements include better handling of position dependencies between sparse inputs and the ability to extrapolate from minimal data. Conditional mechanisms guide the interpolation process based on specific sparse input locations \citep{9914751, zandi2022individualizing}. Some approaches integrate vector quantized variational autoencoders (VQ-VAEs) to learn discrete latent representations of HRTFs, which facilitates more effective subsequent mapping to denser spatial resolutions \citep{zurale2023spatial}. Other researchers have focused on optimizing the latent space by encouraging spatial grouping of HRTFs \citep{10.1121/10.0036032}. These efforts improved interpolation accuracy while preserving crucial spectral details, demonstrating the flexibility of the AE framework for capturing complex HRTF structures and supporting efficient HRTF acquisition from minimal measurements.

\textbf{Convolutional Neural Networks (CNNs).} CNNs, known for their ability to extract local features from structured data like images, have been adopted for HRTF processing. By treating HRTF data as image-like inputs \citep{jiang2023modeling, hogg2024hrtf}, CNN architectures effectively perform spatial upsampling. U-Net architectures with an encoder-decoder structure and skip connections prove particularly effective as they capture both local details and global context in HRTF data \citep{jiang2023modeling}. Recent CNN-based HRTF interpolation advances focus on incorporating physical constraints and using anthropometric measurements as additional input features. These approaches enhance model robustness and generate more realistic HRTFs, especially when interpolating from very sparse measurements or extrapolating to unmeasured spatial regions \citep{zhao2025head}.

Given the spherical nature of HRTF data, researchers have developed specialized architectures. Spherical CNNs (S-CNNs), for instance, perform convolutions directly on the sphere by defining filters in the SH domain, thereby inherently respecting the data's geometric structure \citep{chen2023head}. This approach enables more efficient learning and better generalization. Building on this concept, the Spherical Convolutional Conditional Neural Process (SConvCNP) \citep{10418851} leverages S-CNNs within a meta-learning framework specifically for HRTF error interpolation from sparse measurements. This approach allows the model to not only refine HRTF estimates but also adaptively correct biases and provide well-calibrated uncertainty estimates, leading to significantly improved sample efficiency. These specialized architectures demonstrate the advantages of tailoring network designs to the specific geometric properties of HRTF data.

\textbf{Generative Adversarial Networks (GANs).} GANs provide a powerful data-driven approach for HRTF spatial upsampling, using a generator network to produce realistic HRTFs and a discriminator network to distinguish them. One effective strategy converts spherical HRTF data into 2D image-like representations through projection, enabling standard CNN-based GAN architectures such as the Super-Resolution-based GAN (SR-GAN) \citep{hogg2024hrtf} for upsampling. Other approaches work directly in the SH domain, using specialized GANs to generate high-order coefficients from low-order ones, thereby reconstructing the complete HRTF field \citep{hu2024hrtf}. To address the challenge of sparse and potentially noisy input measurements, the HRTF Denoising U-Net (HRTF-DUNet) \citep{hu2025machine} combines a U-Net-based denoiser with an autoencoding GAN (AE-GAN). This approach effectively upsamples HRTFs even from highly sparse measurements.

\subsubsection{Continuous Spatial Representation}

Implicit neural representations (INRs) \citep{xie2022neural, NEURIPS2020_53c04118} offer a compelling alternative to discrete-domain methods by overcoming the limitations of fixed spatial grids. INRs use a neural network $f_\textbf{w}$ to learn a continuous mapping from spatial coordinates, typically azimuth angles $\theta$ and elevation angles $\phi$, to the corresponding HRTF complex values, denoted as $H(\theta, \phi, f)$:

\begin{equation}
f_\textbf{w} : (\theta, \phi) \rightarrow H(\theta, \phi, f).
\end{equation}

 This approach represents the HRTF as a continuous differentiable function that can be queried at any arbitrary spatial location. Following success in computer vision for image and shape representation \citep{NEURIPS2020_53c04118, NEURIPS2020_55053683, mildenhall2021nerf} and acoustic field modeling \citep{luo2022learning,liang2023neural}, INRs are now being explored for high-fidelity continuous HRTF representation.

Research applying INRs to HRTF modeling has advanced rapidly. Early work demonstrated the potential of conditioning INRs on subject identity \citep{10096144}. Their DL architecture, incorporating Feature-wise Linear Modulation (FiLM) layers and hyperconvolution, dynamically modulated conditional information to accurately predict HRTFs across different datasets and coordinate systems, effectively capturing individual HRTF spatial patterns. Subsequent studies introduced more advanced architectures like the Implicit Gradient Origin Network (IGON) \citep{10095801}, which demonstrated strong capabilities in learning continuous HRTF representations that generalize across individuals. IGON maps limited HRTF samples to continuous representations by learning spatial distributions and employing improved optimization strategies. This leads to better preservation of spectral details and spatial continuity. For extremely sparse measurement scenarios, Retrieval-Augmented Neural Field (RANF) \citep{masuyama2025retrieval} significantly improves personalized HRTF generation from very few data points by combining database retrieval of relevant HRTF exemplars with neural field learning. This approach creates promising paths for fast, low-cost personalized HRTF acquisition. Neural Steerer \citep{di2024neural} further demonstrates how INRs can accurately model array steering vectors by explicitly accounting for important aspects such as inter-channel phase relationships and causality.

To enhance neural field models for HRTFs, researchers have focused on optimization strategies, including prior knowledge integration and improved modeling of key acoustic features. The Neural Infinite Impulse Response Filter Field (NIIRF) \citep{10448477} incorporates infinite impulse response filter structures, allowing the network to predict filter parameters rather than direct HRTF values. This reduces model size and improves efficiency. The study also found that low-rank adaptation (LoRA) techniques effectively balance model efficiency and personalization performance. Adding physical constraints, like the Helmholtz equation as a training regularization term \citep{williams1999fourier}, enhances generalization from sparse data and ensures the physical consistency of the predicted HRTF \citep{ma2023spatial}, particularly for high-frequency components. Other efforts focus on improving prediction accuracy for critical perceptual features. For instance, Lu et al. \cite{lu2025bicg} aim to directly predict specific binaural cues such as ITDs and ILDs, producing HRTFs that are both objectively accurate and perceptually convincing.

\begin{table*}[htbp] 
    \caption{Overview of Publicly Available Human HRTF Databases.} 
    \label{tab:sofa_HRTF}
    \centering

    \begin{tabularx}{\textwidth}{@{} l c c c c c c c @{}} 
        \toprule 
        \textbf{Name} & \textbf{Year} &\textbf{Subjects} & \textbf{Positions} & \textbf{Distance (m)} & \textbf{Spatial Resolution} & \textbf{Sampling Scheme} & \textbf{Morphology} \\ 
        \midrule 
        CIPIC \cite{algazi2001cipic} & 2001 & 45 & 1250 & 1.00 & $\Delta_\theta \geq 5^\circ,\Delta_\varphi = 5.625^\circ$ & Interaural-polar & Anthropometry\\
        Listen \cite{warusfel2003listen} & 2003 & 51 & 187 & 1.95 & $\Delta_\theta \ge 15^\circ,\Delta_\varphi = 15^\circ$ & Geodesic grid & Anthropometry\\
        RIEC \cite{watanabe2014dataset} & 2014 & 105 & 865 & 1.50 & $\Delta_\theta = 5^\circ,\Delta_\varphi = 10^\circ$ & Geodesic grid & 3D meshes\\
        BiLi \cite{carpentier2014measurement} & 2014 & 57 & 1680 & 2.06 & $\Delta_\theta = 6^\circ,\Delta_\varphi = 6^\circ$ & Geodesic grid & No\\
        ARI \cite{majdak2013sound} & 2016 & 250 & 1550 & 1.20 & $\Delta_\theta = 2.5^\circ,\Delta_\varphi \geq 5^\circ$ & Geodesic grid & Anthropometry \\
        ITA \cite{bomhardt2016high} & 2016 & 48 & 2304 & 1.20 & $\Delta_\theta = 5^\circ,\Delta_\varphi = 5^\circ$ & Geodesic grid & Anthropometry; 3D meshes \\
        Aachen \cite{bomhardt2016high} & 2016 & 48 & 2304 & 1.20 & $\Delta_\theta = 5^\circ,\Delta_\varphi = 5^\circ$  & Geodesic grid & Anthropometry; 3D meshes \\
        3D3A \cite{sridhar2017database} & 2017 & 38 & 648 & 0.76 & $\Delta_\theta = 5^\circ,\Delta_\varphi \le 5.625^\circ$ & Geodesic grid & Anthropometry; 3D meshes\\
        SADIE \cite{armstrong2018perceptual} & 2018 & 20 & $\le2818$ & 1.20 & $\Delta_\theta \ge 1^\circ,\Delta_\varphi \le 15^\circ$ & Geodesic grid & Images; 3D meshes\\
        OlHeaD-HRTF \cite{denk2018adapting} & 2018 & 16 & 91 & 2.50-3.00 & $\Delta_\theta = 7.5^\circ,\Delta_\varphi = 30^\circ$ & Interaural-polar & No \\
        HUTUBS \cite{brinkmann2019cross} & 2019 & 96 & 440 & 1.47 & $\Delta_\theta \geq 10^\circ,\Delta_\varphi = 10^\circ$ & Near-Lebedev & Anthropometry; 3D meshes\\
        CHEDAR \cite{ghorbal2020computed} & 2020 & 1253 & $\le2522$ & 0.2,0.5,1.2 & $\Delta_\theta = 5^\circ,\Delta_\varphi = 5^\circ$ & Geodesic grid & Anthropometry; 3D meshes\\
        SONICOM \cite{engel2023sonicom} & 2023 & 200 & 793 & 1.50 & $\Delta_\theta = 5^\circ,\Delta_\varphi \ge 10^\circ$ & Geodesic grid & Images; 3D meshes\\
        \bottomrule 
        \addlinespace[0.5em]
        \multicolumn{8}{@{}p{\textwidth}@{}}{
    \small In this table, for geodesic and Near-Lebedev sampling schemes, $\Delta_\theta$ and $\Delta_\varphi$ generally represent the resolution in elevation and azimuth, respectively. For the interaural-polar scheme, $\Delta_\theta$ typically refers to the lateral angle resolution, and $\Delta_\varphi$ to the polar angle resolution.
} \\
    \end{tabularx}
\end{table*}

\subsection{HRTF Dataset Fusion Strategies}
\label{sec:hrtf}

DL approaches to HRTF personalization depend on high-quality datasets. These datasets contain numerous HRIR samples obtained through acoustic measurements on human subjects or dummy heads, or through numerical simulations based on 3D geometry models. Such data serves as the ground-truth necessary for effective model training. Most publicly available HRTF datasets are stored and shared using the Spatially Oriented Format for Acoustics (SOFA)\footnote{\url{https://www.sofaconventions.org/}} to support distribution and compatibility. Table~\ref{tab:sofa_HRTF} provides details of these datasets, all of which follow this standardized format. Despite adopting a common storage convention and growing in number, these datasets show notable differences. Variations arise from different measurement equipment, anechoic environments, microphone types, processing methods, and spatial sampling patterns \citep{andreopoulou2015inter}. These differences make direct comparisons across datasets difficult \citep{pauwels2023relevance}. More importantly, such diversity limits how well DL models trained on a single dataset can generalize to new situations, creating a significant bottleneck for technological progress in this field.

To address this challenge, INRs offers a promising new approach. By modeling the HRTF as a continuous function of spatial coordinates, INRs naturally accommodate data with different or irregular spatial samplings. This breakthrough enables the integration of data from various sources with different protocols, potentially expanding the scale of effective training data substantially. Effective cross-dataset fusion typically combines coordination at the data level with adaptability at the model level. Researchers are exploring preprocessing techniques, including advanced normalization methods to reduce systematic biases between datasets and provide more consistent input \citep{wen2023mitigating}. Additionally, the continuously differentiable property of INRs naturally accommodates HRTF data with different spatial grid structures. In the SONICOM Listener Acoustic Personalization (LAP) Challenge\footnote{\url{https://www.sonicom.eu/lap-challenge/}}, using neural fields to integrate heterogeneous data for HRTF modeling and upsampling has emerged as a viable strategy. Fusion frameworks also support more precise modeling of key perceptual features. Researchers use fused data and INR architectures to learn how to generate important binaural cues while optimizing data preprocessing to enhance specific cue accuracy \citep{lu2025bicg}. This demonstrates that fusion extends beyond increasing data volume to deepening understanding and achieving precise control over acoustic features. 

In conclusion, addressing HRTF dataset heterogeneity represents a critical step toward advancing personalized spatial audio. Technologies like INRs enable effective fusion of diverse data, overcoming limitations of the traditional methods. These advances lead to more universal, higher-precision HRTF models and establish a solid foundation for the widespread application of personalized spatial audio.

\begin{table*}[htbp] 
    \caption{Summary of Common Objective Evaluation Metrics for HRTF Modeling.}
    \label{tab:objective_metrics}
    \centering
    \begin{tabularx}{\textwidth}{@{} l  | c | c | X @{}} 
        \toprule 
        \textbf{Metric}  &  \textbf{Formula}  &  \textbf{Focus} & \multicolumn{1}{c}{\textbf{Task}}  \\
        \midrule 
        LSD \cite{torres2015personalization} $\downarrow$ &   $\textstyle \sqrt{\frac{1}{N_f} \sum_{f=1}^{N_f} \left( 20 \log_{10} \frac{|H(\theta, \varphi, f)|}{|\hat{H}(\theta, \varphi, f)|} \right)^2}$   &   Spectral Difference  & Measure average log-magnitude error. \\ 
        \midrule
        SDE \cite{zhou2021predictability} $\downarrow$ & $\textstyle \frac{1}{N_d N_f} \sum_{\theta,\varphi} \sum_{f} \left| 20 \log_{10} \frac{|H(\theta, \varphi, f)|}{|\hat{H}(\theta, \varphi, f)|} \right| $  &   Spectral Difference  & Measures mean absolute log-magnitude error. \\
        \midrule
         LRE \cite{thuillier2025hrtf,10418851} $\downarrow$ &  $ \textstyle 20 \log_{10} \left| \frac{\hat{H}_{c,f} - H_{c,f}}{H_{c,f}} \right| $  & Spectral Difference & Measures relative error on log-magnitude spectra. \\
        \midrule
         LMD \cite{thuillier2025hrtf,10418851} $\downarrow$ &  $ \textstyle \left| 20 \log_{10} \left| \frac{\hat{H}_{c,f}}{H_{c,f}} \right| \right|$  & Spectral Difference & Measures mean absolute log-magnitude difference. \\
        \midrule
         RMSE \cite{9746315, zhang2023modelling} $\downarrow$ &  $ \textstyle \sqrt{\frac{1}{N_t} \sum_{t=0}^{N_t-1} (h(t)-\hat{h}(t))^2}$  & Overall Difference & Measures average HRIR difference. \\
        \midrule
         MAE \cite{marggraf2024hrtf} $\downarrow$ &  $\textstyle \frac{1}{N_t} \sum_{t=0}^{N_t-1} \left| h(t) - \hat{h}(t) \right|$   & Overall Difference & Measures average HRIR difference. \\
        \midrule
         SDR \cite{luo2013virtual,kobayashi2023temporal} $\uparrow$ &   $ \textstyle 10 \log_{10}  \frac{\sum_{f=0}^{N_f-1} |H(\theta, \varphi, f)|^2}{\sum_{f=0}^{N_f-1} |H(\theta, \varphi, f) - \hat{H}(\theta, \varphi, f)|^2} $ & Signal-to-Error Ratio & Ratio of HRIR/HRTF signal energy to error energy. \\
        \midrule
        PCC \cite{10.1121/10.0036032} $\uparrow$ & $\rho\left( \{|H(\theta, \varphi, f,)|\}_{\theta,\varphi,f}, \{|\hat{H}(\theta, \varphi, f)|\}_{\theta,\varphi,f} \right)$  & Statistical Correlation & Measures linear correlation of magnitude spectra. \\
        \bottomrule 
        \addlinespace[0.5em]
        \multicolumn{4}{@{}p{\textwidth}@{}}{
    \small \textbf{Abbreviations:} LSD: log-spectral distortion, SDE: spectral distance error, LRE: logarithmic relative error, LMD: logarithmic magnitude distance, RMSE: root mean square error, MAE: mean absolute error, SDR: signal-to-distortion ratio, PCC: Pearson correlation coefficient.

     The symbols used are: $H(\theta,\varphi,f)$ for the true complex HRTF at direction $(\theta,\varphi)$ and frequency $f$; $\hat{H}(\theta,\varphi,f)$ for the predicted complex HRTF; $h(t)$ for the true HRIR at time $t$; $\hat{h}(t)$ for the predicted HRIR. $N_f$ is the number of frequency bins, $N_d$ the number of directions, and $N_t$ the number of time samples. Metrics are typically averaged over directions and/or frequencies as appropriate. Arrows ($\downarrow$ / $\uparrow$) indicate desirable direction.
} \\
    \end{tabularx}
\end{table*}

\subsection{Evaluation Methodologies for HRTF Modeling}
\label{sec:ii-e}

Evaluating DL-based HRTF modeling techniques requires robust assessment methodologies. These methodologies fall into two main categories: subjective perceptual evaluations that directly assess listener experience, and objective evaluations using computational metrics. Objective methods are further divided into signal-based metrics and model-based metrics that use DL to predict perceptual outcomes or simulate human auditory processing.

\subsubsection{Subjective Perceptual Evaluation}

Subjective perceptual evaluation conducted via human listening tests serves as the gold standard for validating the effectiveness of HRTF personalization and the performance of spatial audio systems \citep{blauert1997spatial, annurev:/content/journals/10.1146/annurev.ps.42.020191.001031}. These tests aim to measure how well a modeled HRTF reconstructs key perceptual attributes of an individual's own HRTF, notably sound source localization accuracy, externalization, and timbral naturalness \citep{moller1995head, hartmann1996externalization}.

\textbf{Sound source localization tasks} are fundamental to assessing spatial accuracy. Listeners indicate the perceived direction of sound sources using graphical interfaces or head-pointing. Performance is typically measured using mean absolute error (MAE) or root mean square error (RMSE) in degrees, along with front-back and up-down confusion rates, which provide direct insights into spatial fidelity \citep{wightman1989headphone}. Many HRTF personalization studies use these tasks to demonstrate improvements over generic HRTFs \citep{hu2008hrtf, lee2018personalized, zhang2023modelling}. 

\textbf{Attribute rating of single stimulus using scales} assesses qualitative aspects of the auditory experience for individual stimuli, such as externalization, timbral fidelity, or spatial impression. Listeners typically use Likert or continuous visual analog scales to rate stimuli on specific attributes. The results are often quantified using mean opinion scores (MOS) that reflect overall perceived quality or naturalness \citep{wenzel2017perception, best2020sound}. MOS ratings can be adapted to assess different perceptual dimensions within spatial audio synthesis. For example, studies may report a MOS for overall sample quality, a 'similarity MOS' for likeness to ground-truth, or a 'spatial MOS' for perceived spatial accuracy. 

\textbf{Comparative evaluation tasks} involve direct comparisons between spatial audio stimuli or rendering methods. These tasks help to identify subtle perceptual differences. A basic approach is A/B comparison, where listeners judge differences between a reference and a target stimulus presented alternately. More rigorous methods include forced-choice paradigms such as the two-alternative forced choice (2AFC) test, which requires participants to identify specific stimulus characteristics across intervals \citep{xie2013head,jiang2023modeling}. To assess preference or quality, researchers employ paired comparisons where listeners compare stimuli based on given criteria. Furthermore, standardized tests like the Multiple Stimuli with Hidden Reference and Anchor (MUSHRA) test enable comprehensive audio quality assessment, requiring listeners to rate multiple stimuli against a hidden reference on a continuous scale.

\subsubsection{Objective Evaluation Metrics}

Objective metrics provide quantitative, automated, and repeatable assessments that complement subjective evaluations.

\textbf{Signal-based Metrics.} These metrics quantify physical differences between predicted and ground-truth HRTFs (or HRIRs) based on their signal properties without relying on perceptually trained models. Common metrics are summarized in Table~\ref{tab:objective_metrics}. 

A primary assessment approach involves quantifying spectral differences, which evaluate discrepancies in the frequency domain crucial for localization and timbre. The log-spectral distortion (LSD) is widely used, with lower values indicating closer physical agreement \citep{torres2015personalization}. Other metrics like spectral distance error (SDE) \citep{zhou2021predictability}, logarithmic relative error (LRE), and logarithmic magnitude distance (LMD) \citep{thuillier2025hrtf,10418851}, offer alternative ways to quantify spectral deviations with varying sensitivities to power spectrum disparities. Another class evaluates overall signal differences, often in the time domain or complex spectra. RMSE \citep{9746315, zhang2023modelling} and MAE \citep{marggraf2024hrtf} are common, with MAE being less sensitive to outliers. Metrics assessing signal-to-error ratio, like signal-to-distortion ratio (SDR) \citep{luo2013virtual,kobayashi2023temporal}, quantify the prediction fidelity relative to error magnitude, where higher values indicate better performance. Statistical correlation metrics such as the Pearson correlation coefficient (PCC) \citep{10.1121/10.0036032} measure the linear relationship between features of predicted and target HRTFs, including how well underlying trends are captured. 

\textbf{Model-based Metrics.} These metrics use computational models that simulate aspects of human hearing or are trained on perceptual data to predict spatial audio quality. They aim to bridge the gap between signal-based measures and subjective tests.

Auditory models (AMs) simulate key stages of human hearing to predict perceptual aspects of HRTF-spatialized audio, such as localization accuracy or timbre perception \citep{majdak2022amt,zaar2022predicting}. For spatial audio evaluation, these models process binaural signals to estimate how listeners would perceive them \citep{baumgartner2014modeling, marggraf2024hrtf}. AMs typically employ functional approaches that explicitly model auditory mechanisms, using cue-based analysis and template-matching strategies \citep{reijniers2014ideal}. The Auditory Modeling Toolbox offers accessible resources for implementing them \citep{majdak2022amt}. Recent AM advances focus on predicting sound quality changes caused by HRTF modifications and modeling complete 3D spatial perception through Bayesian methods \citep{barumerli2023bayesian, reijniers2025ideal, 10095152}. While AMs provide consistent metrics useful for early-stage model development, they simplify the complex hearing pathway. They require careful parameter adjustments, and their validation against human perception remains challenging, especially with regard to higher-level cognitive influences.

Data-driven perceptual predictors use DL to directly predict human ratings of spatial audio quality from acoustic features. These approaches offer scalable alternatives to both traditional metrics and resource-intensive listening tests. Inspired by successes in speech quality assessment where models predict MOS for synthesized speech \citep{lo2019mosnet, NEURIPS2021_bc6d7538, zezario2022deep, lian2025apg}, similar methodologies have emerged for spatial audio evaluation \citep{manocha2021dplm,manocha2022saqam,manocha2023spatialization,zheng2025hapg}. Notable developments include the Deep Perceptual Spatial Audio Localization Metric (DPLM) \citep{manocha2021dplm}, which quantifies localization differences using learned embeddings from a direction-of-arrival (DOA) estimation network. The Spatial Audio Quality Assessment Metric (SAQAM) \citep{manocha2022saqam} evaluates both listening quality and spatialization quality through multi-task learning. Furthermore, the Spatialization Quality Metric for Binaural Signals (SQM-BS) \citep{manocha2023spatialization} introduces deep metric learning and multi-task learning to assess the spatialization quality between pairs of binaural signals, designed to be independent of content and duration. Based on SAQAM, the Human Auditory Perception Guided SAQAM (HAPG-SAQAM) \citep{zheng2025hapg} incorporates auditory-guided feature extraction and perceptually weighted loss functions for improved alignment with human judgments across various quality dimensions. However, the performance of these data-driven models depends on their training data quality and diversity, from which they can inherit biases \citep{mehrabi2021survey}. Additionally, their “black box” nature limits the understanding of their decision processes \citep{rudin2019stop}, and they may struggle with entirely new conditions \citep{torralba2011unbiased}. Despite these challenges, ongoing research aims to improve their reliability and interpretability.
\section{Binaural Audio Synthesis with Deep Learning}
\label{sec:iii}

DL has expanded spatial audio capabilities, advancing beyond traditional HRTF-based methods discussed in Section~\ref{sec:ii} towards direct, end-to-end spatial audio synthesis. This advancement builds on recent breakthroughs in audio generation technologies, including text-to-audio (T2A) \citep{kreuk2022audiogen,liu2023audioldm,evans2025stable} and video-to-audio (V2A) \citep{zhou2018visual,NEURIPS2023_98c50f47,li2025tri}. Spatial audio synthesis presents significant challenges that require both sound realism and precise spatial accuracy, aligned with contextual information.

End-to-end neural models excel at this task by implicitly encoding spatial cues without requiring explicit HRTF measurements \citep{gebru2021implicit, richard2021neural}. This approach improves adaptability and reduces reliance on specific HRTF datasets while achieving a better balance between computational demands and perceptual quality than physics-based simulations. Recent approaches leverage advanced architectures, notably including U-Nets \citep{ronneberger2015u}, Transformers \citep{NIPS2017_3f5ee243}, and diffusion models \citep{NEURIPS2020_4c5bcfec}. Combined with self-supervised and multi-task learning \citep{ruder2017overview} techniques, these methods demonstrate enhanced audio quality, better generalization, and potential for real-time applications.

\subsection{Synthesis Paradigms Based on Input Modalities}
\label{sec:synthesis_paradigms}

DL-based spatial audio synthesis can be categorized by the main input modalities that guide the generation process. As shown in Figure~\ref{fig:binaural_synthesis_inputs}, these approaches fall into single-modal or multi-modal categories.

In single-modal synthesis, shown in Figure~\ref{fig:binaural_synthesis_inputs}(a), the main task transforms an input audio signal $\mathbf{x}_{\text{audio}} \in \mathbb{R}^{D \times T} $ into the binaural format. The conditioning information $\mathbf{c}$ consists of explicit spatial parameters such as source/listener position and orientation. The synthesis task can be formulated as:
\begin{equation}
    \mathbf{y}_{\text{binaural}} = f_{\textbf{w}}(\mathbf{x}_{\text{audio}}, \mathbf{c}),
    \label{eq:single_modal_synthesis}
\end{equation}

Multi-modal synthesis, as illustrated in Figure~\ref{fig:binaural_synthesis_inputs}(b), enhances this approach by adding non-auditory information $\mathbf{m}_{\text{non-auditory}}$ alongside the source audio $\mathbf{x}_{\text{audio}}$ (if present). This provides richer contextual cues for spatialization. The general formulation is:
\begin{equation}
    \mathbf{y}_{\text{binaural}} = f_{\textbf{w}}(\mathbf{x}_{\text{audio}}, \mathbf{m}_{\text{non-auditory}}).
    \label{eq:multi_modal_synthesis_general}
\end{equation}
where $\mathbf{x}_{\text{audio}}$ might sometimes be absent if the task is to generate all sound from non-auditory cues. Key multi-modal approaches, differentiated by the nature of the non-auditory information $\mathbf{m}_{\text{non-auditory}}$, include: 

\begin{figure*}[t]
    \centering
    \includegraphics[width=1\linewidth]{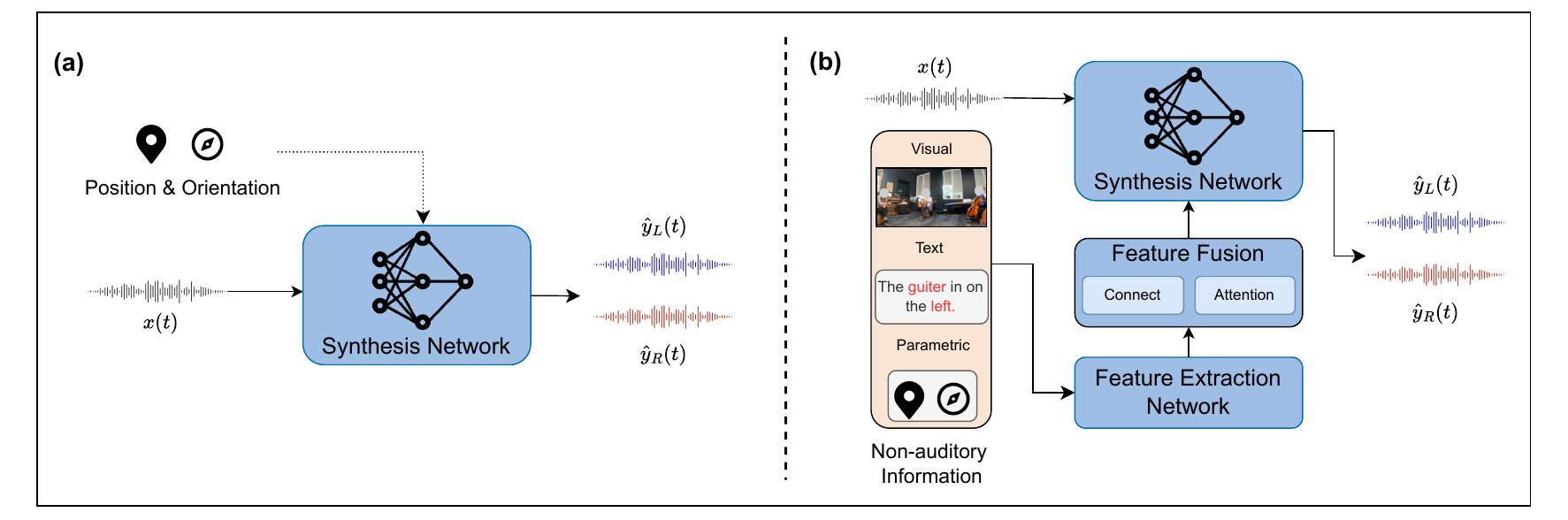}
    \caption{Conceptual illustration of DL-based binaural audio synthesis paradigms: (a) single-modal synthesis driven by audio inputs and spatial parameters, and (b) multi-modal synthesis guided by additional non-auditory information.}
    \label{fig:binaural_synthesis_inputs}
\end{figure*}

\begin{itemize}
    \item \textbf{Visual-guided Synthesis:} In this approach, the non-auditory input is visual data. This can range from 2D images and videos that inform sound source properties and locations \citep{NEURIPS2018_01161aaa, gao20192, zhou2020sep, li2024cyclic}, to detailed 3D scene geometry (e.g., point clouds, meshes) that helps model environmental acoustics and precise spatial relationships \citep{NEURIPS2023_760dff0f, bhosale2024av, gao2024soaf, baek2025av}.
    \item \textbf{Text-guided Synthesis:} Here, the non-auditory input consists of natural language descriptions. These textual prompts are used to control various spatial audio characteristics during generation, such as the type of acoustic environment, source positions, or sound event semantics \citep{li2024tas, zhao2025dualspec, feng2025audiospa}. 
    \item \textbf{Joint Multi-modal Guided Synthesis:} This category leverages a combination of non-auditory modalities, for instance, it can integrate both visual information and textual descriptions to achieve more comprehensive and nuanced control over the spatial audio synthesis \citep{heydari2024immersediffusion, dagli2024see, sun2024both, kim2025visage}.
\end{itemize}

The following sections explore these single-modal and multi-modal synthesis approaches in detail, highlighting key methodologies, architectural innovations, and recent advancements in the field.

\subsection{Single-modal Synthesis: Spatialization from Audio}

Single-modal spatial audio synthesis focuses on generating spatialized audio primarily from audio signals, with spatial parameters such as source and listener position or orientation are typically provided as conditioning metadata. This approach transforms monaural audio inputs into immersive binaural audio by learning the acoustic filtering that occurs as sound interacts with the listener and their environment. Key methods are summarized in Table~\ref{tab:single_modal_synthesis_summary}.

Early research established the feasibility of using DL for this application. A Temporal Convolutional Network (TCN) was shown to be capable of directly synthesizing binaural audio in reverberant environments, with performance matching that of traditional HRTF filtering \citep{gebru2021implicit}. Building on this foundation, the Warping Network (WarpNet) introduced architectural and loss functions improvements to produce realistic and spatially accurate binaural sound in real-time \citep{richard2021neural}. These initial studies confirmed that DL models could learn the intricate mono-to-binaural transformation without requiring explicit HRTF data for every scenario.

Subsequent research focused on enhancing synthesis quality through more powerful generative models. A modified vector-quantized variational autoencoder (VQ-VAE) was developed for speech binauralization, designed to accurately reproduce environmental factors such as background noise and reverberation \citep{huang2022end}. Diffusion models led to significant quality improvement, especially in phase spectrum estimation. BinauralGrad \citep{NEURIPS2022_95f03faf}, for example, employed a two-stage framework that used diffusion models to synthesize the common and specific parts of the binaural audio separately. More recently, DIFFBAS \citep{li2024diffbas} incorporated perceptually motivated interaural phase difference (IPD) losses directly into the diffusion process, which substantially improved realism. 

Researchers have also addressed the challenge of rendering dynamic scenes and improving computational efficiency \citep{liu2022dopplerbas, he2025dual, zhang2025two}. For the synthesis of moving sound sources, the Dual Position Attention Time-Frequency Network (DPATFNet) \citep{he2025dual} uses attention mechanisms to track sound source movement and improve phase estimation. Similarly, Zhang et al. \citep{zhang2025two} proposed a two-stage framework with a position-orientation self-attention (POSA) module to integrate spatial information and capture source motion. To reduce computational demands, Neural Fourier Synthesis (NFS) \citep{lee2023neural} achieved significant reductions in model size and inference time by performing synthesis in the frequency domain, predicting the delays and scales of early reflections based on geometric time delays. These advances demonstrate the growing ability of DL to capture subtle acoustic details that are crucial for convincing spatialization in dynamic environments.

A persistent challenge in supervised mono-to-binaural synthesis is the need for extensive paired monaural and binaural recordings that are costly to acquire. Zero-shot learning approaches offer a promising solution to this data limitation. ZeroBAS \citep{levkovitch2024zero} represents a pioneering effort in this direction, successfully synthesizing binaural audio without paired training data by combining geometric time warping (GTW) techniques with pre-trained generative audio models. This line of research shows potential for developing more adaptable and personalized binaural synthesis systems with reduced data requirements.

\begin{table}[t] 
    \caption{Representative DL-based Approaches for Single-modal Binaural Audio Synthesis.} 
    \label{tab:single_modal_synthesis_summary} 
    \centering
    \setlength{\tabcolsep}{4pt} 

    \begin{tabularx}{\columnwidth}{@{} l c  X c @{}}  
        \toprule

        \textbf{Method} & \textbf{Year} & \textbf{Model} & \textbf{Code} \\ 

        \midrule
        Gebru et al. \cite{gebru2021implicit} & 2021 &  TCN & No\\
        WarpNet \cite{richard2021neural} & 2021 &  Warping + Temporal ConvNet &  \href{https://github.com/facebookresearch/BinauralSpeechSynthesis}{Github} \\
        BinauralGrad \cite{NEURIPS2022_95f03faf} & 2022 &  Diffusion Model &   \href{https://github.com/microsoft/NeuralSpeech/tree/master/BinauralGrad}{Github}\\
        Huang et al. \cite{huang2022end} & 2022 & VQ-VAE & No\\
        DopplerBAS \cite{liu2022dopplerbas} & 2023 & WarpNet / BinauralGrad\textsuperscript{a} & No\\
        NFS \cite{lee2023neural} & 2023 & — & \href{https://github.com/jin-woo-lee/nfs-binaural}{Github} \\
        DIFFBAS \cite{li2024diffbas} & 2024 & WarpNet / NFS\textsuperscript{b} & \href{https://github.com/tongjiRain/DIFFBAS}{Github}\\
        ZeroBAS \cite{levkovitch2024zero} & 2024 & GTW + AS + Denoising Vocoder & No\\
        Zhang et al. \cite{zhang2025two} & 2025 & TW + POSA + GCFM & No\\
        DPATFNet \cite{he2025dual} & 2025 & TDW + DPAB + MPF & No \\
        \bottomrule
        \addlinespace[0.5em]  
        \multicolumn{4}{@{}p{\columnwidth}@{}}{%
        \small \textsuperscript{a}DopplerBAS considers velocity information based on WarpNet and BinauralGrad to simulate the Doppler effect. \textsuperscript{b}DIFFBAS redesigns the loss function based on the models studied in WarpNet and NFS.
    } \\
    \end{tabularx}
\end{table}

\subsection{Multi-modal Guided Spatial Audio Synthesis}

Researchers are exploring multi-modal guided binaural audio synthesis to improve realism, accuracy, and interactive control beyond what audio and spatial parameters alone can achieve. This approach enhances sound reproduction by combining non-auditory information with source audio signals to guide spatial audio generation. Visual and textual cues are the primary additional inputs in this process. Table~\ref{tab:controllable_tts} summarizes key methods, organized by their main guiding input modalities.

\begin{table*}[htbp] 
    \caption{Overview of Multi-modal Guided Spatial Audio Synthesis Methods.} 
    \label{tab:controllable_tts} 
    \centering
    \setlength{\tabcolsep}{4pt} 
    \begin{tabularx}{\textwidth}{@{} l  c ccc ccc cc c @{}} 
        \toprule
        \multirow{2}{*}{\textbf{Method}} &
        \multirow{2}{*}{\parbox{1.2cm}{\centering\textbf{Year}}} & 
        \multicolumn{4}{c}{\textbf{Input Modalities}} & 
        \multicolumn{2}{c}{\textbf{Model Architectures}} & 
        \multirow{2}{*}{\parbox{1.2cm}{\centering\textbf{Feature Fusion}}} &
        \multirow{2}{*}{\parbox{1.2cm}{\centering\textbf{Training Strategy}}} &
        \multirow{2}{*}{\textbf{Code}}  \\ 

        \cmidrule(lr){3-6} \cmidrule(lr){7-8}

        & & 
        \textbf{Aud.} & \textbf{Vis.} & \textbf{Tex.}  & \textbf{Par.} & 
        \textbf{Backbone} & \textbf{Encoder} & 
        &  & \\ 

        \midrule
        \multicolumn{11}{c}{\cellcolor{lightgray}\textbf{Visual-guided Synthesis}} \\
        Morgado et al. \cite{NEURIPS2018_01161aaa} & 2018 & \checkmark  & \checkmark  &     &   &  CNN  & ResNet18 & Connect & Supervised & \href{https://github.com/pedro-morgado/spatialaudiogen}{Github} \\
        Mono2binaural \cite{gao20192} &  2019 &  \checkmark  & \checkmark &   &  & U-Net & ResNet18  & Connect & Self-Supervised &\href{https://github.com/facebookresearch/2.5D-Visual-Sound}{Github} \\
        ASN \cite{lu2019self} & 2019 & \checkmark  & \checkmark  &     &   &  U-Net  & ResNet18 & Connect & Self-Supervised & No \\
        Sep-Stereo/APNet \cite{zhou2020sep} & 2020 &  \checkmark  & \checkmark &   & 
  & U-Net & ResNet18 & APNet & Multi-Task Learning & \href{https://github.com/SheldonTsui/SepStereo_ECCV2020}{Github} \\
        PseudoBinaural \cite{xu2021visually} & 2021 & \checkmark  & \checkmark &   & 
  & U-Net & ResNet18 & Connect & Multi-Task Learning & \href{https://github.com/SheldonTsui/PseudoBinaural_CVPR2021}{Github}\\
        Li et al. \cite{li2021binaural} & 2021 &  \checkmark  & \checkmark &   &  & U-Net & ResNet18  & Attention & Multi-Task Learning & No \\
        L2BNet \cite{rachavarapu2021localize} & 2021 & \checkmark  & \checkmark  &     &   &  U-Net  & ResNet18 & Attention & Semi-Supervised & No\\
        Lin et al. \cite{lin2021exploiting} & 2021 & \checkmark  & \checkmark  &     &   &  U-Net  & ResNet18 & Attention & Semi-Supervised & No \\
       
        MAFNet \cite{zhang2021multi} & 2021 & \checkmark  & \checkmark  &     &   &  U-Net  & ResNet18 & Attention & Self-Supervised & No\\
        Bmonobinaural \cite{parida2022beyond} & 2022 &  \checkmark  & \checkmark  &     &   &  U-Net  & ViT-Large & Attention & Supervised & No \\
        Points2Sound \cite{lluis2022points2sound} & 2022 & \checkmark  & \checkmark  &     &   &  Demucs  & ResNet18 & Conditioning & Supervised & \href{https://github.com/francesclluis/points2sound}{Github} \\
        Garg et al. \cite{garg2023visually} & 2023 &  \checkmark  & \checkmark &   & 
  & U-Net & ResNet18 & Connect & Multi-Task Learning & \href{https://github.com/bigharshrag/geometry-aware-binaural}{Github} \\
        CLUP \cite{li2024cyclic} & 2024 & \checkmark  & \checkmark  &     &   &  Diffusion Model  & ResNet18 & — & Cyclic Learning & No \\
        Liu et al. \cite{liu2024visually} & 2024 & \checkmark  & \checkmark  &     &   &  U-Net  & ResNet18 & Weighting & Contrastive Learning & No \\
        SAGM \cite{li2024cross} & 2024 & \checkmark  & \checkmark  &     &   &  GAN  & C3D & Connect & Supervised & No\\
        CCStereo \cite{chen2025ccstereo} & 2025 & \checkmark  & \checkmark  &     &   &  U-Net  & ResNet & AVAD & Contrastive Learning & No \\
        OmniAudio  \cite{liu2025omniaudio} & 2025 & \checkmark  & \checkmark  &     &   &  DiT  & MetaCLIP-Huge & — & Self-Supervised & \href{https://github.com/liuhuadai/OmniAudio}{Github} \\
        AV-NeRF \cite{NEURIPS2023_760dff0f} & 2023 &  \checkmark  & \checkmark  &    &  & A-NeRF & V-NeRF  & AV-Mapper & Supervised &\href{https://github.com/liangsusan-git/AV-NeRF}{Github}\\
        NeRAF \cite{brunetto2025neraf}  &  2025  & \checkmark  & \checkmark  &    &  & NAcF & NeRF  & Connect & Supervised &\href{https://github.com/AmandineBtto/NeRAF}{Github} \\
        AV-GS \cite{bhosale2024av} & 2024 & \checkmark  & \checkmark  &   &  & Acoustic Field & 3D-GS  & Connect & Supervised & No \\
        AV-Cloud \cite{NEURIPS2024_ff1f4141} & 2024 & \checkmark  & \checkmark  &  & \checkmark & AVCS & AV Anchors  & Attention & Supervised & \href{https://github.com/yoyomimi/AV-Cloud/}{Github}\\
        SOAF \cite{gao2024soaf} & 2024 & \checkmark  & \checkmark &   & \checkmark & NAF & SDFStudio  & Attention & Supervised & No \\
        AV-Surf \cite{baek2025av} & 2025 & \checkmark  & \checkmark &   &  & Transformer & ResNet/PointNet  & Attention & Supervised & No\\
        SoundVista \cite{chen2025soundvista} & 2025 & \checkmark  & \checkmark &   &  \checkmark & Transformer & ResNet18  & Attention & Supervised & No\\
        \midrule
        \multicolumn{11}{c}{\cellcolor{lightgray}\textbf{Text-guided Synthesis}} \\
        TAS \cite{li2024tas} & 2024 & \checkmark  &   & \checkmark &     &  Diffusion Model  & CLIP & SAF & Supervised & No\\
        DualSpec \cite{zhao2025dualspec} & 2025 & \checkmark  &   & \checkmark &     &  Diffusion Model  & FLAN-T5 & — & Semi-Supervised & No \\
        AudioSpa \cite{feng2025audiospa} & 2025 & \checkmark  &   & \checkmark &    &  Residual block & FLAN-T5 & Attention & Supervised & No \\
        SpatialTAS \cite{pan2025wild} & 2025 & \checkmark  &   & \checkmark &    &  Diffusion Model & FLAN-T5 & Attention & Supervised & No \\
        \midrule
        
        \multicolumn{11}{c}{\cellcolor{lightgray}\textbf{Joint Multi-modal Guided Synthesis}} \\
        SEE-2-SOUND \cite{dagli2024see} & 2024 &   & \checkmark  & \checkmark &   &  CoDi  & ViT-H/L & — & Zero-Shot & \href{https://github.com/see2sound/see2sound}{Github}\\
        SpatialSonic \cite{sun2024both} & 2025 & \checkmark & \checkmark  & \checkmark &    &  HTSAT  & Mask-RCNN/T5 & Attention & Supervised & \href{https://github.com/PeiwenSun2000/Both-Ears-Wide-Open}{Github}\\
        ImmerseDiffusion \cite{heydari2024immersediffusion} & 2025 & \checkmark  &   &  \checkmark &  \checkmark  &  DiT  & ELSA / CLAP & — & Supervised & No\\
        Diff-SAGe \cite{kushwaha2025diff} & 2025 &  &   &    & \checkmark  &  SiT  & — & — & Supervised & No \\
        ViSAGe \cite{kim2025visage} & 2025 &  &  \checkmark &   & \checkmark  &  Transformer  & CLIP & Attention & Supervised & \href{https://github.com/jaeyeonkim99/visage}{Github} \\
        ISDrama \cite{zhang2025isdrama} & 2025 & \checkmark & \checkmark & \checkmark & \checkmark & Mamba-Transformer & FLAN-T5 / CLIP & Attention & Supervised & No\\
        \bottomrule
        \addlinespace[0.5em]
        \multicolumn{11}{@{}p{\textwidth}@{}}{%
        \small \textbf{Abbreviations:} Aud(io), Vis(ual), Tex(t), Par(ametric). Parameter information comprises spatial and environmental parameters. Spatial parameters define source-listener relationships through location, orientation, and distance. Environmental parameters include room dimensions and reverberation characteristics.
    } \\
    \end{tabularx}
\end{table*}

\subsubsection{Visual-guided Synthesis}

Visual information from static images, videos, or 3D scene representations provides valuable cues about sound source characteristics, scene layout, and acoustic properties. These visual cues can enhance the spatial accuracy, environmental realism, and scene consistency of synthesized binaural audio. The field has evolved from basic fusion techniques to sophisticated modeling of audiovisual interactions and environmental acoustics. 

Early research demonstrated the effectiveness of combining visual and audio features. Mono2binaural \citep{gao20192} used a CNN to extract global visual features from video frames, which were then combined with audio features to guide the synthesis. To address the limited availability of annotated binaural audiovisual datasets, self-supervised learning became essential. Morgado et al. \cite{NEURIPS2018_01161aaa} developed a self-supervised method for learning audiovisual spatial correspondence from $360^\circ$ video. Similarly, the Audio Spatialization Network (ASN) \citep{lu2019self} employed self-supervision with an auxiliary classifier to learn spatial information implicitly.

Attention mechanisms later became vital for precise audiovisual association and feature fusion. These techniques allow models to focus on visual regions most relevant to current audio events, improving accuracy in complex scenes \citep{zhang2021multi,rachavarapu2021localize}. The Multi-Attention Fusion Network (MAFNet) \citep{zhang2021multi} used both self-attention within the visual modality and cross-modal attention to selectively integrate relevant visual features with audio. Similarly, Li et al. \cite{li2021binaural} applied attention mechanisms to effectively combine visual and audio features, while the Localize-to-Binauralize Network (L2BNet) \citep{rachavarapu2021localize} incorporated attention modules to strengthen the connections between visual cues and inferred source locations prior to synthesis.

Multi-task learning emerged as another effective strategy for enhancing model understanding. By jointly optimizing binaural synthesis with auxiliary tasks such as sound source separation \citep{zhou2020sep,xu2021visually}, models can develop a more comprehensive understanding of sound sources and their spatial layout. Further refinements included flipped audio classification \citep{li2021binaural} to encourage consistent spatial representations and left-right consistency enforcement between audio and visual modalities \citep{lin2021exploiting} to align the generated audio with the visual content both semantically and spatially.

The incorporation of 3D scene structure marked a significant step toward greater realism. Initial efforts used depth maps as additional input, as seen in Bmonobinaural \citep{parida2022beyond}, which leveraged depth as a proxy for distance information. Other approaches utilized explicit 3D geometric representations such as point clouds; for example, Points2Sound \citep{lluis2022points2sound} employed 3D sparse convolutional networks to process such representations. Geometric constraints, such as enforcing spatial consistency between audiovisual streams \citep{garg2023visually}, helped to refine the synthesis. To reduce reliance on paired binaural data, PseudoBinaural \citep{xu2021visually} used only visual information with HRIR models. Advanced self-supervised techniques like contrastive learning enhanced audiovisual representations; for instance, Contextual and Contrastive Stereophonic Learning (CCStereo) \citep{chen2025ccstereo} improved spatial sensitivity through negative-sample mining from shuffled visual features.

Recent trends show the integration of these advancements with powerful generative models and complex cross-modal frameworks. The Cyclic Locating-and-UPmixing (CLUP) model \citep{li2024cyclic} enables visual sound object localization and binaural generation to enhance each other through cyclic learning. The Stereo Audio Generation Model (SAGM) \citep{li2024cross} uses shared spatio-temporal visual information to guide both generator and discriminator components in a GAN. Liu et al. \cite{liu2024visually} proposed generating the left and right audio channels separately with visual guidance and introduced a cross-modal matching loss to explore audiovisual correlations. For $360^\circ$ video, OmniAudio \citep{liu2025omniaudio} uses a Transformer-based dual-branch architecture with self-supervised pre-training to process the complete visual context.

A distinct research direction focuses on modeling environment acoustics using 3D scene geometry derived from visual input. This approach aims for physical realism by simulating sound-environment interactions, going beyond research solely focused on RIR estimation \citep{luo2022learning, su2022inras, liang2023neural, ratnarajah2022mesh2ir,ratnarajah2024listen2scene}. These methods integrate environmental acoustic modeling directly into the spatial audio synthesis pipeline. A key technical approach is to adapt advanced 3D scene representation techniques, such as neural radiance fields (NeRF) and 3D gaussian splatting (GS). These methods excel at learning detailed 3D geometry from multi-view images, which then inform acoustic propagation models to render spatial audio with environment-specific effects. NeRF-based methods \citep{NEURIPS2023_760dff0f, brunetto2025neraf} explore the use of density fields for acoustic rendering. Methods based on GS, including AV-GS \citep{bhosale2024av}, AV-Cloud \citep{NEURIPS2024_ff1f4141}, Scene Occlusion-aware Acoustic Field (SOAF) \citep{gao2024soaf}, and AV-Surfs \citep{baek2025av}, leverage GS representations for acoustic simulations. Notably, AV-Surfs \citep{baek2025av} also estimates surface properties to determine acoustic materials, enabling more physically accurate environmental sound rendering.

\subsubsection{Text-guided Synthesis}

Text-driven spatial audio synthesis offers a more flexible control method compared to visual approaches. It allows users to specify desired sound field characteristics through natural language descriptions, reducing creation barriers and enabling more personalized audio experiences. The main challenge lies in translating unstructured language into structured spatial parameters or effective conditions for generative audio models.

This research area is emerging but shows significant potential. Methods typically follow a two-step process: First, using Natural Language Processing (NLP) techniques, especially Large Language Models (LLMs) or specialized semantic parsing, to extract key spatial information such as source type, location, motion, and environment from text descriptions \citep{tang2024can,devnani2024learning,zheng2024bat}; second, using this parsed structured information to guide spatial audio synthesis models.

While traditional parametric renderers can work with such structured information, recent research increasingly uses deep generative models for improved synthesis quality and flexibility. Text-guided Audio Spatialization (TAS) \citep{li2024tas} demonstrated the conversion monaural audio into spatial audio based on text prompts, offering an adaptable alternative to audiovisual methods. Similarly, SpatialTAS \citep{pan2025wild} employed a latent diffusion model conditioned by text embeddings to achieve flexible audio spatialization, allowing control over source direction, distance, and relative positions. AudioSpa \citep{feng2025audiospa} applied LLMs to process both acoustic and textual information, using fusion multi-head attention to integrate text tokens and enhance multi-modal learning capabilities. DualSpec \citep{zhao2025dualspec} implemented conditional diffusion models that generate high-quality, spatially controllable binaural audio directly from text descriptions.

\subsubsection{Joint Multi-modal Guided Synthesis}

Achieving comprehensive, robust, and interactive spatial audio synthesis requires a framework that can effectively combine multiple modalities, including the audio content itself, visual scene information, textual commands, and potential user interactions. Research in this area explores deep cross-modal learning and advanced generative models capable of handling diverse inputs.

Several recent works demonstrate this trend through advanced generative approaches. SpatialSonic \citep{sun2024both}, pre-trained on large-scale simulated data, shows how diffusion models accept multi-modal conditions for flexible spatial audio generation. SEE-2-SOUND \citep{dagli2024see} aims for zero-shot visual-to-spatial audio mapping, generating plausible spatial sound for novel visual scenes without specific paired training data, requiring strong model generalization capabilities.

Application-focused research is driving further integration across modalities. ImmerseDiffusion \citep{heydari2024immersediffusion} combines spatial, temporal, and environmental conditions within a Diffusion Transformer (DiT) model to generate immersive speech streams for communication contexts. Diff-SAGe \citep{kushwaha2025diff} generates first-order ambisonics conditioned on sound category and sound location, targeting applications that benefit from standardized ambisonic formats. For silent video applications, Video-to-Spatial Audio Generation (ViSAGe) \citep{kim2025visage} produces first-order ambisonics by using Contrastive Language-Image Pre-Training (CLIP) visual features and an autoregressive neural audio codec model that incorporates both directional and visual guidance. In creative applications like spatial drama generation, Immersive Spatial Drama (ISDrama) \citep{zhang2025isdrama} uses rich multi-modal prompts including scripts, video, and character poses to guide a Mamba-Transformer model. This approach includes specific mechanisms for unified pose encoding to address motion effects and aims for detailed prosody control in generated spatial dialogue.

\subsection{Datasets for Binaural Synthesis}

The advancement of DL-based binaural audio synthesis depends on suitable training datasets, with requirements that vary according to the task specifications. Table~\ref{tab:binaural_synthesis_datasets_revised} summarizes key public datasets available to researchers.

\begin{table*}[htbp] 
    \caption{Representative Public Datasets for Binaural Audio Synthesis.} 
    \label{tab:binaural_synthesis_datasets_revised} 
    \centering
    \setlength{\tabcolsep}{4pt} 
 
    \begin{tabularx}{\textwidth}{@{} l  c c l cccc cc c @{}} 
        \toprule
        \multirow{2}{*}{\textbf{Dataset}} &
        \multirow{2}{*}{\textbf{Year}} & 
        \multirow{2}{*}{\textbf{Type}} & 
        \multirow{2}{*}{\textbf{Scene}} & 
        \multicolumn{4}{c}{\parbox{3.0cm}{\centering\textbf{Modality}}} & 
        \multicolumn{2}{c}{\textbf{Scale}} &
        \multirow{2}{*}{\textbf{Link}}  \\ 

        \cmidrule(lr){5-8} \cmidrule(lr){9-10}

        & & & & 
        \textbf{A} & \textbf{V} & \textbf{T}  & \textbf{M} & \textbf{Hours} & \textbf{Samples} &  \\ 

        \midrule
        \multicolumn{11}{c}{\cellcolor{lightgray}\textbf{Audio-driven Synthesis Datasets}} \\
        Binaural Speech \cite{richard2021neural} & 2021 & Real & Regular room, Treated & \checkmark  &   &   &  \checkmark &   $\sim2$h & 8 speakers & \href{https://github.com/facebookresearch/BinauralSpeechSynthesis/releases/tag/v1.0}{Link} \\
        \midrule
        \multicolumn{11}{c}{\cellcolor{lightgray}\textbf{Multi-Model Guided Synthesis Datasets}} \\
        REC-Street \cite{NEURIPS2018_01161aaa} & 2018 &  Real &  Outdoor street  & \checkmark  & \checkmark  &   &    & 3.5h & 43 clips & \href{https://github.com/pedro-morgado/spatialaudiogen}{Link} \\
        YT-All\textsuperscript{a} \cite{NEURIPS2018_01161aaa} & 2018 & Real  &  Real-world   & \checkmark  & \checkmark  &   &  &  113.1h & 1146 clips  & \href{https://github.com/pedro-morgado/spatialaudiogen}{Link} \\
        FAIR-Play \cite{gao20192} & 2019 & Real  &  Music room  &  \checkmark  & \checkmark  &   &  &  5.2h & 1871 clips & \href{https://github.com/facebookresearch/FAIR-Play}{Link} \\ 
        MUSIC-Stereo \cite{xu2021visually} &  2021  &  Real  & Music performance   &\checkmark  & \checkmark  &   &  &  49.7h & 1,120 clips  & \href{https://github.com/roudimit/MUSIC_dataset}{Link}\\
        SimBinaural \cite{garg2023visually}  & 2023 & Sim.  &   —   & \checkmark  & \checkmark  &   &  & 116.1h & 21k clips  & \href{https://utexas.app.box.com/s/6ejm4yydh4963gvq27wrco6iikh1x7ye}{Link} \\ 
        YouTube-Binaural \cite{garg2023visually} & 2023  & Real  &  Real-world  & \checkmark  & \checkmark  &   &   & 27.7h & 426 clips  &  \href{https://utexas.app.box.com/s/9ufbto3d8i6yeoibkkpef0najder9fpp}{Link} \\ 
        BEWO-1M \cite{sun2024both} &  2025  &  Sim., Real  &  —  & \checkmark  &   & \checkmark  &  &  $\sim$2.8k h & $\sim$1M samples  & \href{https://github.com/PeiwenSun2000/Both-Ears-Wide-Open/tree/main/datasets}{Link} \\
         SoundSpaces \cite{chen2020soundspaces, NEURIPS2022_3a48b0ea} &  2022 
 &  Sim.  &  Arbitrary 3D mesh Env.  &  \checkmark  &  \checkmark &   &   &  —  & 1600+ scenes  & \href{https://github.com/facebookresearch/sound-spaces/tree/main}{Link} \\
        GWA \cite{tang2022gwa} & 2022 & Sim. & Diverse pro\text{-}designed houses & \checkmark  &  \checkmark &   & \checkmark  & — &  $>6.8$k scenes &  \href{https://github.com/GAMMA-UMD/GWA}{Link} \\
        RWAVS \cite{NEURIPS2023_760dff0f} &  2023  &  Real  &  Real-world, Multi-env  &  \checkmark  &  \checkmark &   & \checkmark   & 3.8h & $\sim$12k samples & \href{https://huggingface.co/datasets/susanliang/RWAVS}{Link}\\
        Replay \cite{shapovalov2023replay} &  2023  & Real &  Indoor social  & \checkmark  &  \checkmark &   & \checkmark  &  $>4$k min &  $>7$M samples   &  \href{https://github.com/facebookresearch/replay_dataset}{Link} \\
        SoundCam \cite{NEURIPS2023_a4289154} & 2023 & Real & Lab, Living, Meeting & \checkmark  &  \checkmark &   & \checkmark &  —  &  2k samples & \href{https://purl.stanford.edu/xq364hd5023}{Link}\\
        RAF \cite{chen2024real} & 2024 & Real & Real acoustic rooms & \checkmark  &  \checkmark &   & \checkmark &  —  & 47K/39k RIRs\textsuperscript{b} & \href{https://github.com/facebookresearch/real-acoustic-fields/}{Link}\\
        RealMAN \cite{NEURIPS2024_bf8f6f5b} & 2024 & Real & Indoor, Outdoor, Semi, Traffic & \checkmark  &   & \checkmark  & \checkmark  & 83.7/144.5h\textsuperscript{c} & — & \href{https://github.com/Audio-WestlakeU/RealMAN}{Link} \\
        SonicSim \cite{li2025sonicsim} & 2025 & Sim. & 3D scenes & \checkmark  &  \checkmark &   & \checkmark & $\sim$360h  & 90 scenes & \href{https://github.com/JusperLee/SonicSim}{Link} \\
        \bottomrule
        \addlinespace[0.5em]
        \multicolumn{11}{@{}p{\textwidth}@{}}{%
        \small \textbf{Abbreviations:} V: video frame/visual, A: audio/ambisonics, M: masks/metadata/motion, T: text, Sim.: simulated, Real: real-world measurement/recording. 
        
        \textsuperscript{a}The YT-All dataset contains the sub-datasets YT-Music(397 clips) and YT-Clean(496 clips). \textsuperscript{b}47K RIRs for empty rooms, 39K RIRs for furnished rooms. \textsuperscript{c}83.7h voice, 144.5h noise.
    } \\
    \end{tabularx}
\end{table*}

\subsubsection{Audio-driven Synthesis Datasets}

This category primarily encompasses tasks where monaural audio serves as the input for synthesizing binaural audio. High-quality synchronized monaural-binaural audio pairs from diverse acoustic conditions are essential for training robust models. However, publicly available datasets remain somewhat limited. The Binaural Speech dataset \citep{richard2021neural} represents an early contribution, containing approximately two hours of high-quality, real-world recordings of dynamic dialogues with head-tracking information. However, its limited scale constrains model performance across varied scenarios. To address this data scarcity, researchers have resorted to re-recording existing audio corpora under controlled binaural conditions \citep{huang2022end, manocha2023nord} or generating synthetic data through acoustic simulation. It is worth noting, however, that such synthetic datasets are not typically made public.

\subsubsection{Multi-modal Guided Synthesis Datasets}

These tasks require datasets that feature synchronized audio alongside additional guiding modalities such as visual content, textual descriptions, or explicit spatial information.

\textbf{Visual-guided Datasets.} Synthesis methods that leverage video or image content represent a significant area of research. Foundational work utilized real-world $360^\circ$ video datasets including REC-Street, YT-Clean, and YT-Music \citep{NEURIPS2018_01161aaa}, often for self-supervised learning approaches that paired visual content with spatial or binaural audio. Music-focused applications benefit from specialized datasets such as FAIR-Play \citep{gao20192} and MUSIC-Stereo \citep{xu2021visually}. To overcome limitations of real-world data availability, researchers have developed large-scale synthetic or processed datasets. SimBinaural \citep{garg2023visually} provides paired video, binaural audio, and RIRs from simulated 3D indoor scenes. The YouTube-Binaural dataset \citep{garg2023visually} extends accessibility by converting ambisonic audio from real YouTube videos into pseudo-binaural labels for training.

\textbf{Text-guided Datasets.} Text-guided synthesis represents an emerging area that often necessitates custom dataset creation. The Text-guided Audio Spatialization Benchmark (TASBench) dataset \citep{li2024tas} establishes the text-guided audio spatialization task, featuring dense, frame-level text annotations for evaluating fine-grained control capabilities; this dataset is not open source. The large-scale Both Ears Wide Open 1M (BEWO-1M) dataset \citep{sun2024both} provides spatial audio paired with detailed textual descriptions and optional images, supporting broader text and image-to-spatial audio tasks. 

\textbf{Environment-related Datasets.} Research incorporating 3D geometry or environment acoustics requires datasets with richer contextual information. The SoundSpaces dataset \citep{chen2020soundspaces,NEURIPS2022_3a48b0ea} offers navigable 3D environments derived from scans such as Matterport3D \citep{8374622}, providing binaural RIRs for early audio-visual navigation studies. For real-world applications, the Real-world Audio-Visual Scene (RWAVS) dataset \citep{NEURIPS2023_760dff0f} features scenes captured by a moving camera, including video, binaural, and source audio, along with pose data for models requiring geometric awareness. The Replay dataset \citep{shapovalov2023replay} provides over 4000 minutes of real indoor social interactions with multi-view video and multi-channel or binaural audio recordings. Additional relevant datasets support acoustic modeling research, including the Geometric Wave Acoustic (GWA) dataset \citep{tang2022gwa}, SoundCam \citep{NEURIPS2023_a4289154}, Real Acoustic Fields (RAF) \citep{chen2024real}, Real-recorded and Annotated Microphone Array Speech \& Noise (RealMAN) \citep{NEURIPS2024_bf8f6f5b}, and SonicSet \citep{li2025sonicsim}. These resources, while less commonly used in the reviewed synthesis methods, offer valuable data for specialized audio processing tasks.

\subsection{Evaluation Methodologies for Binaural Audio Synthesis}

Evaluating DL-based binaural audio synthesis requires a comprehensive approach that assesses both signal fidelity, spatial accuracy, and the overall perceptual quality of the generated sound field. Evaluation methodologies include subjective listening tests and objective metrics; the latter category encompasses both signal-based measures and model-based approaches that predict perceptual attributes.

\subsubsection{Subjective Listening Tests} 

Subjective listening tests remain the gold standard for evaluating the perceptual quality of synthesized binaural audio. The fundamental methodologies for conducting these assessments were detailed in Section~\ref{sec:ii-e}.

These tests are designed to evaluate key aspects of the synthesized listening experience \citep{he2025dual,richard2021neural}. Listeners typically compare synthesized audio with ground-truth signals, providing ratings on dimensions such as overall audio quality, spatial realism, timbral naturalness, and immersion. Mean Opinion Score (MOS) ratings are used frequently in these evaluations and can be adapted to address specific attributes relevant to the synthesis task, such as 'spatial MOS' or 'comparison MOS'. For multi-modal synthesis, the evaluation must also assess the coherence and plausibility of the synthesized audio relative to the guiding non-auditory cues \citep{li2024tas}, as well as the overall perceived realism of the combined multi-modal scene.

\subsubsection{Objective Metrics for Synthesized Audio}

Objective metrics quantify differences between synthesized and ground-truth signals or assess adherence to acoustic and spatial goals. These metrics vary based on the synthesis task and fall into signal-based measures and model-based perceptual predictors. Table~\ref{tab:binaural_synthesis_objective_metrics_final} presents common metrics used in this field.

\begin{table*}[htbp] 
    \caption{Summary of Common Objective Evaluation Metrics for Binaural Audio Synthesis.}
    \label{tab:binaural_synthesis_objective_metrics_final}
    \centering
    \begin{tabularx}{\textwidth}{@{} l | c | c | X @{}}
        \toprule 
        \textbf{Metric} & \textbf{Formula} & \textbf{Focus}  & \multicolumn{1}{c}{\textbf{Task}}  \\
        \midrule 
          WAV \cite{richard2021neural} $\downarrow$ &  $ \sqrt{\sum_n |\mathbf{x}(n) - \hat{\mathbf{x}}(n)|^2}$ & Similarity & Measures waveform similarity. \\
          \midrule
          STFT Distance \cite{gao20192} $\downarrow$ &  $ \lVert X(t) - \hat{X}(t) \|_2 $ & Similarity & Measures time-frequency similarity. \\
         \midrule
          Mag (MAG) \cite{xu2021visually} $\downarrow$ &  $\sum_{t,f} |\log|X(t,f)| - \log|\hat{X}(t,f)||$  & Similarity & Measures magnitude spectrum similarity. \\
          \midrule
          Phs \cite{xu2021visually} $\downarrow$ & $\sum_{t,f} |\angle X(t,f) - \angle \hat{X}(t,f)|$  & Similarity & Measures phase spectrum similarity. \\
          \midrule
           ENV \cite{NEURIPS2018_01161aaa} $\downarrow$ &  $\left\| E[\mathbf{x}(t)] - E[\hat{\mathbf{x}}(t)] \right\|_2 $   & Similarity & Measures temporal envelope similarity.   \\
         \midrule
          LRE \cite{chen2023novel} $\downarrow$ &  $|10 \log_{10}(\frac{E_L}{E_R}) - 10 \log_{10}(\frac{\hat{E}_L}{\hat{E}_R})|$   & Spatial accuracy & Measures left-right energy balance error. \\
          \midrule
          ITD/ILD Error \cite{zhu2024end} $\downarrow$ &  $\left| \mathrm{ITD}(\hat{\mathbf{x}}) - \mathrm{ITD}(\mathbf{x}) \right|, 10 \log_{10} \left( \frac{\mathrm{ILD}(\hat{\mathbf{x}})}{\mathrm{ILD}(\mathbf{x})} \right)$   & Spatial accuracy & Measures interaural cue (ITD/ILD) accuracy. \\
          \midrule
        FAD \cite{kilgour2018fr} $\downarrow$ &   —   & Perceptual quality & Assesses perceptual realism via embeddings. \\
        \bottomrule 
        \addlinespace[0.5em]
        \multicolumn{4}{@{}p{\textwidth}@{}}{
        \small \textbf{Abbreviations:} WAV: waveform L2, STFT Distance: short-time fourier transform distance, Mag (MAG): magnitude distance, Phs: phase distance, ENV: envelope distance, LRE: left-right energy ratio, ITD/ILD Error: interaural time/level difference error, FAD: Fréchet audio distance. 
        
        The symbols used are: $\mathbf{x}=[x_L(n), x_R(n)]^T$ is the ground-truth discrete-time binaural signal, $\hat{\mathbf{x}}=[\hat{x}_L(n), \hat{x}_R(n)]^T$ is the synthesized signal. $X(t,f)$ and $\hat{X}(t,f)$ are their respective STFTs (time frame $t$, frequency $f$). $E = \sum_n x^2(n)$ is energy. $\mathrm{ITD}(\cdot)$ and $\mathrm{ILD}(\cdot)$ (in dB) are functions for interaural cues. Arrows ($\downarrow$ / $\uparrow$) indicate desirable direction.
} \\
    \end{tabularx}
\end{table*}

A primary category of metrics assesses waveform and spectral similarity. Direct time-domain comparisons often use the L2 distance between waveforms (WAV) \citep{richard2021neural, NEURIPS2022_95f03faf}. In the frequency domain, the short-time Fourier transform (STFT) distance \citep{gao20192} measures the overall dissimilarity in time-frequency representation. More specifically, magnitude spectra distance (Mag/MAG) \citep{xu2021visually} and phase spectra distance (Phs) \citep{xu2021visually} are often evaluated separately, as these components contribute differently to perception. The envelope distance (ENV) \citep{NEURIPS2018_01161aaa} assesses structural similarity over time by comparing temporal energy contours, which relate to perceived dynamics and transients. Metrics evaluating spatial accuracy are critical for binaural synthesis. The left-right energy ratio error (LRE) \citep{chen2023novel} measures discrepancies in the energy balance between channels, which relates to perceived source width or lateral balance. Errors in key binaural cues, such as ITD and ILD (ITD/ILD Error) \citep{zhu2024end}, provide direct measures of how accurately fundamental localization cues are reproduced. For generative models that lack a direct ground-truth comparison, metrics like Fréchet-audio distance (FAD) \citep{kilgour2018fr} are used. FAD compares the statistical distributions of embeddings from synthesized audio and a set of real target examples to assess overall perceptual realism and diversity.

\section{Applications and Impact of DL-powered Spatial Audio}
\label{sec:iv}

Spatial sound provides rich directional, distance, and environmental information, making it a crucial sensory modality. Deep learning (DL) techniques have significantly improved the generation, processing, and integration of spatial audio into computational systems, enhancing performance across numerous applications. These advances enable spatial audio to serve both as a synthesis target for immersive experiences and as an essential input for intelligent systems focused on perception, interaction, and environment understanding. This section explores these advancements through two main categories: applications that directly enhance human experience and interaction, and those that enable intelligent systems and environmental understanding.

\subsection{Enhancing Human Experience and Interaction}

DL-driven spatial audio enhances applications focused on human perception, communication, and immersion by creating more realistic and engaging sound environments.

\subsubsection{Virtual and Augmented Reality} 

High-quality spatial audio is essential for creating immersion and presence in VR/AR \citep{bosman2024effect, correa2023spatial}. DL-powered synthesis contributes significantly to both the “place illusion" (the feeling of being there) and the “plausibility illusion" (the feeling that the scenario is real), making virtual experiences more authentic \citep{slater2009place, bosman2024effect}. Beyond realism, it serves as an attention guidance mechanism, directing users toward important events outside their limited field-of-view \citep{eames2019beyond, ulsamer2020brain}. It provides interaction feedback, confirms user actions, and improves social presence in multi-user applications by facilitating speaker localization and identification \citep{Larsson2010}. In AR applications, DL-based approaches are critical for integrating virtual sounds with real environments by correctly positioning them relative to physical objects \citep{gupta2022augmented}. 

\subsubsection{Hearing Aids and Assistive Technologies} 

For hearing accessibility, DL-driven spatial audio processing offers powerful new solutions. Modern hearing aids can incorporate DL models that leverage spatial cues and perform advanced directional filtering - based on principles like the “cocktail party effect” \citep{hawley2004benefit} - to improve speech clarity in noisy settings. These systems enhance comprehension by spatially separating speech from background noise, potentially restoring spatial hearing abilities for those with hearing loss \citep{zheng2023sixty}. Furthermore, DL-based synthesis enables the creation of realistic virtual auditory environments for audiological assessment and rehabilitation \citep{hohmann2020virtual, pedersen2023virtual, chitra2025ai}, allowing clinicians to create complex listening scenarios for testing and therapy.

\subsubsection{Telepresence and Enhanced Communication} 

In teleconferencing, spatial audio improves speech intelligibility by spatially separating speakers, thus reducing listening effort. This approach mimics real-world conversation dynamics, helping listeners focus on specific speakers in multi-talker scenarios. It enhances shared presence in virtual meetings by accurately representing participant locations within a common virtual acoustic space. This contributes to natural turn-taking cues and a stronger sense of co-location, particularly valuable in VR collaboration platforms \citep{bosman2024effect} and hybrid meetings that bridge remote and physically present participants.

\subsection{Enabling Intelligent Systems and Environmental Understanding}

Beyond human experience, spatial audio cues interpreted by DL models equip intelligent systems with improved capabilities for perception, navigation, and interaction with the physical world. These advances also offer valuable tools for scientific research and design.

\subsubsection{Audio-Visual Navigation and Robotics} 

Spatial sound guides agents or robots through environments, especially toward sound-emitting objects. Current research enables agents to navigate using combined audio-visual inputs. Key challenges that are being addressed with DL include locating static or dynamic sound sources \citep{yu2022pay, younes2023catch}, understanding sound semantic context \citep{tatiya2022knowledge, chen2021semantic}, maintaining robustness amid distractors or noise \citep{YinfengICLR2022saavn}, and bridging the simulation-to-reality gap \citep{chen2024sim2real}. Effective approaches incorporate multi-modal fusion, attention mechanisms, and reinforcement learning \citep{chen2021waypoints} to utilize directional audio cues for localization and path planning.

\subsubsection{Acoustic Scene Understanding and Depth Estimation} 

Sound reflections, echoes, and source properties contain valuable geometric and semantic information about surrounding spaces. Studies show that spatial audio improves visual depth estimation in unclear regions and reveals properties of areas outside direct view \citep{zhang2022stereo, zhu2022beyond, parida2021beyond}. DL models employ cross-modal fusion techniques to combine audio spatial cues with visual information, creating more accurate 3D scene reconstruction \citep{yun2023dense}. Spatial sound also enables audio-based semantic segmentation of environments \citep{vasudevan2020semantic, sokolov20243d}.

\subsubsection{Cross-modal Representation Learning for Perception} 

The physical relationship between sound propagation and spatial environments makes spatial audio an effective supervisory signal for learning robust representations across modalities. Models inspired by echolocation can use sound to develop spatial representations from visual data \citep{gao2020visualechoes}. Training advanced perception models relies on the creation of spatially consistent audio-visual data, a task for which DL-based synthesis is well-suited \citep{roman2025generating}. Additionally, modeling how physical bodies affect sound fields helps create accurate virtual representations and improves understanding of acoustic interactions \citep{NEURIPS2023_8c234d9c}.

\section{Challenges and Future Directions}
\label{sec:v}
DL has significantly advanced spatial audio reproduction technology, demonstrating great potential in HRTF personalization, binaural audio synthesis, and related applications. However, several key challenges remain unresolved. This section summarizes current research bottlenecks and explores future development trends. 

\subsection{Data Availability and Diversity}

The performance of data-driven DL methods is fundamentally limited by the quality and quantity of training data. Acquiring suitable datasets remains a major challenge in spatial audio reproduction research. High-quality HRTF measurement demands specialized equipment and controlled acoustic environments. Consequently, public datasets are limited in number and exhibit significant variations in measurement conditions and spatial sampling protocols. These inconsistencies affect cross-study comparability and model generalization capabilities \citep{andreopoulou2015inter, pauwels2023relevance}. Binaural audio synthesis faces even greater challenges in obtaining large-scale real-world paired data \citep{chen2024real}. This challenge grows more pronounced for complex scenarios with dynamic interactions and multi-modal inputs that need precise timing synchronization and careful annotation.

Future research should focus on developing larger-scale, more diverse datasets that follow standardized protocols and remain openly accessible. Effective techniques for combining heterogeneous data, as discussed in Section~\ref{sec:hrtf}, are essential, with INRs showing promise for handling irregular sampling patterns. Improved data normalization methods are needed to reduce biases between different datasets. While physical acoustic simulations offer a valuable tool for augmenting training data \citep{katz2001boundary, li2025sonicsim}, the field must also address the simulation-to-reality gap to ensure models trained on synthetic data perform robustly in the real world \citep{guezenoc2020wide}. Furthermore, research into less-supervised learning paradigms - including self-supervised, pseudo-supervised and zero-shot learning – is crucial for reducing the reliance on costly annotated data. As data privacy concerns grow, distributed training frameworks like federated learning also warrant exploration \citep{li2020federated, zhang2023fedaudio}.

\subsection{Perceptual Validity and Evaluation}

A critical limitation of current research is the difficulty in accurately evaluating the perceptual effect of synthesized spatial audio. Most objective evaluation metrics reflect signal-level similarity but correlate poorly with human auditory perception \citep{andreopoulou2014evaluating, kim2020investigation, rusk2024comparing}, particularly for complex attributes like immersion and externalization. This disconnect can lead to model optimization that diverges from the actual user experience. While subjective listening experiments provide the most reliable perceptual assessment, they are resource-intensive, often lack standardization, and are subject to significant inter-listener variability. Auditory models (AMs) offer a promising alternative by predicting perceptual outcomes like localization performance, but current models require improved accuracy, generalization, and coverage of complex perceptual phenomena \citep{majdak2022amt}.

Future work requires methodological breakthroughs in evaluation. Developing objective metrics that correlate strongly with key dimensions of auditory perception is essential. This might involve integrating more sophisticated psychoacoustic principles or using DL to directly predict subjective ratings from audio signals \citep{manocha2023spatialization, manocha2022saqam}. The field would benefit from standardized subjective evaluation protocols and more efficient assessment methods, such as online crowdsourcing platforms or immersive virtual environments \citep{cuevas20193d}. Enhanced AMs that can simulate individual differences and perception in complex multi-modal scenarios represent another important research direction \citep{baumgartner2014modeling, barumerli2023bayesian}. Finally, incorporating explainable artificial intelligence (XAI) techniques can help clarify relationships between a model's internal behavior and its perceptual outcomes \citep{de2025data}.

\subsection{Generalization Ability and Robustness}

DL models often exhibit decreased performance when deployed in acoustic environments that differ from their training data. This issue is a major barrier to the practical deployment of spatial audio technologies. Models must generalize not only to new users and acoustic conditions but also handle reverberation, ambient noise, diverse sound sources, and variations in playback devices. They should also respond effectively to dynamic changes, such as head and source movements.

Addressing the gap between training data and real-world applications will likely require a combination of domain adaptation and transfer learning techniques \citep{ganin2016domain, wilson2020survey}. Well-designed data augmentation strategies help simulate a wider range of real-world conditions during training. Robust optimization methods, like adversarial training, can enhance model stability \citep{salamon2017deep, NIPS2014_f033ed80}. A particularly promising direction is the integration of physical acoustic laws as prior knowledge; physics-informed neural networks (PINNs) \citep{raissi2019physics,karniadakis2021physics} have demonstrated potential in HRTF modeling for improved physical realism and generalization from sparse data \citep{ma2023spatial,nair2024physics,olivieri2024physics,pezzoli2023implicit,10819673,chen2025spatial,luan2025acoustic}. Research on continual learning and online adaptation could enable models to adjust their parameters after deployment based on user feedback or changing environmental acoustics. Neural field models offer efficient representation and rendering of dynamic acoustic scenes \citep{luo2022learning,brunetto2025neraf}.

\subsection{Controllability, Interpretability, and Interactivity}

Many advanced DL models, particularly end-to-end generative models, operate as "black boxes". Their internal mechanisms lack transparency, and users have limited means of exercising detailed control over the generated output. These limitations restrict their application in creative applications, complicate debugging, and hinder the development of truly personalized interaction systems.

XAI techniques are essential for revealing the decision-making processes of these models. Methods such as feature attribution and concept analysis can help diagnose model behaviour by examining how key acoustic features are represented internally \citep{samek2017explainable, de2025data}. Designing modular or structured generative architectures that disentangle different auditory attributes could enable more precise user control \citep{higgins2017beta, NEURIPS2020_4c5bcfec}. More intuitive user interfaces - ones that allow interaction through speech, text, gestures, or physiological signals - are also necessary to bridge the gap between user intent and model output. Advanced conditional generative models are needed to respond to higher-level semantic commands, enabling truly personalized and creative spatial audio \citep{li2024tas, zhao2025dualspec, sun2024both}.

\subsection{Computational Efficiency and Real-time Capability}

High-performance DL models, including large-scale Transformers and diffusion models, often have substantial computational resources and memory requirements. This conflicts with the need for real-time processing and energy efficiency, which is particularly critical for deployment on mobile devices, VR/AR headsets, and hearing aids. 

Model compression techniques, including knowledge distillation \citep{hinton2015distilling}, network pruning, and quantization \citep{NIPS2015_ae0eb3ee}, are effective for reducing deployment requirements. Designing lightweight network architectures optimized for the specific characteristics of audio signals has also shown significant promise \citep{lee2023neural}. For generative models, improved sampling algorithms remain essential for real-time performance, especially for iterative models like diffusion models \citep{salimans2022progressive, song2020denoising}. While specialized hardware accelerators like graphics processing units (GPUs) can enhance processing speed, future advances will likely depend on hardware-software co-design to create high-performance, real-time spatial audio systems that balance computational costs with perceptual quality.
\section{Conclusion}

This survey has reviewed recent advances in the application of Deep Learning (DL) to spatial audio reproduction, with a particular focus on personalized binaural techniques. Our analysis demonstrates that DL is not merely an incremental improvement but is fundamentally reshaping core spatial audio technologies. In particular, DL has transformed HRTF modeling by enabling data-driven personalization at a scale previously unattainable. Simultaneously, significant progress in end-to-end binaural audio synthesis has facilitated robust spatial cue recovery and the sophisticated integration of multi-modal information. These technological advancements have a profound dual impact: they are creating more immersive and interactive environments for human listeners while also empowering intelligent systems with a more sophisticated understanding of the acoustical world. Despite this considerable progress, critical challenges remain in data availability, perceptual evaluation, and model performance. Addressing these bottlenecks is vital for the next generation of research, which will push spatial audio technologies towards greater realism, personalization, and accessibility.

\section*{Acknowledgments}
This work is funded by the Special Fund for Talent Development in AI-Enabled Interdisciplinary Programs at East China Normal University (2024JCRC-08) and the National Natural Science Foundation of China (12411530075).

\bibliographystyle{IEEEtran}
\bibliography{IEEEabrv, reference}

\begin{thebibliography}{100}
\providecommand{\url}[1]{#1}
\csname url@samestyle\endcsname
\providecommand{\newblock}{\relax}
\providecommand{\bibinfo}[2]{#2}
\providecommand{\BIBentrySTDinterwordspacing}{\spaceskip=0pt\relax}
\providecommand{\BIBentryALTinterwordstretchfactor}{4}
\providecommand{\BIBentryALTinterwordspacing}{\spaceskip=\fontdimen2\font plus
\BIBentryALTinterwordstretchfactor\fontdimen3\font minus
  \fontdimen4\font\relax}
\providecommand{\BIBforeignlanguage}[2]{{%
\expandafter\ifx\csname l@#1\endcsname\relax
\typeout{** WARNING: IEEEtran.bst: No hyphenation pattern has been}%
\typeout{** loaded for the language `#1'. Using the pattern for}%
\typeout{** the default language instead.}%
\else
\language=\csname l@#1\endcsname
\fi
#2}}
\providecommand{\BIBdecl}{\relax}
\BIBdecl

\bibitem{moore2012introduction}
B.~C. Moore, \emph{An introduction to the psychology of hearing}.\hskip 1em
  plus 0.5em minus 0.4em\relax Leiden: Brill, 2012.

\bibitem{blauert1997spatial}
J.~Blauert, \emph{Spatial hearing: the psychophysics of human sound
  localization}.\hskip 1em plus 0.5em minus 0.4em\relax Cambridge: MIT press,
  1997.

\bibitem{rajguru2020spatial}
C.~Rajguru, M.~Obrist, and G.~Memoli, ``Spatial soundscapes and virtual worlds:
  Challenges and opportunities,'' \emph{Frontiers in Psychology}, vol.~11, p.
  569056, 2020.

\bibitem{xie2020spatial}
B.~Xie, ``Spatial sound-history, principle, progress and challenge,''
  \emph{Chinese Journal of Electronics}, vol.~29, no.~3, pp. 397--416, 2020.

\bibitem{wendt2014computationally}
T.~Wendt, S.~Van De~Par, and S.~D. Ewert, ``A computationally-efficient and
  perceptually-plausible algorithm for binaural room impulse response
  simulation,'' \emph{Journal of the Audio Engineering Society}, vol.~62,
  no.~11, pp. 748--766, 2014.

\bibitem{kirsch2021low}
C.~Kirsch and S.~D. Ewert, ``Low-order filter approximation of diffraction for
  virtual acoustics,'' in \emph{2021 IEEE Workshop on Applications of Signal
  Processing to Audio and Acoustics (WASPAA)}.\hskip 1em plus 0.5em minus
  0.4em\relax IEEE, 2021, pp. 341--345.

\bibitem{ewert2022filter}
S.~D. Ewert, ``A filter representation of diffraction at infinite and finite
  wedges,'' \emph{JASA Express Letters}, vol.~2, no.~9, 2022.

\bibitem{kirsch2023universal}
C.~Kirsch and S.~D. Ewert, ``A universal filter approximation of edge
  diffraction for geometrical acoustics,'' \emph{IEEE/ACM Transactions on
  Audio, Speech, and Language Processing}, vol.~31, pp. 1636--1651, 2023.

\bibitem{pulkki2019machine}
V.~Pulkki and U.~P. Svensson, ``Machine-learning-based estimation and rendering
  of scattering in virtual reality,'' \emph{The Journal of the Acoustical
  Society of America}, vol. 145, no.~4, pp. 2664--2676, 2019.

\bibitem{kirsch2023computationally}
C.~Kirsch, T.~Wendt, S.~Van De~Par, H.~Hu, and S.~D. Ewert,
  ``Computationally-efficient simulation of late reverberation for
  inhomogeneous boundary conditions and coupled rooms,'' \emph{Journal of the
  Audio Engineering Society}, vol.~71, no.~4, pp. 186--201, 2023.

\bibitem{pelzer2014integrating}
S.~Pelzer, L.~Asp{\"o}ck, D.~Schr{\"o}der, and M.~Vorl{\"a}nder, ``Integrating
  real-time room acoustics simulation into a cad modeling software to enhance
  the architectural design process,'' \emph{Buildings}, vol.~4, no.~2, pp.
  113--138, 2014.

\bibitem{brinkmann2019round}
F.~Brinkmann, L.~Asp{\"o}ck, D.~Ackermann, S.~Lepa, M.~Vorl{\"a}nder, and
  S.~Weinzierl, ``A round robin on room acoustical simulation and
  auralization,'' \emph{The Journal of the Acoustical Society of America}, vol.
  145, no.~4, pp. 2746--2760, 2019.

\bibitem{savioja2015overview}
L.~Savioja and U.~P. Svensson, ``Overview of geometrical room acoustic modeling
  techniques,'' \emph{The Journal of the Acoustical Society of America}, vol.
  138, no.~2, pp. 708--730, 2015.

\bibitem{seeber2017interactive}
B.~U. Seeber and S.~W. Clapp, ``Interactive simulation and free-field
  auralization of acoustic space with the {rtSOFE},'' \emph{The Journal of the
  Acoustical Society of America}, vol. 141, no. 5\_Supplement, p. 3974, 2017.

\bibitem{schroder2011physically}
D.~Schr{\"o}der, \emph{Physically based real-time auralization of interactive
  virtual environments}.\hskip 1em plus 0.5em minus 0.4em\relax Berlin: Logos
  Verlag Berlin GmbH, 2011, vol.~11.

\bibitem{schissler2016efficient}
C.~Schissler, A.~Nicholls, and R.~Mehra, ``Efficient {HRTF-based} spatial audio
  for area and volumetric sources,'' \emph{IEEE Transactions on Visualization
  and Computer Graphics}, vol.~22, no.~4, pp. 1356--1366, 2016.

\bibitem{ranjan2015natural}
R.~Ranjan and W.-S. Gan, ``Natural listening over headphones in augmented
  reality using adaptive filtering techniques,'' \emph{IEEE/ACM Transactions on
  Audio, Speech, and Language Processing}, vol.~23, no.~11, pp. 1988--2002,
  2015.

\bibitem{Larsson2010}
P.~Larsson, A.~V{\"a}ljam{\"a}e, D.~V{\"a}stfj{\"a}ll, A.~Tajadura-Jim{\'e}nez,
  and M.~Kleiner, \emph{Auditory-Induced Presence in Mixed Reality Environments
  and Related Technology}.\hskip 1em plus 0.5em minus 0.4em\relax London:
  Springer London, 2010, pp. 143--163.

\bibitem{berkhout1993acoustic}
A.~J. Berkhout, D.~de~Vries, and P.~Vogel, ``Acoustic control by wave field
  synthesis,'' \emph{The Journal of the Acoustical Society of America},
  vol.~93, no.~5, pp. 2764--2778, 1993.

\bibitem{daniel2003further}
J.~Daniel, S.~Moreau, and R.~Nicol, ``Further investigations of high-order
  ambisonics and wavefield synthesis for holophonic sound imaging,'' in
  \emph{Audio Engineering Society Convention 114}.\hskip 1em plus 0.5em minus
  0.4em\relax Audio Engineering Society, 2003.

\bibitem{rumsey2012spatial}
F.~Rumsey, \emph{Spatial audio}.\hskip 1em plus 0.5em minus 0.4em\relax
  Routledge, 2012.

\bibitem{wightman1989headphone}
F.~L. Wightman and D.~J. Kistler, ``Headphone simulation of free-field
  listening. {I: Stimulus} synthesis,'' \emph{The Journal of the Acoustical
  Society of America}, vol.~85, no.~2, pp. 858--867, 1989.

\bibitem{annurev:/content/journals/10.1146/annurev.ps.42.020191.001031}
J.~C. Middlebrooks and D.~M. Green, ``Sound localization by human listeners,''
  \emph{Annual Review of Psychology}, vol.~42, no. 1991, pp. 135--159, 1991.

\bibitem{xie2013head}
B.~Xie, \emph{Head-related transfer function and virtual auditory
  display}.\hskip 1em plus 0.5em minus 0.4em\relax J. Ross Publishing, 2013.

\bibitem{moller1992fundamentals}
H.~M{\o}ller, ``Fundamentals of binaural technology,'' \emph{Applied
  Acoustics}, vol.~36, no. 3-4, pp. 171--218, 1992.

\bibitem{middlebrooks1999individual}
J.~C. Middlebrooks, ``Individual differences in external-ear transfer functions
  reduced by scaling in frequency,'' \emph{The Journal of the Acoustical
  Society of America}, vol. 106, no.~3, pp. 1480--1492, 1999.

\bibitem{wenzel1993localization}
E.~M. Wenzel, M.~Arruda, D.~J. Kistler, and F.~L. Wightman, ``Localization
  using nonindividualized head-related transfer functions,'' \emph{The Journal
  of the Acoustical Society of America}, vol.~94, no.~1, pp. 111--123, 1993.

\bibitem{moller1996binaural}
H.~M{\o}ller, M.~F. S{\o}rensen, C.~B. Jensen, and D.~Hammersh{\o}i, ``Binaural
  technique: Do we need individual recordings?'' \emph{Journal of the Audio
  Engineering Society}, vol.~44, no.~6, pp. 451--469, 1996.

\bibitem{lee2018personalized}
G.~W. Lee and H.~K. Kim, ``Personalized {HRTF} modeling based on deep neural
  network using anthropometric measurements and images of the ear,''
  \emph{Applied Sciences}, vol.~8, no.~11, p. 2180, 2018.

\bibitem{li2020measurement}
S.~Li and J.~Peissig, ``Measurement of head-related transfer functions: A
  review,'' \emph{Applied Sciences}, vol.~10, no.~14, p. 5014, 2020.

\bibitem{algazi2001cipic}
V.~R. Algazi, R.~O. Duda, D.~M. Thompson, and C.~Avendano, ``The {CIPIC HRTF}
  database,'' in \emph{Proceedings of the 2001 IEEE Workshop on the
  Applications of Signal Processing to Audio and Acoustics (Cat. No.
  01TH8575)}.\hskip 1em plus 0.5em minus 0.4em\relax IEEE, 2001, pp. 99--102.

\bibitem{katz2001boundary}
B.~F. Katz, ``Boundary element method calculation of individual head-related
  transfer function. {I. Rigid} model calculation,'' \emph{The Journal of the
  Acoustical Society of America}, vol. 110, no.~5, pp. 2440--2448, 2001.

\bibitem{brinkmann2019cross}
F.~Brinkmann, M.~Dinakaran, R.~Pelzer, P.~Grosche, D.~Voss, and S.~Weinzierl,
  ``A cross-evaluated database of measured and simulated {HRTFs} including {3D}
  head meshes, anthropometric features, and headphone impulse responses,''
  \emph{Journal of the Audio Engineering Society}, vol.~67, no.~9, pp.
  705--718, 2019.

\bibitem{pulkki1997virtual}
V.~Pulkki, ``Virtual sound source positioning using vector base amplitude
  panning,'' \emph{Journal of the Audio Engineering Society}, vol.~45, no.~6,
  pp. 456--466, 1997.

\bibitem{bruschi2023new}
V.~Bruschi, N.~Dourou, A.~Carini, and S.~Cecchi, ``A new {HRTF} interpolation
  approach for nonlinear {3D} audio systems,'' in \emph{2023 Immersive and 3D
  Audio: from Architecture to Automotive (I3DA)}.\hskip 1em plus 0.5em minus
  0.4em\relax IEEE, 2023, pp. 1--9.

\bibitem{kistler1992model}
D.~J. Kistler and F.~L. Wightman, ``A model of head-related transfer functions
  based on principal components analysis and minimum-phase reconstruction,''
  \emph{The Journal of the Acoustical Society of America}, vol.~91, no.~3, pp.
  1637--1647, 1992.

\bibitem{gebru2021implicit}
I.~D. Gebru, D.~Markovi{\'c}, A.~Richard, S.~Krenn, G.~A. Butler, F.~De~la
  Torre, and Y.~Sheikh, ``Implicit {HRTF} modeling using temporal convolutional
  networks,'' in \emph{ICASSP 2021-2021 IEEE International Conference on
  Acoustics, Speech and Signal Processing (ICASSP)}.\hskip 1em plus 0.5em minus
  0.4em\relax IEEE, 2021, pp. 3385--3389.

\bibitem{richard2021neural}
A.~Richard, D.~Markovic, I.~D. Gebru, S.~Krenn, G.~A. Butler, F.~Torre, and
  Y.~Sheikh, ``Neural synthesis of binaural speech from mono audio,'' in
  \emph{International Conference on Learning Representations}, 2021.

\bibitem{mcmullen2022machine}
K.~McMullen and Y.~Wan, ``A machine learning tutorial for spatial auditory
  display using head-related transfer functions,'' \emph{The Journal of the
  Acoustical Society of America}, vol. 151, no.~2, pp. 1277--1293, 2022.

\bibitem{bruschi2024review}
V.~Bruschi, L.~Grossi, N.~A. Dourou, A.~Quattrini, A.~Vancheri, T.~Leidi, and
  S.~Cecchi, ``A review on head-related transfer function generation for
  spatial audio,'' \emph{Applied Sciences}, vol.~14, no.~23, p. 11242, 2024.

\bibitem{fantini2025survey}
D.~Fantini, M.~Geronazzo, F.~Avanzini, and S.~Ntalampiras, ``A survey on
  machine learning techniques for head-related transfer function
  individualization,'' \emph{IEEE Open Journal of Signal Processing}, 2025.

\bibitem{cobos2022overview}
M.~Cobos, J.~Ahrens, K.~Kowalczyk, and A.~Politis, ``An overview of machine
  learning and other data-based methods for spatial audio capture, processing,
  and reproduction,'' \emph{EURASIP Journal on Audio, Speech, and Music
  Processing}, vol. 2022, no.~1, p.~10, 2022.

\bibitem{katz2012perceptually}
B.~F. Katz and G.~Parseihian, ``Perceptually based head-related transfer
  function database optimization,'' \emph{The Journal of the Acoustical Society
  of America}, vol. 131, no.~2, pp. EL99--EL105, 2012.

\bibitem{michele2010estimation}
G.~Michele, S.~Spagnol, A.~Federico \emph{et~al.}, ``Estimation and modeling of
  pinna-related transfer functions,'' in \emph{Proceedings of the 13th
  International Conference on Digital Audio Effects, DAFx 2010}.\hskip 1em plus
  0.5em minus 0.4em\relax Institute of Electronic Music and Acoustics (IEM),
  University of Music and Performing Arts, 2010, pp. 431--438.

\bibitem{raykar2005extracting}
V.~C. Raykar, R.~Duraiswami, and B.~Yegnanarayana, ``Extracting the frequencies
  of the pinna spectral notches in measured head related impulse responses,''
  \emph{The Journal of the Acoustical Society of America}, vol. 118, no.~1, pp.
  364--374, 2005.

\bibitem{guezenoc2020wide}
C.~Guezenoc and R.~Seguier, ``A wide dataset of ear shapes and pinna-related
  transfer functions generated by random ear drawings,'' \emph{The Journal of
  the Acoustical Society of America}, vol. 147, no.~6, pp. 4087--4096, 2020.

\bibitem{romigh2015efficient}
G.~D. Romigh, D.~S. Brungart, R.~M. Stern, and B.~D. Simpson, ``Efficient real
  spherical harmonic representation of head-related transfer functions,''
  \emph{IEEE Journal of Selected Topics in Signal Processing}, vol.~9, no.~5,
  pp. 921--930, 2015.

\bibitem{chen2019autoencoding}
T.-Y. Chen, T.-H. Kuo, and T.-S. Chi, ``Autoencoding {HRTFs} for {DNN} based
  {HRTF} personalization using anthropometric features,'' in \emph{ICASSP
  2019-2019 IEEE International Conference on Acoustics, Speech and Signal
  Processing (ICASSP)}.\hskip 1em plus 0.5em minus 0.4em\relax IEEE, 2019, pp.
  271--275.

\bibitem{baldi2012autoencoders}
P.~Baldi, ``Autoencoders, unsupervised learning, and deep architectures,'' in
  \emph{Proceedings of ICML workshop on unsupervised and transfer
  learning}.\hskip 1em plus 0.5em minus 0.4em\relax JMLR Workshop and
  Conference Proceedings, 2012, pp. 37--49.

\bibitem{hu2008hrtf}
H.~Hu, L.~Zhou, H.~Ma, and Z.~Wu, ``{HRTF} personalization based on artificial
  neural network in individual virtual auditory space,'' \emph{Applied
  Acoustics}, vol.~69, no.~2, pp. 163--172, 2008.

\bibitem{wang2020global}
Y.~Wang, Y.~Zhang, Z.~Duan, and M.~Bocko, ``Global {HRTF} personalization using
  anthropometric measures,'' in \emph{Audio Engineering Society Conference:
  2020 AES International Conference on Audio for Virtual and Augmented
  Reality}.\hskip 1em plus 0.5em minus 0.4em\relax Audio Engineering Society,
  2020.

\bibitem{zhang2020modeling}
M.~Zhang, Z.~Ge, T.~Liu, X.~Wu, and T.~Qu, ``Modeling of individual {HRTFs}
  based on spatial principal component analysis,'' \emph{IEEE/ACM Transactions
  on Audio, Speech, and Language Processing}, vol.~28, pp. 785--797, 2020.

\bibitem{miccini2020hrtf}
R.~Miccini and S.~Spagnol, ``{HRTF} individualization using deep learning,'' in
  \emph{2020 IEEE Conference on Virtual Reality and 3D User Interfaces
  Abstracts and Workshops (VRW)}.\hskip 1em plus 0.5em minus 0.4em\relax IEEE,
  2020, pp. 390--395.

\bibitem{yao2022individualization}
D.~Yao, J.~Zhao, L.~Cheng, J.~Li, X.~Li, X.~Guo, and Y.~Yan, ``An
  individualization approach for head-related transfer function in arbitrary
  directions based on deep learning,'' \emph{JASA Express Letters}, vol.~2,
  no.~6, p. 064401, 2022.

\bibitem{zhang2023modelling}
R.~Zhang, R.~Meng, J.~Sang, Y.~Hu, X.~Li, and C.~Zheng, ``Modelling individual
  head-related transfer function ({HRTF}) based on anthropometric parameters
  and generic {HRTF} amplitudes,'' \emph{CAAI Transactions on Intelligence
  Technology}, vol.~8, no.~2, pp. 364--378, 2023.

\bibitem{sanchez2025towards}
J.~C.~A. S{\'a}nchez, L.~Comanducci, M.~Pezzoli, and F.~Antonacci, ``Towards
  {HRTF} personalization using denoising diffusion models,'' in \emph{ICASSP
  2025-2025 IEEE International Conference on Acoustics, Speech and Signal
  Processing (ICASSP)}.\hskip 1em plus 0.5em minus 0.4em\relax IEEE, 2025, pp.
  1--5.

\bibitem{miccini2021hybrid}
R.~Miccini and S.~Spagnol, ``A hybrid approach to structural modeling of
  individualized {HRTFs},'' in \emph{2021 IEEE Conference on Virtual Reality
  and 3D User Interfaces Abstracts and Workshops (VRW)}.\hskip 1em plus 0.5em
  minus 0.4em\relax IEEE, 2021, pp. 80--85.

\bibitem{zhao2022magnitude}
M.~Zhao, Z.~Sheng, and Y.~Fang, ``Magnitude modeling of personalized {HRTF}
  based on ear images and anthropometric measurements,'' \emph{Applied
  Sciences}, vol.~12, no.~16, p. 8155, 2022.

\bibitem{ko2023prtfnet}
B.-Y. Ko, G.-T. Lee, H.~Nam, and Y.-H. Park, ``{PRTFNet: HRTF}
  individualization for accurate spectral cues using a compact prtf,''
  \emph{IEEE Access}, vol.~11, pp. 96\,119--96\,130, 2023.

\bibitem{javeri2023machine}
N.~Javeri, P.~B. Dutta, K.~Sunder, and K.~Jain, ``A machine learning approach
  to predicting personalized head related transfer functions and headphone
  equalization from video capture data,'' in \emph{2023 Immersive and 3D Audio:
  from Architecture to Automotive (I3DA)}.\hskip 1em plus 0.5em minus
  0.4em\relax IEEE, 2023, pp. 1--9.

\bibitem{fantini2021hrtf}
D.~Fantini, F.~Avanzini, S.~Ntalampiras, and G.~Presti, ``{HRTF}
  individualization based on anthropometric measurements extracted from {3D}
  head meshes,'' in \emph{2021 Immersive and 3D Audio: from Architecture to
  Automotive (I3DA)}.\hskip 1em plus 0.5em minus 0.4em\relax IEEE, 2021, pp.
  1--10.

\bibitem{zhou2021predictability}
Y.~Zhou, H.~Jiang, and V.~K. Ithapu, ``On the predictability of {HRTFs} from
  ear shapes using deep networks,'' in \emph{ICASSP 2021-2021 IEEE
  International Conference on Acoustics, Speech and Signal Processing
  (ICASSP)}.\hskip 1em plus 0.5em minus 0.4em\relax IEEE, 2021, pp. 441--445.

\bibitem{wang2022predicting}
Y.~Wang, Y.~Zhang, Z.~Duan, and M.~Bocko, ``Predicting global head-related
  transfer functions from scanned head geometry using deep learning and compact
  representations,'' \emph{arXiv preprint arXiv:2207.14352}, 2022.

\bibitem{zhao2024efficient}
J.~Zhao, D.~Yao, J.~Gu, and J.~Li, ``Efficient prediction of individual
  head-related transfer functions based on {3D} meshes,'' \emph{Applied
  Acoustics}, vol. 219, p. 109938, 2024.

\bibitem{huang2023audioear}
X.~Huang, Y.~Wang, Y.~Liu, B.~Ni, W.~Zhang, J.~Liu, and T.~Li, ``{AudioEar:}
  single-view ear reconstruction for personalized spatial audio,'' in
  \emph{Proceedings of the AAAI Conference on Artificial Intelligence},
  vol.~37, no.~1, 2023, pp. 944--952.

\bibitem{di2024denoising}
F.~Di~Giusto, F.~Llu{\'\i}s, S.~van Ophem, and E.~Deckers, ``Denoising of
  photogrammetric dummy head ear point clouds for individual head-related
  transfer functions computation,'' \emph{arXiv preprint arXiv:2408.16410},
  2024.

\bibitem{jayaram2023hrtf}
V.~Jayaram, I.~Kemelmacher-Shlizerman, and S.~M. Seitz, ``{HRTF} estimation in
  the wild,'' in \emph{Proceedings of the 36th Annual ACM Symposium on User
  Interface Software and Technology}, 2023, pp. 1--9.

\bibitem{thuillier2025hrtf}
E.~Thuillier, J.-M. Lemercier, E.~Moliner, T.~Gerkmann, and
  V.~V{\"a}lim{\"a}ki, ``{HRTF} estimation using a score-based prior,'' in
  \emph{ICASSP 2025-2025 IEEE International Conference on Acoustics, Speech and
  Signal Processing (ICASSP)}.\hskip 1em plus 0.5em minus 0.4em\relax IEEE,
  2025, pp. 1--5.

\bibitem{xie2012recovery}
B.-S. Xie, ``Recovery of individual head-related transfer functions from a
  small set of measurements,'' \emph{The Journal of the Acoustical Society of
  America}, vol. 132, no.~1, pp. 282--294, 2012.

\bibitem{chen2008head}
L.~Chen, H.~Hu, and Z.~Wu, ``Head-related impulse response interpolation in
  virtual sound system,'' in \emph{2008 Fourth International Conference on
  Natural Computation}, vol.~6.\hskip 1em plus 0.5em minus 0.4em\relax IEEE,
  2008, pp. 162--166.

\bibitem{9914751}
Y.~Ito, T.~Nakamura, S.~Koyama, and H.~Saruwatari, ``Head-related transfer
  function interpolation from spatially sparse measurements using autoencoder
  with source position conditioning,'' in \emph{2022 International Workshop on
  Acoustic Signal Enhancement (IWAENC)}, 2022, pp. 1--5.

\bibitem{zandi2022individualizing}
N.~H. Zandi, A.~M. El-Mohandes, and R.~Zheng, ``Individualizing head-related
  transfer functions for binaural acoustic applications,'' in \emph{2022 21st
  ACM/IEEE International Conference on Information Processing in Sensor
  Networks (IPSN)}.\hskip 1em plus 0.5em minus 0.4em\relax IEEE, 2022, pp.
  105--117.

\bibitem{zurale2023spatial}
D.~Zurale and S.~Dubnov, ``Spatial upsampling of sparse head related transfer
  functions-a {VQ-VAE} \& {Transformer} based approach,'' in \emph{Audio
  Engineering Society Conference: AES 2023 International Conference on Spatial
  and Immersive Audio}.\hskip 1em plus 0.5em minus 0.4em\relax Audio
  Engineering Society, 2023.

\bibitem{10.1121/10.0036032}
K.-W. Chang, Y.-L. Shen, and T.-S. Chi, ``Spatial grouping as a method to
  improve personalized head-related transfer function prediction,'' \emph{JASA
  Express Letters}, vol.~5, no.~3, p. 034801, 03 2025.

\bibitem{zurale2022deep}
D.~Zurale, S.~Yadegari, and S.~Dubnov, ``Deep {HRTF} encoding \& interpolation:
  Exploring spatial correlations using convolutional neural networks,'' in
  \emph{19th Sound and Music Computing Conference, SMC 2022}.\hskip 1em plus
  0.5em minus 0.4em\relax Sound and Music Computing Network, 2022, pp.
  350--357.

\bibitem{jiang2023modeling}
Z.~Jiang, J.~Sang, C.~Zheng, A.~Li, and X.~Li, ``Modeling individual
  head-related transfer functions from sparse measurements using a
  convolutional neural network,'' \emph{The Journal of the Acoustical Society
  of America}, vol. 153, no.~1, pp. 248--259, 2023.

\bibitem{chen2023head}
X.~Chen, F.~Ma, Y.~Zhang, A.~Bastine, and P.~N. Samarasinghe, ``Head-related
  transfer function interpolation with a spherical {CNN},'' \emph{arXiv
  preprint arXiv:2309.08290}, 2023.

\bibitem{10418851}
E.~Thuillier, C.~T. Jin, and V.~Välimäki, ``{HRTF} interpolation using a
  spherical neural process meta-learner,'' \emph{IEEE/ACM Transactions on
  Audio, Speech, and Language Processing}, vol.~32, pp. 1790--1802, 2024.

\bibitem{zhao2025head}
J.~Zhao, D.~Yao, and J.~Li, ``Head-related transfer function upsampling with
  spatial extrapolation features,'' \emph{IEEE Transactions on Audio, Speech
  and Language Processing}, vol.~33, pp. 1034--1048, 2025.

\bibitem{hogg2024hrtf}
A.~O. Hogg, M.~Jenkins, H.~Liu, I.~Squires, S.~J. Cooper, and L.~Picinali,
  ``{HRTF} upsampling with a generative adversarial network using a gnomonic
  equiangular projection,'' \emph{IEEE/ACM Transactions on Audio, Speech, and
  Language Processing}, 2024.

\bibitem{hu2024hrtf}
X.~Hu, L.~Picinali, J.~Li, A.~Hogg \emph{et~al.}, ``{HRTF} spatial upsampling
  in the spherical harmonics domain employing a generative adversarial
  network,'' in \emph{Proceedings of the 27th International Conference on
  Digital Audio Effects, DAFx 2024}, 2024.

\bibitem{hu2025machine}
X.~Hu, J.~Li, L.~Picinali, and A.~O. Hogg, ``A machine learning approach for
  denoising and upsampling {HRTFs},'' \emph{arXiv preprint arXiv:2504.17586},
  2025.

\bibitem{10096144}
J.~W. Lee, S.~Lee, and K.~Lee, ``Global {HRTF} interpolation via learned affine
  transformation of hyper-conditioned features,'' in \emph{ICASSP 2023 - 2023
  IEEE International Conference on Acoustics, Speech and Signal Processing
  (ICASSP)}, 2023, pp. 1--5.

\bibitem{10095801}
Y.~Zhang, Y.~Wang, and Z.~Duan, ``{HRTF Field:} unifying measured {HRTF}
  magnitude representation with neural fields,'' in \emph{ICASSP 2023 - 2023
  IEEE International Conference on Acoustics, Speech and Signal Processing
  (ICASSP)}, 2023, pp. 1--5.

\bibitem{ma2023spatial}
F.~Ma, T.~D. Abhayapala, P.~N. Samarasinghe, and X.~Chen, ``Spatial upsampling
  of head-related transfer functions using a physics-informed neural network,''
  \emph{arXiv preprint arXiv:2307.14650}, 2023.

\bibitem{10448477}
Y.~Masuyama, G.~Wichern, F.~G. Germain, Z.~Pan, S.~Khurana, C.~Hori, and
  J.~Le~Roux, ``{NIIRF:} neural {IIR} filter field for {HRTF} upsampling and
  personalization,'' in \emph{ICASSP 2024 - 2024 IEEE International Conference
  on Acoustics, Speech and Signal Processing (ICASSP)}, 2024, pp. 1016--1020.

\bibitem{di2024neural}
D.~Di~Carlo, A.~A. Nugraha, M.~Fontaine, Y.~Bando, and K.~Yoshii, ``{Neural
  Steerer:} novel steering vector synthesis with a causal neural field over
  frequency and direction,'' in \emph{2024 IEEE International Conference on
  Acoustics, Speech and Signal Processing Workshops (ICASSPW)}, 2024, pp.
  740--744.

\bibitem{masuyama2025retrieval}
Y.~Masuyama, G.~Wichern, F.~G. Germain, C.~Ick, and J.~Le~Roux,
  ``Retrieval-augmented neural field for {HRTF} upsampling and
  personalization,'' in \emph{ICASSP 2025-2025 IEEE International Conference on
  Acoustics, Speech and Signal Processing (ICASSP)}.\hskip 1em plus 0.5em minus
  0.4em\relax IEEE, 2025, pp. 1--5.

\bibitem{lu2025bicg}
X.~Lu, Y.~Wang, J.~Sang, and C.~Zheng, ``{BiCG:} binaural cue generation from
  unified {HRTF} datasets,'' in \emph{ICASSP 2025-2025 IEEE International
  Conference on Acoustics, Speech and Signal Processing (ICASSP)}.\hskip 1em
  plus 0.5em minus 0.4em\relax IEEE, 2025, pp. 1--5.

\bibitem{xie2022neural}
Y.~Xie, T.~Takikawa, S.~Saito, O.~Litany, S.~Yan, N.~Khan, F.~Tombari,
  J.~Tompkin, V.~sitzmann, and S.~Sridhar, ``Neural fields in visual computing
  and beyond,'' \emph{Computer Graphics Forum}, vol.~41, no.~2, pp. 641--676,
  2022.

\bibitem{NEURIPS2020_53c04118}
V.~Sitzmann, J.~Martel, A.~Bergman, D.~Lindell, and G.~Wetzstein, ``Implicit
  neural representations with periodic activation functions,'' in
  \emph{Advances in Neural Information Processing Systems}, vol.~33.\hskip 1em
  plus 0.5em minus 0.4em\relax Curran Associates, Inc., 2020, pp. 7462--7473.

\bibitem{NEURIPS2020_55053683}
M.~Tancik, P.~Srinivasan, B.~Mildenhall, S.~Fridovich-Keil, N.~Raghavan,
  U.~Singhal, R.~Ramamoorthi, J.~Barron, and R.~Ng, ``Fourier features let
  networks learn high frequency functions in low dimensional domains,'' in
  \emph{Advances in Neural Information Processing Systems}, vol.~33.\hskip 1em
  plus 0.5em minus 0.4em\relax Curran Associates, Inc., 2020, pp. 7537--7547.

\bibitem{mildenhall2021nerf}
B.~Mildenhall, P.~P. Srinivasan, M.~Tancik, J.~T. Barron, R.~Ramamoorthi, and
  R.~Ng, ``{NeRF}: Representing scenes as neural radiance fields for view
  synthesis,'' \emph{Communications of the ACM}, vol.~65, no.~1, pp. 99--106,
  2021.

\bibitem{luo2022learning}
A.~Luo, Y.~Du, M.~Tarr, J.~Tenenbaum, A.~Torralba, and C.~Gan, ``Learning
  neural acoustic fields,'' in \emph{Advances in Neural Information Processing
  Systems}, vol.~35.\hskip 1em plus 0.5em minus 0.4em\relax Curran Associates,
  Inc., 2022, pp. 3165--3177.

\bibitem{liang2023neural}
S.~Liang, C.~Huang, Y.~Tian, A.~Kumar, and C.~Xu, ``Neural acoustic context
  field: Rendering realistic room impulse response with neural fields,''
  \emph{arXiv preprint arXiv:2309.15977}, 2023.

\bibitem{williams1999fourier}
E.~G. Williams, \emph{Fourier acoustics: sound radiation and nearfield
  acoustical holography}.\hskip 1em plus 0.5em minus 0.4em\relax Elsevier,
  1999.

\bibitem{warusfel2003listen}
O.~Warusfel, ``Listen {HRTF} database,'' \emph{online, IRCAM and AK, Available:
  http://recherche. ircam. fr/equipes/salles/listen/index. html}, 2003.

\bibitem{watanabe2014dataset}
K.~Watanabe, Y.~Iwaya, Y.~Suzuki, S.~Takane, and S.~Sato, ``Dataset of
  head-related transfer functions measured with a circular loudspeaker array,''
  \emph{Acoustical Science and Technology}, vol.~35, no.~3, pp. 159--165, 2014.

\bibitem{carpentier2014measurement}
T.~Carpentier, H.~Bahu, M.~Noisternig, and O.~Warusfel, ``Measurement of a
  head-related transfer function database with high spatial resolution,'' in
  \emph{7th forum acusticum (EAA)}, 2014.

\bibitem{majdak2013sound}
P.~Majdak, B.~Masiero, and J.~Fels, ``Sound localization in individualized and
  non-individualized crosstalk cancellation systems,'' \emph{The Journal of the
  Acoustical Society of America}, vol. 133, no.~4, pp. 2055--2068, 2013.

\bibitem{bomhardt2016high}
R.~Bomhardt, M.~de~la Fuente~Klein, and J.~Fels, ``A high-resolution
  head-related transfer function and three-dimensional ear model database,''
  \emph{Proceedings of Meetings on Acoustics}, vol.~29, no.~1, 2016.

\bibitem{sridhar2017database}
R.~Sridhar, J.~G. Tylka, and E.~Y. Choueiri, ``A database of head-related
  transfer function and morphological measurements,'' in \emph{143rd Audio
  Engineering Society Convention 2017}, 2017, pp. 851 -- 855.

\bibitem{armstrong2018perceptual}
C.~Armstrong, L.~Thresh, D.~Murphy, and G.~Kearney, ``A perceptual evaluation
  of individual and non-individual {HRTFs}: A case study of the {SADIE II}
  database,'' \emph{Applied Sciences}, vol.~8, no.~11, p. 2029, 2018.

\bibitem{denk2018adapting}
F.~Denk, S.~M. Ernst, S.~D. Ewert, and B.~Kollmeier, ``Adapting hearing devices
  to the individual ear acoustics: Database and target response correction
  functions for various device styles,'' \emph{Trends in Hearing}, vol.~22, p.
  2331216518779313, 2018.

\bibitem{ghorbal2020computed}
S.~Ghorbal, X.~Bonjour, and R.~S{\'e}guier, ``Computed {HRIRs} and ears
  database for acoustic research,'' in \emph{Audio Engineering Society
  Convention 148}.\hskip 1em plus 0.5em minus 0.4em\relax Audio Engineering
  Society, 2020.

\bibitem{engel2023sonicom}
I.~Engel, R.~Daugintis, T.~Vicente, A.~O. Hogg, J.~Pauwels, A.~J. Tournier, and
  L.~Picinali, ``The {SONICOM HRTF} dataset,'' \emph{Journal of the Audio
  Engineering Society}, vol.~71, no.~5, pp. 241--253, 2023.

\bibitem{andreopoulou2015inter}
A.~Andreopoulou, D.~R. Begault, and B.~F. Katz, ``Inter-laboratory round robin
  {HRTF} measurement comparison,'' \emph{IEEE Journal of Selected Topics in
  Signal Processing}, vol.~9, no.~5, pp. 895--906, 2015.

\bibitem{pauwels2023relevance}
J.~Pauwels and L.~Picinali, ``On the relevance of the differences between
  {HRTF} measurement setups for machine learning,'' in \emph{ICASSP 2023-2023
  IEEE International Conference on Acoustics, Speech and Signal Processing
  (ICASSP)}.\hskip 1em plus 0.5em minus 0.4em\relax IEEE, 2023, pp. 1--5.

\bibitem{wen2023mitigating}
Y.~Wen, Y.~Zhang, and Z.~Duan, ``Mitigating cross-database differences for
  learning unified {HRTF} representation,'' in \emph{2023 IEEE Workshop on
  Applications of Signal Processing to Audio and Acoustics (WASPAA)}.\hskip 1em
  plus 0.5em minus 0.4em\relax IEEE, 2023, pp. 1--5.

\bibitem{torres2015personalization}
E.~A. Torres-Gallegos, F.~Orduna-Bustamante, and F.~Ar{\'a}mbula-Cos{\'\i}o,
  ``Personalization of head-related transfer functions ({HRTF}) based on
  automatic photo-anthropometry and inference from a database,'' \emph{Applied
  Acoustics}, vol.~97, pp. 84--95, 2015.

\bibitem{9746315}
B.~Zhi, D.~N. Zotkin, and R.~Duraiswami, ``Towards fast and convenient
  end-to-end {HRTF} personalization,'' in \emph{ICASSP 2022 - 2022 IEEE
  International Conference on Acoustics, Speech and Signal Processing
  (ICASSP)}, 2022, pp. 441--445.

\bibitem{marggraf2024hrtf}
N.~Marggraf-Turley, M.~Lovedee-Turner, and E.~De~Sena, ``{HRTF} recommendation
  based on the predicted binaural colouration model,'' in \emph{ICASSP
  2024-2024 IEEE International Conference on Acoustics, Speech and Signal
  Processing (ICASSP)}.\hskip 1em plus 0.5em minus 0.4em\relax IEEE, 2024, pp.
  1106--1110.

\bibitem{luo2013virtual}
Y.~Luo, D.~N. Zotkin, and R.~Duraiswami, ``Virtual autoencoder based
  recommendation system for individualizing head-related transfer functions,''
  in \emph{2013 IEEE Workshop on Applications of Signal Processing to Audio and
  Acoustics (WASPAA)}.\hskip 1em plus 0.5em minus 0.4em\relax IEEE, 2013, pp.
  1--4.

\bibitem{kobayashi2023temporal}
T.~Kobayashi, Y.~Maruyama, I.~Nambu, S.~Yano, and Y.~Wada, ``Temporal
  convolutional neural networks to generate a head-related impulse response
  from one direction to another,'' \emph{arXiv preprint arXiv:2310.14018},
  2023.

\bibitem{moller1995head}
H.~M{\o}ller, M.~F. S{\o}rensen, D.~Hammersh{\o}i, and C.~B. Jensen,
  ``Head-related transfer functions of human subjects,'' \emph{Journal of the
  Audio Engineering Society}, vol.~43, no.~5, pp. 300--321, 1995.

\bibitem{hartmann1996externalization}
W.~M. Hartmann and A.~Wittenberg, ``On the externalization of sound images,''
  \emph{The Journal of the Acoustical Society of America}, vol.~99, no.~6, pp.
  3678--3688, 1996.

\bibitem{wenzel2017perception}
E.~M. Wenzel, D.~R. Begault, and M.~Godfroy-Cooper, ``Perception of spatial
  sound,'' in \emph{Immersive Sound}.\hskip 1em plus 0.5em minus 0.4em\relax
  Routledge, 2017, pp. 5--39.

\bibitem{best2020sound}
V.~Best, R.~Baumgartner, M.~Lavandier, P.~Majdak, and N.~Kop{\v{c}}o, ``Sound
  externalization: A review of recent research,'' \emph{Trends in Hearing},
  vol.~24, p. 2331216520948390, 2020.

\bibitem{majdak2022amt}
P.~Majdak, C.~Hollomey, and R.~Baumgartner, ``{AMT 1. x:} a toolbox for
  reproducible research in auditory modeling,'' \emph{Acta Acustica}, vol.~6,
  p.~19, 2022.

\bibitem{zaar2022predicting}
J.~Zaar and L.~H. Carney, ``Predicting speech intelligibility in
  hearing-impaired listeners using a physiologically inspired auditory model,''
  \emph{Hearing Research}, vol. 426, p. 108553, 2022.

\bibitem{baumgartner2014modeling}
R.~Baumgartner, P.~Majdak, and B.~Laback, ``Modeling sound-source localization
  in sagittal planes for human listeners,'' \emph{The Journal of the Acoustical
  Society of America}, vol. 136, no.~2, pp. 791--802, 2014.

\bibitem{reijniers2014ideal}
J.~Reijniers, D.~Vanderelst, C.~Jin, S.~Carlile, and H.~Peremans, ``An
  ideal-observer model of human sound localization,'' \emph{Biological
  Cybernetics}, vol. 108, pp. 169--181, 2014.

\bibitem{barumerli2023bayesian}
R.~Barumerli, P.~Majdak, M.~Geronazzo, D.~Meijer, F.~Avanzini, and
  R.~Baumgartner, ``A bayesian model for human directional localization of
  broadband static sound sources,'' \emph{Acta Acustica}, vol.~7, p.~12, 2023.

\bibitem{reijniers2025ideal}
J.~Reijniers, G.~McLachlan, B.~Partoens, and H.~Peremans, ``Ideal-observer
  model of human sound localization of sources with unknown spectrum,''
  \emph{Scientific Reports}, vol.~15, no.~1, p. 7289, 2025.

\bibitem{10095152}
R.~Daugintis, R.~Barumerli, L.~Picinali, and M.~Geronazzo, ``Classifying
  non-individual head-related transfer functions with a computational auditory
  model: Calibration and metrics,'' in \emph{ICASSP 2023 - 2023 IEEE
  International Conference on Acoustics, Speech and Signal Processing
  (ICASSP)}, 2023, pp. 1--5.

\bibitem{lo2019mosnet}
C.-C. Lo, S.-W. Fu, W.-C. Huang, X.~Wang, J.~Yamagishi, Y.~Tsao, and H.-M.
  Wang, ``{MOSNet}: Deep learning based objective assessment for voice
  conversion,'' in \emph{Proc. Interspeech 2019}, 2019.

\bibitem{NEURIPS2021_bc6d7538}
P.~Manocha, B.~Xu, and A.~Kumar, ``{NORESQA}: A framework for speech quality
  assessment using non-matching references,'' in \emph{Advances in Neural
  Information Processing Systems}, vol.~34.\hskip 1em plus 0.5em minus
  0.4em\relax Curran Associates, Inc., 2021, pp. 22\,363--22\,378.

\bibitem{zezario2022deep}
R.~E. Zezario, S.-W. Fu, F.~Chen, C.-S. Fuh, H.-M. Wang, and Y.~Tsao, ``Deep
  learning-based non-intrusive multi-objective speech assessment model with
  cross-domain features,'' \emph{IEEE/ACM Transactions on Audio, Speech, and
  Language Processing}, vol.~31, pp. 54--70, 2022.

\bibitem{lian2025apg}
Z.~Lian, L.~Wang, and H.~Huang, ``{APG-MOS}: Auditory perception guided-mos
  predictor for synthetic speech,'' \emph{arXiv preprint arXiv:2504.20447},
  2025.

\bibitem{manocha2021dplm}
P.~Manocha, A.~Kumar, B.~Xu, A.~Menon, I.~D. Gebru, V.~K. Ithapu, and
  P.~Calamia, ``{DPLM}: A deep perceptual spatial-audio localization metric,''
  in \emph{2021 IEEE Workshop on Applications of Signal Processing to Audio and
  Acoustics (WASPAA)}.\hskip 1em plus 0.5em minus 0.4em\relax IEEE, 2021, pp.
  6--10.

\bibitem{manocha2022saqam}
------, ``{SAQAM}: Spatial audio quality assessment metric,'' in \emph{Proc.
  Interspeech 2022}, 2022, pp. 649--653.

\bibitem{manocha2023spatialization}
P.~Manocha, I.~D. Gebru, A.~Kumar, D.~Markovic, and A.~Richard,
  ``Spatialization quality metric for binaural speech,'' in \emph{Proc.
  Interspeech 2023}, 2023, pp. 5426--5430.

\bibitem{zheng2025hapg}
Y.~Zheng, J.~Yao, X.~Deng, Y.~Yang, R.~Liao, W.~Tu, and C.~Lin, ``{HAPG-SAQAM}:
  Human auditory perception guided spatial audio quality assessment metric,''
  in \emph{ICASSP 2025-2025 IEEE International Conference on Acoustics, Speech
  and Signal Processing (ICASSP)}.\hskip 1em plus 0.5em minus 0.4em\relax IEEE,
  2025, pp. 1--5.

\bibitem{mehrabi2021survey}
N.~Mehrabi, F.~Morstatter, N.~Saxena, K.~Lerman, and A.~Galstyan, ``A survey on
  bias and fairness in machine learning,'' \emph{ACM Computing Surveys (CSUR)},
  vol.~54, no.~6, pp. 1--35, 2021.

\bibitem{rudin2019stop}
C.~Rudin, ``Stop explaining black box machine learning models for high stakes
  decisions and use interpretable models instead,'' \emph{Nature Machine
  Intelligence}, vol.~1, no.~5, pp. 206--215, 2019.

\bibitem{torralba2011unbiased}
A.~Torralba and A.~A. Efros, ``Unbiased look at dataset bias,'' in \emph{CVPR
  2011}.\hskip 1em plus 0.5em minus 0.4em\relax IEEE, 2011, pp. 1521--1528.

\bibitem{kreuk2022audiogen}
F.~Kreuk, G.~Synnaeve, A.~Polyak, U.~Singer, A.~Défossez, J.~Copet, D.~Parikh,
  Y.~Taigman, and Y.~Adi, ``{AudioGen}: Textually guided audio generation,'' in
  \emph{The Eleventh International Conference on Learning Representations},
  2023.

\bibitem{liu2023audioldm}
H.~Liu, Z.~Chen, Y.~Yuan, X.~Mei, X.~Liu, D.~Mandic, W.~Wang, and M.~D.
  Plumbley, ``{AudioLDM}: Text-to-audio generation with latent diffusion
  models,'' in \emph{Proceedings of the International Conference on Machine
  Learning}, 2023, pp. 21\,450--21\,474.

\bibitem{evans2025stable}
Z.~Evans, J.~D. Parker, C.~Carr, Z.~Zukowski, J.~Taylor, and J.~Pons, ``Stable
  audio open,'' in \emph{ICASSP 2025-2025 IEEE International Conference on
  Acoustics, Speech and Signal Processing (ICASSP)}.\hskip 1em plus 0.5em minus
  0.4em\relax IEEE, 2025, pp. 1--5.

\bibitem{zhou2018visual}
Y.~Zhou, Z.~Wang, C.~Fang, T.~Bui, and T.~L. Berg, ``Visual to sound:
  Generating natural sound for videos in the wild,'' in \emph{Proceedings of
  the IEEE/CVF Conference on Computer Vision and Pattern Recognition}, 2018,
  pp. 3550--3558.

\bibitem{NEURIPS2023_98c50f47}
S.~Luo, C.~Yan, C.~Hu, and H.~Zhao, ``{Diff-Foley}: Synchronized video-to-audio
  synthesis with latent diffusion models,'' in \emph{Advances in Neural
  Information Processing Systems}, vol.~36.\hskip 1em plus 0.5em minus
  0.4em\relax Curran Associates, Inc., 2023, pp. 48\,855--48\,876.

\bibitem{li2025tri}
B.~Li, F.~Yang, Y.~Mao, Q.~Ye, H.~Chen, and Y.~Zhong, ``{Tri-Ergon:}
  fine-grained video-to-audio generation with multi-modal conditions and lufs
  control,'' in \emph{Proceedings of the AAAI Conference on Artificial
  Intelligence}, vol.~39, no.~5, 2025, pp. 4616--4624.

\bibitem{ronneberger2015u}
O.~Ronneberger, P.~Fischer, and T.~Brox, ``{U-Net}: Convolutional networks for
  biomedical image segmentation,'' in \emph{Medical Image Computing and
  Computer-assisted Intervention--MICCAI 2015: 18th International
  Conference}.\hskip 1em plus 0.5em minus 0.4em\relax Springer, 2015, pp.
  234--241.

\bibitem{NIPS2017_3f5ee243}
A.~Vaswani, N.~Shazeer, N.~Parmar, J.~Uszkoreit, L.~Jones, A.~N. Gomez, L.~u.
  Kaiser, and I.~Polosukhin, ``Attention is all you need,'' in \emph{Advances
  in Neural Information Processing Systems}, vol.~30.\hskip 1em plus 0.5em
  minus 0.4em\relax Curran Associates, Inc., 2017.

\bibitem{NEURIPS2020_4c5bcfec}
J.~Ho, A.~Jain, and P.~Abbeel, ``Denoising diffusion probabilistic models,'' in
  \emph{Advances in Neural Information Processing Systems}, vol.~33.\hskip 1em
  plus 0.5em minus 0.4em\relax Curran Associates, Inc., 2020, pp. 6840--6851.

\bibitem{ruder2017overview}
S.~Ruder, ``An overview of multi-task learning in deep neural networks,''
  \emph{arXiv preprint arXiv:1706.05098}, 2017.

\bibitem{NEURIPS2018_01161aaa}
P.~Morgado, N.~Nvasconcelos, T.~Langlois, and O.~Wang, ``Self-supervised
  generation of spatial audio for 360\textdegree video,'' in \emph{Advances in
  Neural Information Processing Systems}, vol.~31.\hskip 1em plus 0.5em minus
  0.4em\relax Curran Associates, Inc., 2018.

\bibitem{gao20192}
R.~Gao and K.~Grauman, ``{2.5 D} visual sound,'' in \emph{Proceedings of the
  IEEE/CVF Conference on Computer Vision and Pattern Recognition}, 2019, pp.
  324--333.

\bibitem{zhou2020sep}
H.~Zhou, X.~Xu, D.~Lin, X.~Wang, and Z.~Liu, ``Sep-stereo: Visually guided
  stereophonic audio generation by associating source separation,'' in
  \emph{Computer Vision--ECCV 2020: 16th European Conference}.\hskip 1em plus
  0.5em minus 0.4em\relax Springer, 2020, pp. 52--69.

\bibitem{li2024cyclic}
Z.~Li, B.~Zhao, and Y.~Yuan, ``Cyclic learning for binaural audio generation
  and localization,'' in \emph{Proceedings of the IEEE/CVF Conference on
  Computer Vision and Pattern Recognition}, 2024, pp. 26\,669--26\,678.

\bibitem{NEURIPS2023_760dff0f}
S.~Liang, C.~Huang, Y.~Tian, A.~Kumar, and C.~Xu, ``{AV-NeRF:} learning neural
  fields for real-world audio-visual scene synthesis,'' in \emph{Advances in
  Neural Information Processing Systems}, vol.~36.\hskip 1em plus 0.5em minus
  0.4em\relax Curran Associates, Inc., 2023, pp. 37\,472--37\,490.

\bibitem{bhosale2024av}
S.~Bhosale, H.~Yang, D.~Kanojia, J.~Deng, and X.~Zhu, ``{AV-GS}: Learning
  material and geometry aware priors for novel view acoustic synthesis,'' in
  \emph{Advances in Neural Information Processing Systems}.\hskip 1em plus
  0.5em minus 0.4em\relax Curran Associates, Inc., 2024.

\bibitem{gao2024soaf}
H.~Gao, J.~Ma, D.~Ahmedt-Aristizabal, C.~Nguyen, and M.~Liu, ``{SOAF}: Scene
  occlusion-aware neural acoustic field,'' \emph{arXiv preprint
  arXiv:2407.02264}, 2024.

\bibitem{baek2025av}
H.~Baek, H.~Shin, J.~Seo, C.~Kim, S.~Kim, H.~Kim, and S.~Kim, ``{AV-Surf}:
  Surface-enhanced geometry-aware novel-view acoustic synthesis,'' \emph{arXiv
  preprint arXiv:2503.12806}, 2025.

\bibitem{li2024tas}
Z.~Li, B.~Zhao, and Y.~Yuan, ``{TAS}: Personalized text-guided audio
  spatialization,'' in \emph{Proceedings of the 32nd ACM International
  Conference on Multimedia}, 2024, pp. 9029--9037.

\bibitem{zhao2025dualspec}
L.~Zhao, S.~Chen, L.~Feng, X.-L. Zhang, and X.~Li, ``{DualSpec}:
  Text-to-spatial-audio generation via dual-spectrogram guided diffusion
  model,'' \emph{arXiv preprint arXiv:2502.18952}, 2025.

\bibitem{feng2025audiospa}
L.~Feng, L.~Zhao, B.~Zhu, X.-L. Zhang, and X.~Li, ``{AudioSpa}: Spatializing
  sound events with text,'' \emph{arXiv preprint arXiv:2502.11219}, 2025.

\bibitem{heydari2024immersediffusion}
M.~Heydari, M.~Souden, B.~Conejo, and J.~Atkins, ``{ImmerseDiffusion}: A
  generative spatial audio latent diffusion model,'' in \emph{ICASSP 2025-2025
  IEEE International Conference on Acoustics, Speech and Signal Processing
  (ICASSP)}.\hskip 1em plus 0.5em minus 0.4em\relax IEEE, 2025, pp. 1--5.

\bibitem{dagli2024see}
R.~Dagli, S.~Prakash, R.~Wu, and H.~Khosravani, ``{SEE-2-SOUND}: Zero-shot
  spatial environment-to-spatial sound,'' \emph{arXiv preprint
  arXiv:2406.06612}, 2024.

\bibitem{sun2024both}
P.~Sun, S.~Cheng, X.~Li, Z.~Ye, H.~Liu, H.~Zhang, W.~Xue, and Y.~Guo, ``Both
  ears wide open: Towards language-driven spatial audio generation,'' in
  \emph{International Conference on Learning Representations}, 2024.

\bibitem{kim2025visage}
J.~Kim, H.~Yun, and G.~Kim, ``Vi{SAG}e: Video-to-spatial audio generation,'' in
  \emph{The Thirteenth International Conference on Learning Representations},
  2025.

\bibitem{huang2022end}
W.~C. Huang, D.~Markovic, A.~Richard, I.~D. Gebru, and A.~Menon, ``End-to-end
  binaural speech synthesis,'' in \emph{Proc. Interspeech 2022}, 2022.

\bibitem{NEURIPS2022_95f03faf}
Y.~Leng, Z.~Chen, J.~Guo, H.~Liu, J.~Chen, X.~Tan, D.~Mandic, L.~He, X.~Li,
  T.~Qin, s.~zhao, and T.-Y. Liu, ``{BinauralGrad}: A two-stage conditional
  diffusion probabilistic model for binaural audio synthesis,'' in
  \emph{Advances in Neural Information Processing Systems}, vol.~35.\hskip 1em
  plus 0.5em minus 0.4em\relax Curran Associates, Inc., 2022, pp.
  23\,689--23\,700.

\bibitem{li2024diffbas}
Y.~Li, Y.~Shen, and D.~Wang, ``{DIFFBAS}: An advanced binaural audio synthesis
  model focusing on binaural differences recovery,'' \emph{Applied Sciences},
  vol.~14, no.~8, p. 3385, 2024.

\bibitem{liu2022dopplerbas}
J.~Liu, Z.~Ye, Q.~Chen, S.~Zheng, W.~Wang, Q.~Zhang, and Z.~Zhao,
  ``{DopplerBAS}: Binaural audio synthesis addressing doppler effect,'' in
  \emph{Findings of the Association for Computational Linguistics: ACL
  2023}.\hskip 1em plus 0.5em minus 0.4em\relax Association for Computational
  Linguistics, 2023, pp. 11\,905--11\,912.

\bibitem{he2025dual}
C.~He, W.~Chen, and M.~Wang, ``Dual position attention time-frequency network
  for binaural audio synthesis,'' in \emph{ICASSP 2025-2025 IEEE International
  Conference on Acoustics, Speech and Signal Processing (ICASSP)}.\hskip 1em
  plus 0.5em minus 0.4em\relax IEEE, 2025, pp. 1--5.

\bibitem{zhang2025two}
W.~Zhang, C.~He, Y.~Cao, S.~Xu, and M.~Wang, ``Two-stage unet with {Gated-Conv}
  fusion for binaural audio synthesis,'' \emph{Sensors}, vol.~25, no.~6, p.
  1790, 2025.

\bibitem{lee2023neural}
J.~W. Lee and K.~Lee, ``Neural fourier shift for binaural speech rendering,''
  in \emph{ICASSP 2023-2023 IEEE International Conference on Acoustics, Speech
  and Signal Processing (ICASSP)}.\hskip 1em plus 0.5em minus 0.4em\relax IEEE,
  2023, pp. 1--5.

\bibitem{levkovitch2024zero}
A.~Levkovitch, J.~Salazar, S.~Mariooryad, R.~Skerry-Ryan, N.~Bar, B.~Kleijn,
  and E.~Nachmani, ``Zero-shot mono-to-binaural speech synthesis,'' \emph{arXiv
  preprint arXiv:2412.08356}, 2024.

\bibitem{lu2019self}
Y.-D. Lu, H.-Y. Lee, H.-Y. Tseng, and M.-H. Yang, ``Self-supervised audio
  spatialization with correspondence classifier,'' in \emph{2019 IEEE
  International Conference on Image Processing (ICIP)}.\hskip 1em plus 0.5em
  minus 0.4em\relax IEEE, 2019, pp. 3347--3351.

\bibitem{xu2021visually}
X.~Xu, H.~Zhou, Z.~Liu, B.~Dai, X.~Wang, and D.~Lin, ``Visually informed
  binaural audio generation without binaural audios,'' in \emph{Proceedings of
  the IEEE/CVF Conference on Computer Vision and Pattern Recognition}, 2021,
  pp. 15\,485--15\,494.

\bibitem{li2021binaural}
S.~Li, S.~Liu, and D.~Manocha, ``Binaural audio generation via multi-task
  learning,'' \emph{ACM Transactions on Graphics (TOG)}, vol.~40, no.~6, pp.
  1--13, 2021.

\bibitem{rachavarapu2021localize}
K.~K. Rachavarapu, V.~Sundaresha, A.~Rajagopalan \emph{et~al.}, ``Localize to
  binauralize: Audio spatialization from visual sound source localization,'' in
  \emph{Proceedings of the IEEE/CVF International Conference on Computer
  Vision}, 2021, pp. 1930--1939.

\bibitem{lin2021exploiting}
Y.-B. Lin and Y.-C.~F. Wang, ``Exploiting audio-visual consistency with partial
  supervision for spatial audio generation,'' in \emph{Proceedings of the AAAI
  Conference on Artificial Intelligence}, vol.~35, no.~3, 2021, pp. 2056--2063.

\bibitem{zhang2021multi}
W.~Zhang and J.~Shao, ``Multi-attention audio-visual fusion network for audio
  spatialization,'' in \emph{Proceedings of the 2021 International Conference
  on Multimedia Retrieval}, 2021, pp. 394--401.

\bibitem{parida2022beyond}
K.~K. Parida, S.~Srivastava, and G.~Sharma, ``Beyond mono to binaural:
  Generating binaural audio from mono audio with depth and cross modal
  attention,'' in \emph{Proceedings of the IEEE/CVF Winter Conference on
  Applications of Computer Vision}, 2022, pp. 3347--3356.

\bibitem{lluis2022points2sound}
F.~Llu{\'\i}s, V.~Chatziioannou, and A.~Hofmann, ``{Points2Sound:} from mono to
  binaural audio using {3D} point cloud scenes,'' \emph{EURASIP Journal on
  Audio, Speech, and Music Processing}, vol. 2022, no.~1, p.~33, 2022.

\bibitem{garg2023visually}
R.~Garg, R.~Gao, and K.~Grauman, ``Visually-guided audio spatialization in
  video with geometry-aware multi-task learning,'' \emph{International Journal
  of Computer Vision}, vol. 131, no.~10, pp. 2723--2737, 2023.

\bibitem{liu2024visually}
M.~Liu, J.~Wang, X.~Qian, and X.~Xie, ``Visually guided binaural audio
  generation with cross-modal consistency,'' in \emph{ICASSP 2024-2024 IEEE
  International Conference on Acoustics, Speech and Signal Processing
  (ICASSP)}.\hskip 1em plus 0.5em minus 0.4em\relax IEEE, 2024, pp. 7980--7984.

\bibitem{li2024cross}
Z.~Li, B.~Zhao, and Y.~Yuan, ``Cross-modal generative model for visual-guided
  binaural stereo generation,'' \emph{Knowledge-Based Systems}, vol. 296, p.
  111814, 2024.

\bibitem{chen2025ccstereo}
Y.~Chen, K.~Shimada, C.~Simon, Y.~Ikemiya, T.~Shibuya, and Y.~Mitsufuji,
  ``{CCStereo}: Audio-visual contextual and contrastive learning for binaural
  audio generation,'' \emph{arXiv preprint arXiv:2501.02786}, 2025.

\bibitem{liu2025omniaudio}
H.~Liu, T.~Luo, K.~Luo, Q.~Jiang, P.~Sun, J.~Wang, R.~Huang, Q.~Chen, W.~Wang,
  X.~Li, S.~Zhang, Z.~Yan, Z.~Zhao, and W.~Xue, ``{OmniAudio:} generating
  spatial audio from 360-degree video,'' in \emph{Forty-second International
  Conference on Machine Learning}, 2025.

\bibitem{brunetto2025neraf}
A.~Brunetto, S.~Hornauer, and F.~Moutarde, ``Ne{RAF}: {3D} scene infused neural
  radiance and acoustic fields,'' in \emph{The Thirteenth International
  Conference on Learning Representations}, 2025.

\bibitem{NEURIPS2024_ff1f4141}
M.~Chen and E.~Shlizerman, ``{AV-Cloud}: Spatial audio rendering through
  audio-visual cloud splatting,'' in \emph{Advances in Neural Information
  Processing Systems}, vol.~37.\hskip 1em plus 0.5em minus 0.4em\relax Curran
  Associates, Inc., 2024, pp. 141\,021--141\,044.

\bibitem{chen2025soundvista}
M.~Chen, I.~D. Gebru, I.~Ananthabhotla, C.~Richardt, D.~Markovic, J.~Sandakly,
  S.~Krenn, T.~Keebler, E.~Shlizerman, and A.~Richard, ``{SoundVista}:
  Novel-view ambient sound synthesis via visual-acoustic binding,'' in
  \emph{Proceedings of the IEEE/CVF Conference on Computer Vision and Pattern
  Recognition}, 2025, pp. 8331--8341.

\bibitem{pan2025wild}
T.~Pan, J.~Liu, Z.~Huang, J.~Tang, and G.~Wu, ``In-the-wild audio
  spatialization with flexible text-guided localization,'' in \emph{Proceedings
  of the Annual Meeting of the Association for Computational Linguistics},
  2025.

\bibitem{kushwaha2025diff}
S.~S. Kushwaha, J.~Ma, M.~R. Thomas, Y.~Tian, and A.~Bruni, ``{Diff-SAGe}:
  End-to-end spatial audio generation using diffusion models,'' in \emph{ICASSP
  2025-2025 IEEE International Conference on Acoustics, Speech and Signal
  Processing (ICASSP)}.\hskip 1em plus 0.5em minus 0.4em\relax IEEE, 2025, pp.
  1--5.

\bibitem{zhang2025isdrama}
Y.~Zhang, W.~Guo, C.~Pan, Z.~Zhu, T.~Jin, and Z.~Zhao, ``{ISDrama}: Immersive
  spatial drama generation through multimodal prompting,'' in \emph{ACM
  International Conference on Multimedia (ACM MM)}, 2025.

\bibitem{su2022inras}
K.~Su, M.~Chen, and E.~Shlizerman, ``{INRAS}: Implicit neural representation
  for audio scenes,'' in \emph{Advances in Neural Information Processing
  Systems}, vol.~35.\hskip 1em plus 0.5em minus 0.4em\relax Curran Associates,
  Inc., 2022, pp. 8144--8158.

\bibitem{ratnarajah2022mesh2ir}
A.~Ratnarajah, Z.~Tang, R.~Aralikatti, and D.~Manocha, ``{MESH2IR}: Neural
  acoustic impulse response generator for complex {3D} scenes,'' in
  \emph{Proceedings of the 30th ACM International Conference on Multimedia},
  2022, pp. 924--933.

\bibitem{ratnarajah2024listen2scene}
A.~Ratnarajah and D.~Manocha, ``{Listen2Scene}: Interactive material-aware
  binaural sound propagation for reconstructed {3D} scenes,'' in \emph{2024
  IEEE Conference Virtual Reality and 3D User Interfaces (VR)}.\hskip 1em plus
  0.5em minus 0.4em\relax IEEE, 2024, pp. 254--264.

\bibitem{tang2024can}
C.~Tang, W.~Yu, G.~Sun, X.~Chen, T.~Tan, W.~Li, J.~Zhang, L.~Lu, Z.~Ma, Y.~Wang
  \emph{et~al.}, ``Can large language models understand spatial audio?'' in
  \emph{Proc. Interspeech 2024}, 2024.

\bibitem{devnani2024learning}
B.~Devnani, S.~Seto, Z.~Aldeneh, A.~Toso, E.~Menyaylenko, B.-J. Theobald,
  J.~Sheaffer, and M.~Sarabia, ``Learning spatially-aware language and audio
  embeddings,'' in \emph{Advances in Neural Information Processing Systems},
  vol.~37.\hskip 1em plus 0.5em minus 0.4em\relax Curran Associates, Inc.,
  2024, pp. 33\,505--33\,537.

\bibitem{zheng2024bat}
Z.~Zheng, P.~Peng, Z.~Ma, X.~Chen, E.~Choi, and D.~Harwath, ``{BAT:} learning
  to reason about spatial sounds with large language models,'' in
  \emph{Proceedings of the 41st International Conference on Machine
  Learning}.\hskip 1em plus 0.5em minus 0.4em\relax JMLR.org, 2024.

\bibitem{chen2020soundspaces}
C.~Chen, U.~Jain, C.~Schissler, S.~V.~A. Gari, Z.~Al-Halah, V.~K. Ithapu,
  P.~Robinson, and K.~Grauman, ``{SoundSpaces}: Audio-visual navigation in {3D}
  environments,'' in \emph{Computer Vision--ECCV 2020: 16th European
  Conference}.\hskip 1em plus 0.5em minus 0.4em\relax Springer, 2020, pp.
  17--36.

\bibitem{NEURIPS2022_3a48b0ea}
C.~Chen, C.~Schissler, S.~Garg, P.~Kobernik, A.~Clegg, P.~Calamia, D.~Batra,
  P.~Robinson, and K.~Grauman, ``{SoundSpaces 2.0:} a simulation platform for
  visual-acoustic learning,'' in \emph{Advances in Neural Information
  Processing Systems}, vol.~35.\hskip 1em plus 0.5em minus 0.4em\relax Curran
  Associates, Inc., 2022, pp. 8896--8911.

\bibitem{tang2022gwa}
Z.~Tang, R.~Aralikatti, A.~J. Ratnarajah, and D.~Manocha, ``{GWA:} a large
  high-quality acoustic dataset for audio processing,'' in \emph{ACM SIGGRAPH
  2022 Conference Proceedings}, ser. SIGGRAPH '22.\hskip 1em plus 0.5em minus
  0.4em\relax New York, NY, USA: Association for Computing Machinery, 2022.

\bibitem{shapovalov2023replay}
R.~Shapovalov, Y.~Kleiman, I.~Rocco, D.~Novotny, A.~Vedaldi, C.~Chen,
  F.~Kokkinos, B.~Graham, and N.~Neverova, ``Replay: Multi-modal multi-view
  acted videos for casual holography,'' in \emph{Proceedings of the IEEE/CVF
  International Conference on Computer Vision}, 2023, pp. 20\,338--20\,348.

\bibitem{NEURIPS2023_a4289154}
M.~Wang, S.~Clarke, J.-H. Wang, R.~Gao, and J.~Wu, ``{SoundCam}: A dataset for
  finding humans using room acoustics,'' in \emph{Advances in Neural
  Information Processing Systems}, vol.~36.\hskip 1em plus 0.5em minus
  0.4em\relax Curran Associates, Inc., 2023, pp. 52\,238--52\,264.

\bibitem{chen2024real}
Z.~Chen, I.~D. Gebru, C.~Richardt, A.~Kumar, W.~Laney, A.~Owens, and
  A.~Richard, ``{Real Acoustic Fields:} an audio-visual room acoustics dataset
  and benchmark,'' in \emph{Proceedings of the IEEE/CVF Conference on Computer
  Vision and Pattern Recognition}, 2024, pp. 21\,886--21\,896.

\bibitem{NEURIPS2024_bf8f6f5b}
B.~Yang, C.~Quan, Y.~Wang, P.~Wang, Y.~Yang, Y.~Fang, N.~Shao, H.~Bu, X.~Xu,
  and X.~Li, ``{RealMAN:} a real-recorded and annotated microphone array
  dataset for dynamic speech enhancement and localization,'' in \emph{Advances
  in Neural Information Processing Systems}, vol.~37.\hskip 1em plus 0.5em
  minus 0.4em\relax Curran Associates, Inc., 2024, pp. 105\,997--106\,019.

\bibitem{li2025sonicsim}
K.~Li, W.~Sang, C.~Zeng, R.~Yang, G.~Chen, and X.~Hu, ``{SonicSim:} a
  customizable simulation platform for speech processing in moving sound source
  scenarios,'' in \emph{The Thirteenth International Conference on Learning
  Representations}, 2025.

\bibitem{manocha2023nord}
P.~Manocha, I.~D. Gebru, A.~Kumar, D.~Markovic, and A.~Richard, ``Nord:
  Non-matching reference based relative depth estimation from binaural
  speech,'' in \emph{ICASSP 2023-2023 IEEE International Conference on
  Acoustics, Speech and Signal Processing (ICASSP)}.\hskip 1em plus 0.5em minus
  0.4em\relax IEEE, 2023, pp. 1--5.

\bibitem{8374622}
A.~Chang, A.~Dai, T.~Funkhouser, M.~Halber, M.~Niebner, M.~Savva, S.~Song,
  A.~Zeng, and Y.~Zhang, ``{Matterport3D:} learning from {RGB-D} data in indoor
  environments,'' in \emph{2017 International Conference on 3D Vision (3DV)},
  2017, pp. 667--676.

\bibitem{chen2023novel}
C.~Chen, A.~Richard, R.~Shapovalov, V.~K. Ithapu, N.~Neverova, K.~Grauman, and
  A.~Vedaldi, ``Novel-view acoustic synthesis,'' in \emph{Proceedings of the
  IEEE/CVF Conference on Computer Vision and Pattern Recognition}, 2023, pp.
  6409--6419.

\bibitem{zhu2024end}
Y.~Zhu, Q.~Kong, J.~Shi, S.~Liu, X.~Ye, J.-C. Wang, H.~Shan, and J.~Zhang,
  ``End-to-end paired ambisonic-binaural audio rendering,'' \emph{IEEE/CAA
  Journal of Automatica Sinica}, vol.~11, no.~2, pp. 502--513, 2024.

\bibitem{kilgour2018fr}
K.~Kilgour, M.~Zuluaga, D.~Roblek, and M.~Sharifi, ``Fr\'echet audio distance:
  A metric for evaluating music enhancement algorithms,'' in \emph{Proc.
  Interspeech 2019}, 2019.

\bibitem{bosman2024effect}
I.~d.~V. Bosman, O.~O. Buruk, K.~J{\o}rgensen, and J.~Hamari, ``The effect of
  audio on the experience in virtual reality: a scoping review,''
  \emph{Behaviour \& Information Technology}, vol.~43, no.~1, pp. 165--199,
  2024.

\bibitem{correa2023spatial}
G.~Corr{\^e}a De~Almeida, V.~Costa~de Souza, L.~G. Da~Silveira~J{\'u}nior, and
  M.~R. Veronez, ``Spatial audio in virtual reality: A systematic review,'' in
  \emph{Proceedings of the 25th Symposium on Virtual and Augmented Reality},
  2023, pp. 264--268.

\bibitem{slater2009place}
M.~Slater, ``Place illusion and plausibility can lead to realistic behaviour in
  immersive virtual environments,'' \emph{Philosophical Transactions of the
  Royal Society B: Biological Sciences}, vol. 364, no. 1535, pp. 3549--3557,
  2009.

\bibitem{eames2019beyond}
A.~Eames, ``Beyond reality,'' in \emph{Proceedings of the 17th ACM SIGGRAPH
  International Conference on Virtual-Reality Continuum and its Applications in
  Industry}, 2019, pp. 1--2.

\bibitem{ulsamer2020brain}
P.~Ulsamer, K.~Pfeffel, and N.~H. M{\"u}ller, ``Brain activation in virtual
  reality for attention guidance,'' in \emph{International Conference on
  Human-Computer Interaction}.\hskip 1em plus 0.5em minus 0.4em\relax Springer,
  2020, pp. 190--200.

\bibitem{gupta2022augmented}
R.~Gupta, J.~He, R.~Ranjan, W.-S. Gan, F.~Klein, C.~Schneiderwind,
  A.~Neidhardt, K.~Brandenburg, and V.~V{\"a}lim{\"a}ki, ``Augmented/mixed
  reality audio for hearables: Sensing, control, and rendering,'' \emph{IEEE
  Signal Processing Magazine}, vol.~39, no.~3, pp. 63--89, 2022.

\bibitem{hawley2004benefit}
M.~L. Hawley, R.~Y. Litovsky, and J.~F. Culling, ``The benefit of binaural
  hearing in a cocktail party: Effect of location and type of interferer,''
  \emph{The Journal of the Acoustical Society of America}, vol. 115, no.~2, pp.
  833--843, 2004.

\bibitem{zheng2023sixty}
C.~Zheng, H.~Zhang, W.~Liu, X.~Luo, A.~Li, X.~Li, and B.~C. Moore, ``Sixty
  years of frequency-domain monaural speech enhancement: From traditional to
  deep learning methods,'' \emph{Trends in Hearing}, vol.~27, p.
  23312165231209913, 2023.

\bibitem{hohmann2020virtual}
V.~Hohmann, R.~Paluch, M.~Krueger, M.~Meis, and G.~Grimm, ``The virtual reality
  lab: Realization and application of virtual sound environments,'' \emph{Ear
  and Hearing}, vol.~41, pp. 31S--38S, 2020.

\bibitem{pedersen2023virtual}
R.~L. Pedersen, L.~Picinali, N.~Kajs, and F.~Patou, ``Virtual-reality-based
  research in hearing science: a platforming approach,'' \emph{Journal of the
  Audio Engineering Society}, vol.~71, no.~6, pp. 374--389, 2023.

\bibitem{chitra2025ai}
S.~Chitra~Thara, K.~Vidhya~Lekshmi, and N.~Venkateswaramurthy, ``{AI}-driven
  innovations in hearing health: A review of artificial intelligence
  applications in audiology and hearing technologies,'' \emph{Current Aging
  Science}, 2025.

\bibitem{yu2022pay}
Y.~Yu, L.~Cao, F.~Sun, X.~Liu, and L.~Wang, ``Pay self-attention to
  audio-visual navigation,'' in \emph{British Machine Vision Conference
  (BMVC)}.\hskip 1em plus 0.5em minus 0.4em\relax British Machine Vision
  Association, 2022.

\bibitem{younes2023catch}
A.~Younes, D.~Honerkamp, T.~Welschehold, and A.~Valada, ``Catch me if you hear
  me: Audio-visual navigation in complex unmapped environments with moving
  sounds,'' \emph{IEEE Robotics and Automation Letters}, vol.~8, no.~2, pp.
  928--935, 2023.

\bibitem{tatiya2022knowledge}
G.~Tatiya, J.~Francis, L.~Bondi, I.~Navarro, E.~Nyberg, J.~Sinapov, and J.~Oh,
  ``Knowledge-driven scene priors for semantic audio-visual embodied
  navigation,'' \emph{arXiv preprint arXiv:2212.11345}, 2022.

\bibitem{chen2021semantic}
C.~Chen, Z.~Al-Halah, and K.~Grauman, ``Semantic audio-visual navigation,'' in
  \emph{Proceedings of the IEEE/CVF Conference on Computer Vision and Pattern
  Recognition}, 2021, pp. 15\,516--15\,525.

\bibitem{YinfengICLR2022saavn}
Y.~Yu, W.~Huang, F.~Sun, C.~Chen, Y.~Wang, and X.~Liu, ``Sound adversarial
  audio-visual navigation,'' in \emph{The Tenth International Conference on
  Learning Representations}, 2022.

\bibitem{chen2024sim2real}
C.~Chen, J.~Ramos, A.~Tomar, and K.~Grauman, ``{Sim2Real} transfer for
  audio-visual navigation with frequency-adaptive acoustic field prediction,''
  in \emph{2024 IEEE/RSJ International Conference on Intelligent Robots and
  Systems (IROS)}.\hskip 1em plus 0.5em minus 0.4em\relax IEEE, 2024, pp.
  8595--8602.

\bibitem{chen2021waypoints}
C.~Chen, S.~Majumder, A.-H. Ziad, R.~Gao, S.~Kumar~Ramakrishnan, and
  K.~Grauman, ``Learning to set waypoints for audio-visual navigation,'' in
  \emph{The Ninth International Conference on Learning Representations}, 2021.

\bibitem{zhang2022stereo}
C.~Zhang, K.~Tian, B.~Ni, G.~Meng, B.~Fan, Z.~Zhang, and C.~Pan, ``Stereo depth
  estimation with echoes,'' in \emph{European Conference on Computer
  Vision}.\hskip 1em plus 0.5em minus 0.4em\relax Springer, 2022, pp. 496--513.

\bibitem{zhu2022beyond}
L.~Zhu, E.~Rahtu, and H.~Zhao, ``Beyond visual field of view: Perceiving {3D}
  environment with echoes and vision,'' \emph{arXiv preprint arXiv:2207.01136},
  2022.

\bibitem{parida2021beyond}
K.~K. Parida, S.~Srivastava, and G.~Sharma, ``Beyond image to depth: Improving
  depth prediction using echoes,'' in \emph{Proceedings of the IEEE/CVF
  Conference on Computer Vision and Pattern Recognition}, 2021, pp. 8268--8277.

\bibitem{yun2023dense}
H.~Yun, J.~Na, and G.~Kim, ``Dense {2D-3D} indoor prediction with sound via
  aligned cross-modal distillation,'' in \emph{Proceedings of the IEEE/CVF
  International Conference on Computer Vision}, 2023, pp. 7863--7872.

\bibitem{vasudevan2020semantic}
A.~B. Vasudevan, D.~Dai, and L.~Van~Gool, ``Semantic object prediction and
  spatial sound super-resolution with binaural sounds,'' in \emph{European
  Conference on Computer Vision}.\hskip 1em plus 0.5em minus 0.4em\relax
  Springer, 2020, pp. 638--655.

\bibitem{sokolov20243d}
A.~Sokolov, S.~Bhosale, and X.~Zhu, ``{3D} audio-visual segmentation,'' in
  \emph{NeurIPS 2024 Workshop on Audio Imagination}, 2024.

\bibitem{gao2020visualechoes}
R.~Gao, C.~Chen, Z.~Al-Halah, C.~Schissler, and K.~Grauman, ``{VisualEchoes}:
  Spatial image representation learning through echolocation,'' in
  \emph{Computer Vision--ECCV 2020: 16th European Conference}.\hskip 1em plus
  0.5em minus 0.4em\relax Springer, 2020, pp. 658--676.

\bibitem{roman2025generating}
A.~S. Roman, A.~Chang, G.~Meza, and I.~R. Roman, ``Generating diverse
  audio-visual 360 soundscapes for sound event localization and detection,''
  \emph{arXiv preprint arXiv:2504.02988}, 2025.

\bibitem{NEURIPS2023_8c234d9c}
X.~XU, D.~Markovic, J.~Sandakly, T.~Keebler, S.~Krenn, and A.~Richard,
  ``{Sounding Bodies:} modeling {3D} spatial sound of humans using body pose
  and audio,'' in \emph{Advances in Neural Information Processing Systems},
  vol.~36.\hskip 1em plus 0.5em minus 0.4em\relax Curran Associates, Inc.,
  2023, pp. 44\,740--44\,752.

\bibitem{li2020federated}
T.~Li, A.~K. Sahu, A.~Talwalkar, and V.~Smith, ``Federated learning:
  Challenges, methods, and future directions,'' \emph{IEEE Signal Processing
  Magazine}, vol.~37, no.~3, pp. 50--60, 2020.

\bibitem{zhang2023fedaudio}
T.~Zhang, T.~Feng, S.~Alam, S.~Lee, M.~Zhang, S.~S. Narayanan, and
  S.~Avestimehr, ``Fedaudio: A federated learning benchmark for audio tasks,''
  in \emph{ICASSP 2023-2023 IEEE International Conference on Acoustics, Speech
  and Signal Processing (ICASSP)}.\hskip 1em plus 0.5em minus 0.4em\relax IEEE,
  2023, pp. 1--5.

\bibitem{andreopoulou2014evaluating}
A.~Andreopoulou and A.~Roginska, ``Evaluating {HRTF} similarity through
  subjective assessments: Factors that can affect judgment,'' in
  \emph{Proceedings - 40th International Computer Music Conference, ICMC 2014
  and 11th Sound and Music Computing Conference, SMC 2014 - Music Technology
  Meets Philosophy}.\hskip 1em plus 0.5em minus 0.4em\relax National and
  Kapodistrian University of Athens, 2014, pp. 1375--1381.

\bibitem{kim2020investigation}
C.~Kim, V.~Lim, and L.~Picinali, ``Investigation into consistency of subjective
  and objective perceptual selection of non-individual head-related transfer
  functions,'' \emph{Journal of the Audio Engineering Society}, vol.~68,
  no.~11, pp. 819--831, 2020.

\bibitem{rusk2024comparing}
Z.~T. Rusk, M.~Neal, and M.~C. Vigeant, ``Comparing subjective similarity
  ratings and quantitative errors for the evaluation of free-field binaural
  panning techniques,'' \emph{The Journal of the Acoustical Society of
  America}, vol. 155, no. 3\_Supplement, pp. A215--A215, 2024.

\bibitem{cuevas20193d}
M.~Cuevas-Rodr{\'\i}guez, L.~Picinali, D.~Gonz{\'a}lez-Toledo, C.~Garre,
  E.~de~la Rubia-Cuestas, L.~Molina-Tanco, and A.~Reyes-Lecuona, ``{3D Tune-In
  Toolkit}: An open-source library for real-time binaural spatialisation,''
  \emph{PloS one}, vol.~14, no.~3, p. e0211899, 2019.

\bibitem{de2025data}
J.~A. De~Rus, M.~Montagud, J.~Lopez-Ballester, F.~J. Ferri, and M.~Cobos, ``A
  data-driven exploration of elevation cues in {HRTFs}: An explainable {AI}
  perspective across multiple datasets,'' \emph{arXiv preprint
  arXiv:2503.11312}, 2025.

\bibitem{ganin2016domain}
Y.~Ganin, E.~Ustinova, H.~Ajakan, P.~Germain, H.~Larochelle, F.~Laviolette,
  M.~March, and V.~Lempitsky, ``Domain-adversarial training of neural
  networks,'' \emph{Journal of Machine Learning Research}, vol.~17, no.~59, pp.
  1--35, 2016.

\bibitem{wilson2020survey}
G.~Wilson and D.~J. Cook, ``A survey of unsupervised deep domain adaptation,''
  \emph{ACM Transactions on Intelligent Systems and Technology (TIST)},
  vol.~11, no.~5, pp. 1--46, 2020.

\bibitem{salamon2017deep}
J.~Salamon and J.~P. Bello, ``Deep convolutional neural networks and data
  augmentation for environmental sound classification,'' \emph{IEEE Signal
  Processing Letters}, vol.~24, no.~3, pp. 279--283, 2017.

\bibitem{NIPS2014_f033ed80}
I.~J. Goodfellow, J.~Pouget-Abadie, M.~Mirza, B.~Xu, D.~Warde-Farley, S.~Ozair,
  A.~Courville, and Y.~Bengio, ``Generative adversarial nets,'' in
  \emph{Advances in Neural Information Processing Systems}, vol.~27.\hskip 1em
  plus 0.5em minus 0.4em\relax Curran Associates, Inc., 2014.

\bibitem{raissi2019physics}
M.~Raissi, P.~Perdikaris, and G.~E. Karniadakis, ``Physics-informed neural
  networks: A deep learning framework for solving forward and inverse problems
  involving nonlinear partial differential equations,'' \emph{Journal of
  Computational Physics}, vol. 378, pp. 686--707, 2019.

\bibitem{karniadakis2021physics}
G.~E. Karniadakis, I.~G. Kevrekidis, L.~Lu, P.~Perdikaris, S.~Wang, and
  L.~Yang, ``Physics-informed machine learning,'' \emph{Nature Reviews
  Physics}, vol.~3, no.~6, pp. 422--440, 2021.

\bibitem{nair2024physics}
S.~Nair, T.~F. Walsh, G.~Pickrell, and F.~Semperlotti, ``Physics and geometry
  informed neural operator network with application to acoustic scattering,''
  \emph{arXiv preprint arXiv:2406.03407}, 2024.

\bibitem{olivieri2024physics}
M.~Olivieri, X.~Karakonstantis, M.~Pezzoli, F.~Antonacci, A.~Sarti, and
  E.~Fernandez-Grande, ``Physics-informed neural network for volumetric sound
  field reconstruction of speech signals,'' \emph{EURASIP Journal on Audio,
  Speech, and Music Processing}, vol. 2024, no.~1, p.~42, 2024.

\bibitem{pezzoli2023implicit}
M.~Pezzoli, F.~Antonacci, and A.~Sarti, ``Implicit neural representation with
  physics-informed neural networks for the reconstruction of the early part of
  room impulse responses,'' in \emph{10th Convention of the European Acoustics
  Association}, 2023, pp. 2127--2184.

\bibitem{10819673}
S.~Koyama, J.~G.~C. Ribeiro, T.~Nakamura, N.~Ueno, and M.~Pezzoli,
  ``Physics-informed machine learning for sound field estimation: Fundamentals,
  state of the art, and challenges,'' \emph{IEEE Signal Processing Magazine},
  vol.~41, no.~6, pp. 60--71, 2024.

\bibitem{chen2025spatial}
W.~Chen and X.~Wei, ``Spatial interpolation of head-related transfer functions
  using a physics-informed autoencoder,'' \emph{Multimedia Systems}, vol.~31,
  no.~3, p. 247, 2025.

\bibitem{luan2025acoustic}
X.~Luan, K.~Yokota, and G.~Scavone, ``Acoustic field reconstruction in tubes
  via physics-informed neural networks,'' \emph{arXiv preprint
  arXiv:2505.12557}, 2025.

\bibitem{samek2017explainable}
W.~Samek, T.~Wiegand, and K.-R. M{\"u}ller, ``Explainable artificial
  intelligence: Understanding, visualizing and interpreting deep learning
  models,'' \emph{arXiv preprint arXiv:1708.08296}, 2017.

\bibitem{higgins2017beta}
I.~Higgins, L.~Matthey, A.~Pal, C.~Burgess, X.~Glorot, M.~Botvinick,
  S.~Mohamed, and A.~Lerchner, ``{beta-VAE:} learning basic visual concepts
  with a constrained variational framework,'' in \emph{International Conference
  on Learning Representations}, 2017.

\bibitem{hinton2015distilling}
G.~Hinton, O.~Vinyals, and J.~Dean, ``Distilling the knowledge in a neural
  network,'' \emph{arXiv preprint arXiv:1503.02531}, 2015.

\bibitem{NIPS2015_ae0eb3ee}
S.~Han, J.~Pool, J.~Tran, and W.~Dally, ``Learning both weights and connections
  for efficient neural network,'' in \emph{Advances in Neural Information
  Processing Systems}, vol.~28.\hskip 1em plus 0.5em minus 0.4em\relax Curran
  Associates, Inc., 2015.

\bibitem{salimans2022progressive}
T.~Salimans and J.~Ho, ``Progressive distillation for fast sampling of
  diffusion models,'' in \emph{International Conference on Learning
  Representations}, 2022.

\bibitem{song2020denoising}
J.~Song, C.~Meng, and S.~Ermon, ``Denoising diffusion implicit models,'' in
  \emph{International Conference on Learning Representations}, 2021.

\end{thebibliography}
\vspace{12pt}

\vfill

\end{document}